\documentclass[tighten, twocolumn, 12pt]{aastex62}
\def\apjl{{ApJ}}                
\def\apjs{{ApJS}}               
\def\ao{{Appl.~Opt.}}           
\def\aap{{A\&A}}                

\usepackage{listings}
\usepackage{color}
\usepackage{graphicx,times}
\usepackage{natbib}
\usepackage{amssymb}
\usepackage{newtxtext}
\usepackage[T1]{fontenc}
\usepackage{ae,aecompl}
\usepackage{threeparttable}
\usepackage{ulem}
\usepackage{txfonts}
\usepackage{amssymb}	

\def\kms  {km~s$^{-1}$}

\bibpunct{(}{)}{;}{a}{}{,}
\usepackage{hyperref}
\usepackage{longtable}
\usepackage{booktabs}
\definecolor{dkgreen}{rgb}{0,0.6,0}
\definecolor{gray}{rgb}{0.5,0.5,0.5}
\definecolor{mauve}{rgb}{0.58,0,0.82}
\definecolor{golden}{rgb}{0.86,0.65,0.01}

\usepackage{CJK}
\usepackage{graphicx}
\usepackage{hyperref}  
\begin{document}
\begin{CJK*}{UTF8}{gbsn}
%
%
%
\title[]{Survey for Distant Stellar Aggregates in Galactic Disk: Detecting Two Thousand Star Clusters and Candidates, along with the Dwarf Galaxy IC~10}
\correspondingauthor{Zhihong He}
\email{hezh@cwnu.edu.cn}
\author[0000-0002-6989-8192]{Zhihong He (何治宏)}
\affil{School of Physics and Astronomy, China West Normal University, No. 1 Shida Road, Nanchong 637002, China }
\author{Yangping Luo (罗杨平)}
\affil{School of Physics and Astronomy, China West Normal University, No. 1 Shida Road, Nanchong 637002, China }
\author{Kun Wang (王坤)}
\affil{School of Physics and Astronomy, China West Normal University, No. 1 Shida Road, Nanchong 637002, China }
\author{Anbing Ren (任安炳)}
\affil{School of Physics and Astronomy, China West Normal University, No. 1 Shida Road, Nanchong 637002, China }
\author{Liming Peng (彭黎明)}
\affil{School of Physics and Astronomy, China West Normal University, No. 1 Shida Road, Nanchong 637002, China }
\author{Qian Cui (崔倩)}
\affil{School of Physics and Astronomy, China West Normal University, No. 1 Shida Road, Nanchong 637002, China }
\author{Xiaochen Liu (刘效臣)}
\affil{School of Physics and Astronomy, China West Normal University, No. 1 Shida Road, Nanchong 637002, China }
\author{Qingquan Jiang (蒋青权)}
\affil{School of Physics and Astronomy, China West Normal University, No. 1 Shida Road, Nanchong 637002, China }
\vspace{10pt}


\begin{abstract}
Despite having data for over 10$^9$ stars from Gaia, only less than 10$^4$ star clusters and candidates have been discovered. Particularly, distant star clusters are rarely identified, due to the challenges posed by heavy extinction and great distance. However, Gaia data has continued to improve, enabling even fainter cluster members to be distinguished from field stars. In this work, we will introduce a star cluster search method based on the DBSCAN algorithm; we have made improvements to make it better suited for identifying clusters on dimmer and more distant stars. After removing member stars of known Gaia-based clusters, we have identified 2086 objects with |$b$|~<~10~deg, of which 1488 are highly reliable open star clusters, along with 569 candidates, 28 globular cluster candidates and 1 irregular galaxy IC~10 at low Galactic latitudes. We found that the proper motion of IC~10 is similar yet slightly different from the water maser observations, which is an important result for the comparison with Gaia and VLBA. Besides, when compared with the star clusters appearing in Gaia DR2/EDR3, we have found nearly three times as many new objects above a distance of 5~kpc, including hundreds of them above A$_v$ > 5~mag. And it has enabled us to detect a higher number of old clusters, over a billion years old, that are difficult to detect due to observational limitations. Our findings significantly expand the remote cluster sample and enhance our understanding of the limits of Gaia DR3 data in stellar aggregates research. The full figure set for 2085 clusters can be seen in \url{https://nadc.china-vo.org/res/r101258/}.
\end{abstract}
\keywords{Galaxy: stellar content - star clusters: general - surveys: Gaia}

\section{Introduction}\label{sec:intro}

Gaia is a billion-star surveyor that can achieve astrometric uncertainty of tens of micro-arcseconds~\citep[][]{Gaia16}. This makes it an invaluable tool for studying the Milky Way and stellar objects in it, especially open clusters (hereafter OCs)~\citep{Cantat22}. OCs provide accurate distance and kinematics information through the astrometric data of cluster members and a valuable wide age range (from 1~Myr to <~10~Gyr), which makes them ideal tracers for studying the Galactic structure and stellar evolution~\citep[e.g.][]{Kuhn19,cg20arm,Castro21}, as intended by the Gaia mission~\citep{Perryman01}.

In the past five years, researchers have been actively searching for undiscovered star clusters, with more than 5500 OCs and candidates detected using various methods after the second Gaia data release ~\citep[][DR2]{Gaia18-Brown}. Three-quarters of these are newly found based on Gaia data, leading to a wealth of research on Gaia OCs~\citep[e.g.][]{CG18,Liu19,Sim19,CG19-0,Castro19,Kounkel20,Castro20,Ferreira20,ferreira21,he21,he22a,hao22,Hunt21,castro22,he22b,Perren22,he23a}. However, not all possible clusters in Gaia data have been found, and recent reports show that new clusters are still being discovered~\citep[e.g.][]{hao22c,casado23,chi23,qin23,li2023,hunt23}. 

Moreover, most of the Gaia OCs and candidates found in the Galactic disk are within 5~kpc, with only $\sim$370 having a parallax lower than 0.2~mas based on Gaia DR2/EDR3~\citep[][EDR3]{gaia2021}. The primary limitation for finding distant OCs in Gaia data has been Galactic extinction, with most found OCs having an extinction value of less than A$_v$ = $\sim$3~mag~\citep[][hereafter CG20]{CG20}. OCs with an A$_v$ greater than 5~mag were considered not be easy to detected, which has led to less focus on the search for those OCs. This absence of distant findings makes it challenging to trace the Galactic structures beyond 5~kpc based on OCs~\citep[e.g.][]{CG18,he211}. However, although the optical observation limitations exist in the Galactic disk, we believe that the vast astrometric dataset in Gaia DR3 is still not being fully utilized. 

Based on Gaia data, the DBSCAN algorithm~\citep{Ester96} shows it is an efficient clustering method that has contributed significantly to new OC searches. Combined with the k$_{th}$ nearest neighbor distance (hereafter k$_{th}$NND) algorithm, the pioneering work of ~\citet[][]{Castro18} used them in Gaia data to identify 1214 new OCs and candidates in Gaia DR2 and EDR3~\citep{Castro20,castro22}. Inspired by ~\citet[][]{Castro18}, our previous research work used a two-Gaussian fit to k$_{th}$NND histogram to obtain the clustering coefficient and included linear velocity/distance in the clustering vector. This improved method is more efficient in OC searches, as it is insensitive to the variable parallax and has identified 2541 new OC/candidates~\citep{he21,he22a,he23a}. and 616 known nearby clusters~\citep[][]{he22b}. However, to study further regions, we need to use the angular scale and take narrower parallax cuts to reduce the impact of parallax errors.

For those OC objects present in Gaia data, ~\citet{CG20_0} presented an effective way to distinguish between a star cluster and an asterism based on the apparent radius and proper motion dispersions of cluster members. However, we found some clusters have low dispersion in astrometric data but poor  Color-Magnitude Diagrams (hereafter CMDs) and fewer members. We found the presence of possible asterisms in the clustering results is due to its nearby stellar aggregates or the surrounding dense star fields. To avoid such impacts, here we introduce an improved method called "Two-Gaussian Fitting for Isolated Groups" (TGFIG) to obtain a more accurate clustering coefficient and distinguish possible star clusters from field stars more clearly. In this method, we clip stars in dense fields such as existing clusters and unclustered dense regions, which reduces the pseudo-cluster signal in DBSCAN clustering.

This work focuses on finding new star clusters in the Galactic disk. In Section~\ref{sec:data}, we introduce the data and star cluster catalogs we used, and in Section~\ref{sec:method}, we outline our TGFIG method and search steps, including cluster search, cross-match, and isochrone fitting. In Section~\ref{sec:result}, we present the newly found results, including different types of objects and some interesting examples. Finally, we present our conclusions and prospects in Section~\ref{sec:summary}.

\section{Data}\label{sec:data}

\subsection{Gaia astrometric data}
Compared with Gaia DR2, the astrometric data in Gaia EDR3/DR3 are superior. For example, at G = 17 mag, the average uncertainty of parallax in Gaia DR3 is 0.07~mas, compared to 0.1~mas in Gaia DR2. Additionally, the proper motions in Gaia EDR3/DR3 have improved threefold, from 0.2 to 0.07 mas yr$^{-1}$.
It should be noted that for distant cluster searches, the improvement of parallax values may not change significantly when separating two distant stars. For example, comparing two stars at 1~kpc and 2~kpc, the parallax differential is 0.5~mas. For stars at 5~kpc and 10~kpc, the differential is only 0.1~mas, which approaches the mean parallax error for G~=~17 mag. Since farther clusters are usually fainter than nearby objects, the member stars on the main-sequence are mostly under 17 to 18~mag. Therefore, the new parallax determination may not improve significantly for distant cluster searches, particularly for old clusters. However, the increased accuracy of proper motion can improve comparisons between cluster members and field stars. Thus, some undiscovered faint/distant clusters in Gaia DR2 could be found in Gaia EDR3/DR3.

For the search for new star clusters, we utilized Gaia EDR3 data, taking advantage of the astrometric data $(l, b, \varpi, \mu_\alpha^\ast, \mu_\delta)$ to identify clusters and the photometric data (G, BP-RP) for isochrone fitting. To identify clusters located in the Galactic disk, we selected sources with |$b$|~<~10~deg. Once we had identified all clusters, we cross-matched member stars with Gaia DR3 data to obtain radial velocity values. The cross-matching radius was 0.1 arc-second. 

Unlike in most cluster searches, we did not impose an apparent magnitude cut here to include the entire database and to identify fainter and more distant stars. However, researchers can impose criteria for each member in the final tables. It should be noted that for some red faint sources, their magnitude may be overestimated, especially in the BP band~\citep{Riello21}. Therefore, when doing the isochrone fits, we imposed a magnitude cut of 19.2~mag, and we carefully checked each result. For input data near $l$ = 0~deg, we correctly accounted for the transition from 360 to 0 deg near the Galactic centre.
\subsection{Cluster catalogs}\label{sec:catalog}

To build a catalog of new star clusters and to reduce the impact of high-density regions on clustering parameters~(see Section~\ref{sec:TGFIG}), we needed to eliminate all known cluster (and candidate) members (Section~\ref{sec:TGFIG}), so we relied on previously reported OCs and globular cluster (hereafter GC) catalogs, including the works of ~\citet[][]{CG20,Liu19,Ferreira19,Ferreira20,ferreira21,Qin20,he21,he22a,he22b,Hunt21,Casado21,Jaehnig21, li22,hao22,castro22,he23a,Vasiliev21}. 

However, not all member stars of clusters have been reported, so we used dispersion (sigma) measures (e.g. CG20, He22) in $(l,b, \varpi, \mu_\alpha^\ast, \mu_\delta)$  or standard deviations to extract possible cluster members, removing stars within 5-sigma range of $(l, b)$ and 3-sigma range of $(\varpi, \mu_\alpha^\ast, \mu_\delta)$ from all cataloged objects. For catalogs with no dispersion measures or member star information, we used the median dispersion depending on the parallax of those clusters. Specifically, we selected clusters with $\varpi$ $\pm$~0.2~mas in CG20 and calculated the median dispersion of those clusters as the clip range. To extract most GC members, we used the fixed dispersion range of (0.1~deg, 0.2~mas, 0.2~mas~yr$^{-1}$ ).

Although most cluster members were clipped, these cuts may neglect clusters' tidal tails, stellar streams, or some extended structures. During the preparation of this work, several new cluster catalogs were published~\citep[][]{hao22c,chi23,qin23}, and we were unable to clip those cluster members in our search procedures. However, we also cross-matched those cluster catalogs after the search. In the days leading up to our submission, ~\citet[][hereafter HR23]{hunt23} reported a catalog of 7200 cluster objects, with 4114 of these classified as reliable open clusters. We also cross-matched the results with HR23 and present an online table with matched cluster identifications.

Additionally, we checked a total list of 129 GC candidates (hereafter GCCs) found in the last two decades that were not cataloged in ~\citet{Vasiliev21}, including ~\citet{Willman2005,Belokurov2010,Minniti2011,Munoz2012,Laevens2014,Laevens2015,Minniti2017A,Minniti2017,Ryu2018,Camargo2018,Barb2019,Palma2019,Camargo2019,Barba2019, Villanova2019,Bellazzini2020,Garro2020,Obasi2021,Minniti2021,Minniti2021b,Gatto2021,Buzzo2021,Garro2021,Dias2022,Gran2022,Garro2022}. Although all of those clusters were not found depending only on Gaia data, some member stars of those GCCs could be detected in Gaia data. Nonetheless, only a few of those GCCs have parallax and proper motion values, so we checked them through CMDs and positions after the search. In the end, we also cross-matched Gaia-based cluster catalog~\citet{dias21}, pre-Gaia clusters in ~\citet{Dias02,Kharchenko13} and present the newly identified pre-Gaia OCs in this work.

\section{Method}\label{sec:method}

\subsection{The ~\textbf{TGFIG} method}\label{sec:TGFIG}
The main steps of the TGFIG method were introduced in ~\citet{he21,he23a}, but we have made improvements to make it suitable for distant OC searches. The method is based on the clustering algorithm DBSCAN. In an input data vector, such as Gaia astrometric data $(l, b, \varpi, \mu_\alpha^\ast, \mu_\delta)$, the algorithm calculates the distances between different points in the vector, and the number of neighbors $n_{point}$ within radius $\varepsilon$ is calculated. Based on two input parameters ($\varepsilon$, $MinPts$), which are the radius and minimum number, respectively, each point of the vector is labeled in three subsets: core member ($n_{point} \geqslant MinPts$), outer member ($n_{point} < MinPts$, but the point is a neighbor of a core member), or noise ($n_{point} < MinPts$, and it is not a neighbor for any core member). Here, we improved our previous searches, since some of the steps came from our previous studies, but to search for distant clusters, we made changes to the method.

\begin{figure*}
\begin{center}
\includegraphics[width=0.8\linewidth]{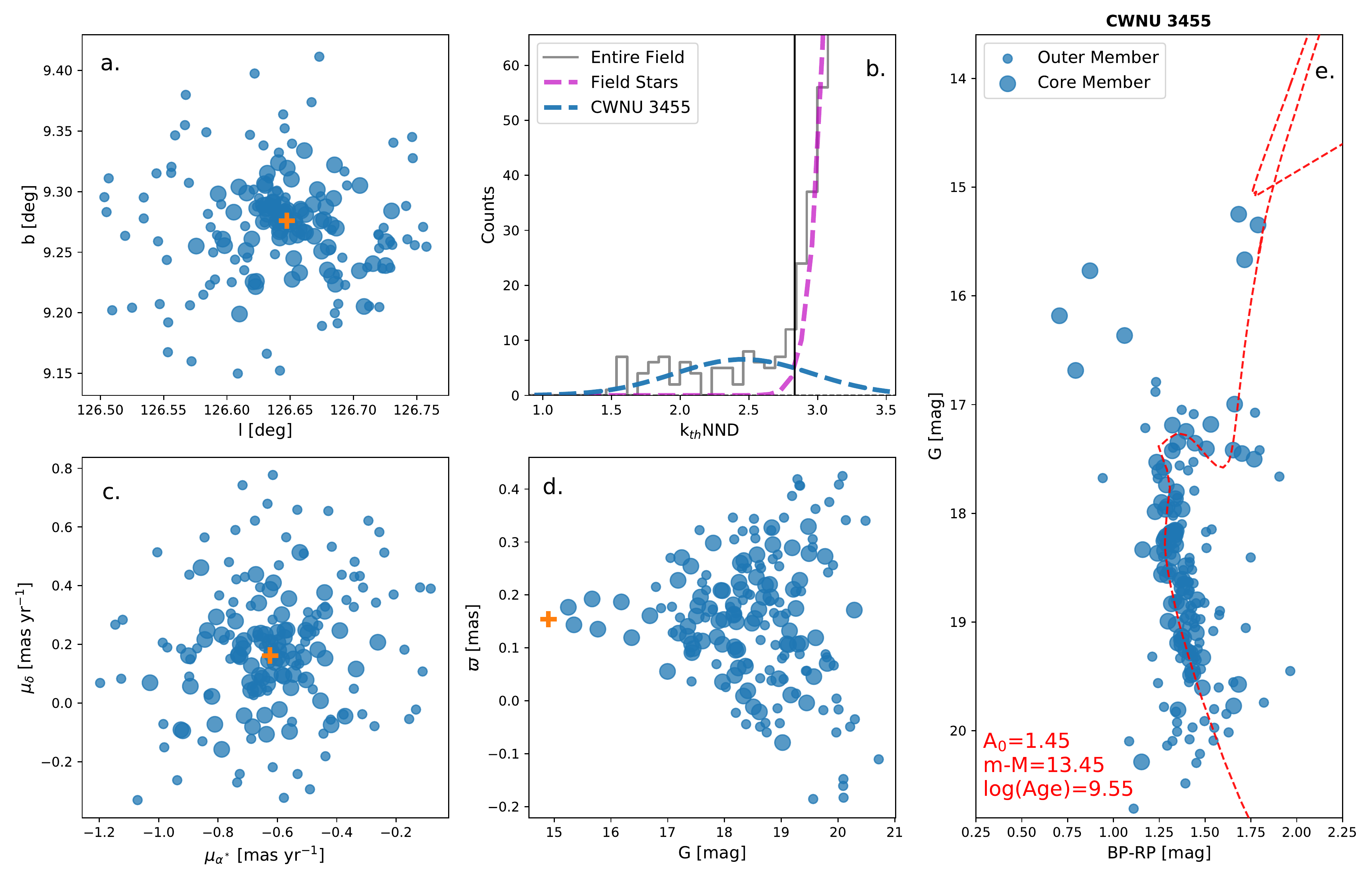}
\caption{Example of two-Gaussian fitting in k$_{th}$NND histogram and the cluster result. CWNU~3455 is detected as a reliable cluster with bound member stars (a.); the k$_{th}$NND histogram fittings shows clearly separate between field stars and member stars of the cluster (b.), along with its low dispersion proper motions (c.), magnitude-parallax distribution (d.), and clearly CMD with reliable isochrone fitting (e.). The cluster also may have many blue straggler stars (hereafter BSSs), that could makes it more interesting for researchers. Due to overestimation of BP band photometry~\citep{Riello21} in the fainter end of the CMD (especially for G < 19 to 19.5~mag), the CMD shifts to the bluer side, resulting in unreliable parameters. Therefore, we excluded G > 19.2~mag members when ran the systematic isochrone fits (Section~\ref{sec:isochrone}).}
\label{fig:kds}
\end{center}
\end{figure*}

Inspired by the work of ~\citet{Castro18}, which adopts k$_{th}$NND histograms to help get the $\varepsilon$ values, we conducted a two-Gaussian fits method to get the $\varepsilon$ values in different input vectors. In our studies, we considered that when a bound star cluster exists in the search region, the k$_{th}$NND of the vector could be divided into two approximate Gaussian distributions, as shown in Figure~\ref{fig:kds}. We use two Gaussian curves to fit such a signature:
\begin{equation}\label{eq2}
N_{kNND}=\sum_{i=1}^{2}a_i \cdot e^{\frac{-(\mathbf{d}_{k,nn}-\mu_i)^2}{2\sigma_i^2}}
\end{equation}
  \begin{equation}\label{eq3}
D_{ik}=\sqrt{(x_{i1}-x_{k1})^2+(x_{i2}-x_{k2})^2+...+(x_{i5}-x_{k5})^2}
\end{equation}
where $(a, \mu, \sigma)_{i/j}$ are the parameter of the Gaussian function, $D_{ik}$ is the distance between data points $\bf{x}_i$ and $\bf{x}_k$. and the possible cluster will have a lower k$_{th}$NND in the histogram, and once the fits have a real solution, it may catch a cluster signature. Such an estimate depends on the physically bound groups, so it is more reliable when $\varepsilon$ is in the right profile of the Gaussian curve~1.

However, for the real situation in OC searches, if there are some huge cluster/stream/large asterism/dense stellar regions in the input dataset, it may increase the $\varepsilon$ value and cause the plausible signal for some unbound stellar groups, which may lead to unreal cluster candidates in the result. To reduce the effects of the problem, in this work, we take two-fold clipping procedures and clustering steps in the searches.

Firstly, we clipped all known clusters in the search field and ran the clustering. For each once-detected group, we recorded it as a possible target group. 
As the star clusters recorded in the articles are not the complete lists, we then re-clustered it (target group) in a 100 to 500~pc region in $(l, b)$, at least 0.5~deg. In the 2nd clustering, we clipped other once-detected groups.
Secondly, considering the field star contamination, especially for brighter star contamination to the further OCs, we clipped all stars beyond 5 times of the uncertainty in ($\varpi_{0} \pm \sigma_{\varpi},  \mu_{0} \pm \sigma_{\mu}$), where ($\varpi_{0}, \mu_{0}$) are the median astrometric values, $\sigma_{\varpi}$, $\sigma_{\mu}$ are parallax and proper motion dispersions of the target group. After these clips, the densest data points are the target group in the input dataset, which will not be affected by any other dense groups, and the contamination could be reduced to the least level.


In this study, we divided the Galactic disk into 6480 3 $\times$ 3~deg ($l_0 \pm 1.5$~deg, $b_0 \pm 1.5$~deg) cells under Galactic coordinates, where ($l_0,b_0$) represents the center of the cell. For each cell, there is an overlap of $\pm$ 0.5~deg. Within each cell, we applied the parallax cut of 0.2~mas, with a step of 0.1~mas. We considered the results within ($l_0 \pm 1.25$~deg, $b_0 \pm 1.25$~deg, $\varpi_0 \pm 0.1$~mas) in each cell, with a (0.25~deg, 0.05~mas) offset to ensure that the results stayed within the boundary. We used the 3-sigma range in $(l, b, \varpi, \mu_\alpha^\ast, \mu_\delta)$ to cross-match the results and extract the duplicate results in the boundary. We applied the DBSCAN clustering procedure on each field with 21 different $MinPts$ values (6 to 25) and extracted tiny groups with a cut of group numbers (N$_{core} \geq 5$ | N$_{all} \geq 30$) in each step. We selected the results with the maximum N$_{core}$ as the final result to ensure unbiased $MinPts$ values. 

In contrast to our previous works, we used $(l, b, \varpi, \mu_\alpha^\ast, \mu_\delta)$ as input data as $(d\cdot\sin\theta_{l}\cdot\cos b,d\cdot\sin\theta_b,d\cdot\mu_{\alpha^*},d\cdot\mu_{\delta},\varpi)$ is only suitable for nearby clusters, which limits distant cluster searches. However, liner distance/velocity remains valuable for nearby cluster searches. It is noteworthy that the core member of the result is mostly located in the core part of the vector, and the outer member also has some large uncertainties but may still be member stars, particularly for faint stars down to $\sim$ 20~mag. This is limited by their low astrometric accuracy and will be better evaluated in future Gaia data releases.

\subsection{Visual inspection}\label{sec:visual}

To supplement the above automatic steps, we performed visual inspections on all groups. We kept a few un-redetected groups as the final result if they had bound stars and reliable CMDs, which were not detected in the second DBSCAN either due to the dense search fields or the field range (100 to 500~pc) being insufficient to compare them with field stars. Besides, we cross-matched all resulted clusters with GC candidates and OC catalogs, visually checked all cross-matches, and extracted some substructures~\citep[as described in][]{Piatti23} in the study. 

Moreover, based on our visual inspection that considered k$_{th}$NND histogram Gaussian fits, positions, and CMDs, we removed some ($\sim$200) results those with unbound shapes and unreliable CMDs. We kept those clusters that showed separate structures in k$_{th}$NND histograms and the main sequence is clearly shown in the core members, even though some member stars, particularly the outer members of the cluster, were not on the evolution sequence. For some clusters, their space distribution $(l, b)$ was not bound or their CMDs showed heavy field star contamination, but they were still separated in the k$_{th}$NND histogram fits, and we kept them as candidate OC objects (see Section~\ref{sec:newoc}).

\subsection{isochrone fitting}\label{sec:isochrone}

For each OC or candidate, we performed isochrone fitting by first automatically fitting the data using our previous methods as described in ~\citet{he21,he23a}. We used the function:~\footnote{This function is the same as the one used in ~\citet{Liu19}. However, we independently produced the python fitting codes, which will be accessible to anyone interested in using them.}
\begin{equation}\label{eqage1}
\mathbf{x}_k = (G + m - M, BP-RP)_{k}
\end{equation} 

\begin{equation}\label{eqage2}
\mathbf{x}_{kN} = [G_0 + c_G \cdot A_0, (BP-RP)_0 + (c_{BP_0}-c_{RP_0}) \cdot A_0]_{kN}
\end{equation}

\begin{equation}\label{eqage3}
\bar{d^2}= \frac{\sum_{k=1}^{n}(\mathbf{x}_k-\mathbf{x}_{kN})^2}{n}
\end{equation}
where $c$ is the extinction coefficient, $\mathbf{x}_k$ is the photometric data of the $k_{th}$ member star,  $\mathbf{x}_{kN}$ is the nearest star to $\mathbf{x}_k$ in the theoretical isochrone,  A$_0$ is V band extinction (A$_v$, hereafter we use A$_0$ instead) and $\bar{d^2}$ is the average value of $n$ input $\mathbf{x}_k$ vectors. 

We fit the Gaia EDR3 band~\footnote{\url{https://www.cosmos.esa.int/web/gaia/edr3-passbands}} \footnote{\url{http://stev.oapd.inaf.it/cgi-bin/cmd_3.6}} theoretical isochrone lines~\citep{Bressan12} to the CMD of each cluster. The isochrone lines' age range was [6.0, 10.1]~dex, with a step of 0.05~dex. We should note that, as we did in ~\citet{he23a}, we neglected the Galactic metallicity gradient~\citep{Magrini09} and used the solar metallicity as a fixed isochrone parameter. This may only have a minor impact on the fitting process~\citep{Salaris04,cg20arm}, and a more improved fitting that considering metallicity would be come out in the future. Besides, We determined the range of the distance modulus by median parallax $\pm$ 0.1~mas of the member stars, with a step of 0.05. We also used an extinction A$_0$ step of 0.05~mag, from 0.05 to 15~mag. The extinction coefficient was derived from the polynomial function:
$c = c_1+c_2 \ast bp\_rp_0+c_3 \ast bp\_rp_0^2+c_4  \ast bp\_rp_0^3+c_5 \ast A_0+
c6 \ast A_0^2+c_7  \ast A_0^3+c_8  \ast bp\_rp_0  \ast A_0+ c_9  \ast A_0  \ast bp\_rp_0^2+
c_{10}  \ast bp\_rp_0  \ast A_0^2$
, here $c_1$ to $c_{10}$ values were adopted from the public auxiliary data provided by ESA/Gaia/DPAC/CU5 and prepared by Carine Babusiaux.

We then carefully checked each result visually and manually removed stars that were obviously far from the evolutionary sequence, such as stragglers showing up in the CMD (see Section~\ref{sec:comparison}). Note that we only removed these stars from the isochrone fits, but we still kept them in the member star list. Besides, for poor fits with low core-membership, we used all member stars in order to obtain valuable fits.

\section{Results and Discussion}\label{sec:result}

\subsection{Genaral}
\begin{figure*}
\begin{center}
\includegraphics[width=1.0\linewidth]{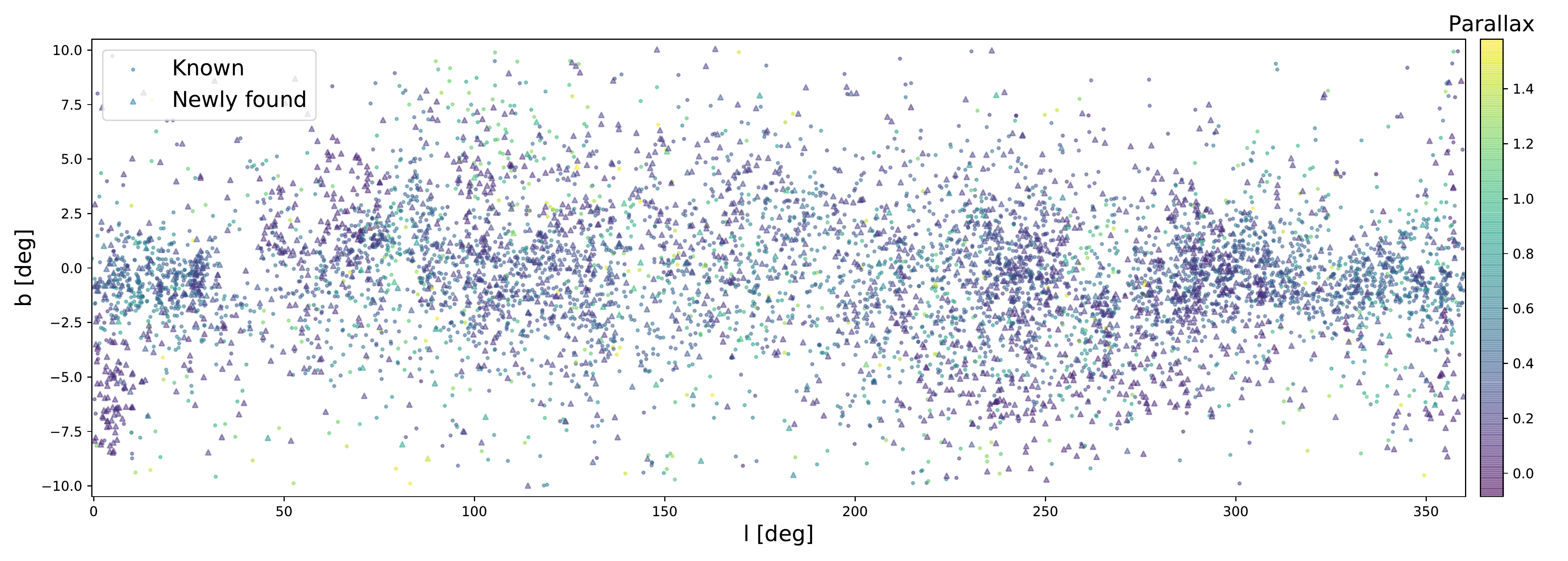}
\caption{Locations of all newly detected clusters (triangles) and all reported Gaia-based star clusters (dots) in Galactic coordinates with an absolute latitude of less than 10~deg. The color of the symbols represents the parallax value of the objects.}
\label{fig:lb}
\end{center}
\end{figure*}

\begin{figure*}
\begin{center}
\includegraphics[width=0.245\linewidth]{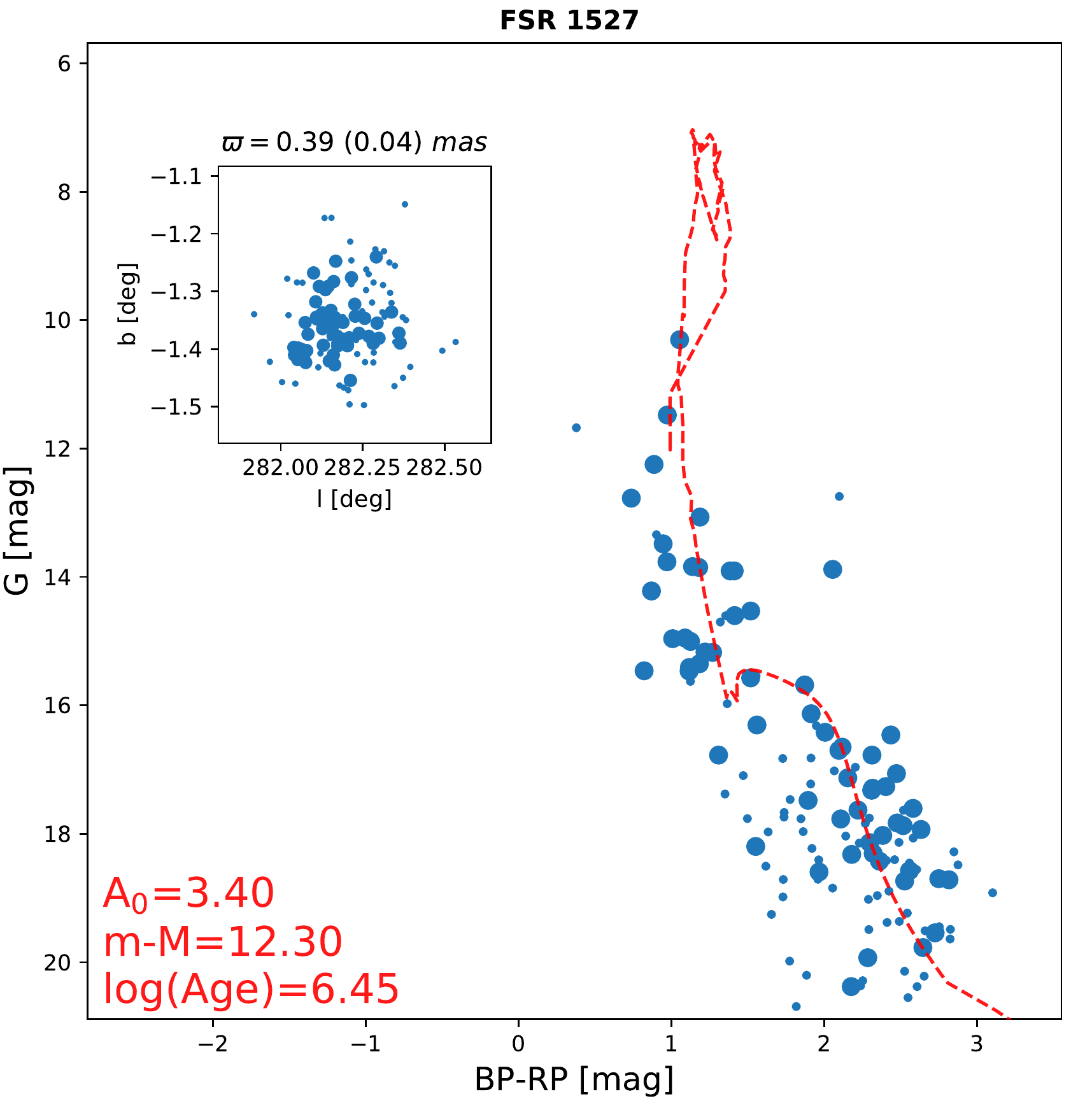}
\includegraphics[width=0.245\linewidth]{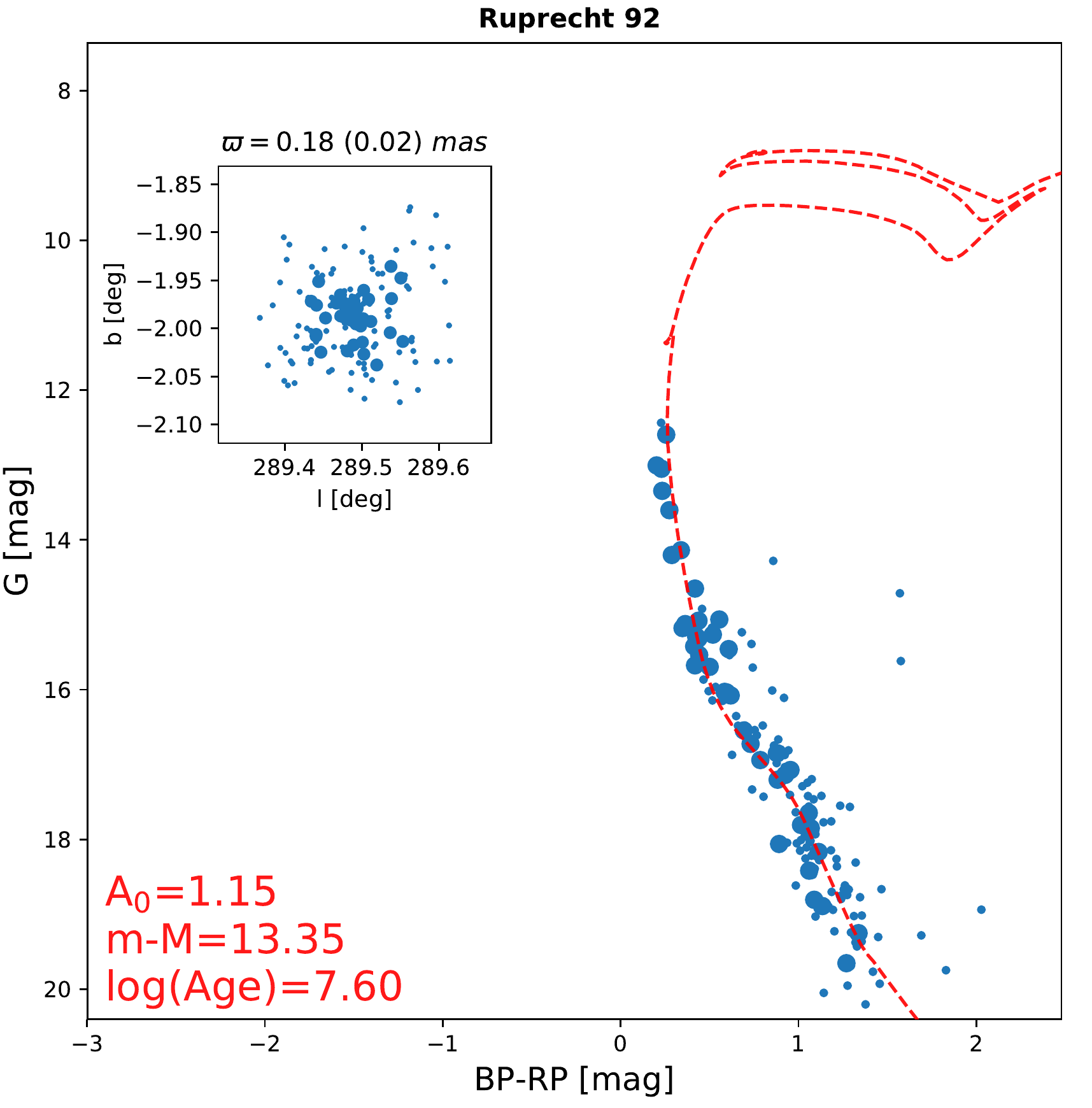}
\includegraphics[width=0.245\linewidth]{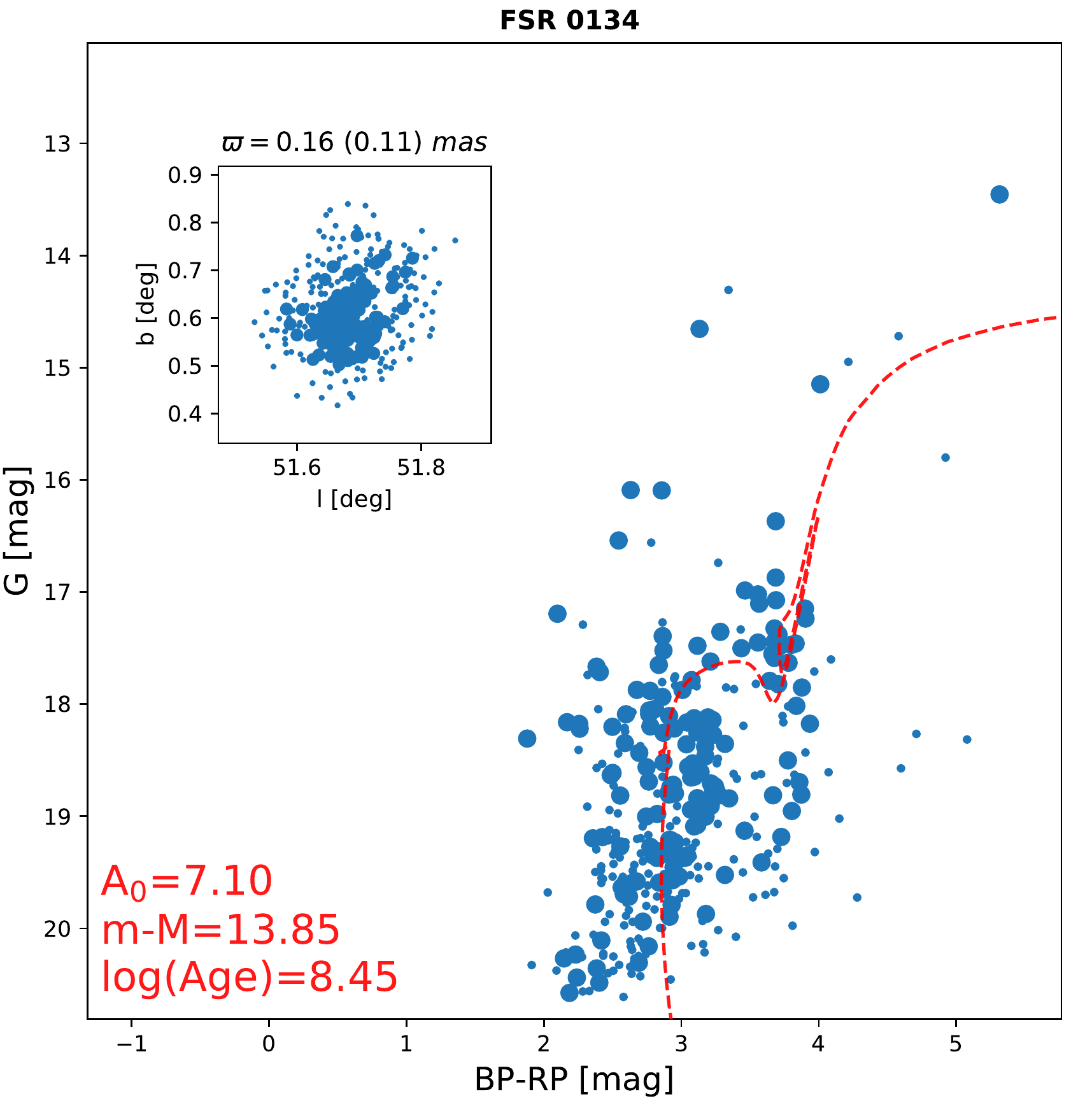}
\includegraphics[width=0.245\linewidth]{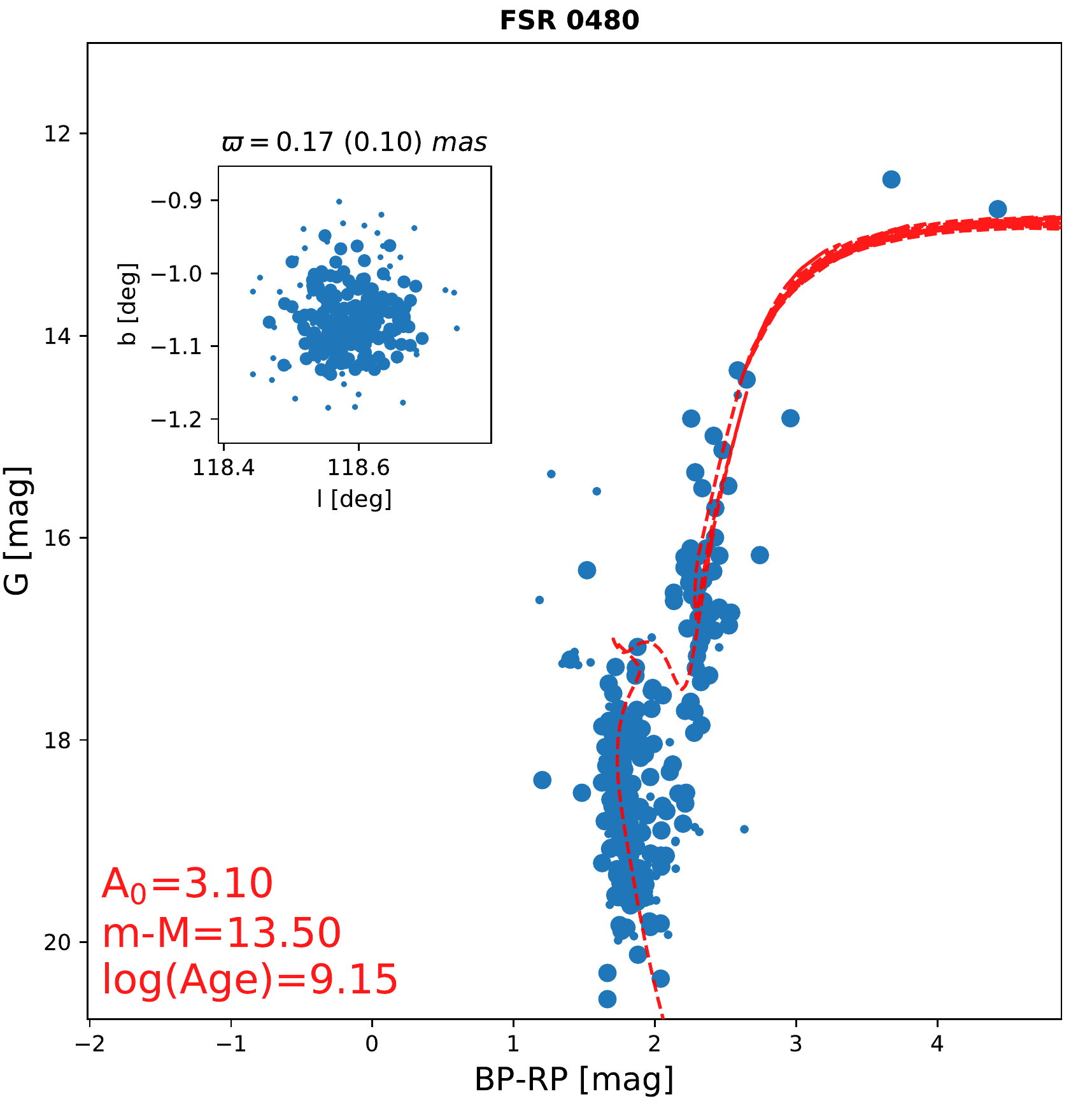}
\includegraphics[width=0.245\linewidth]{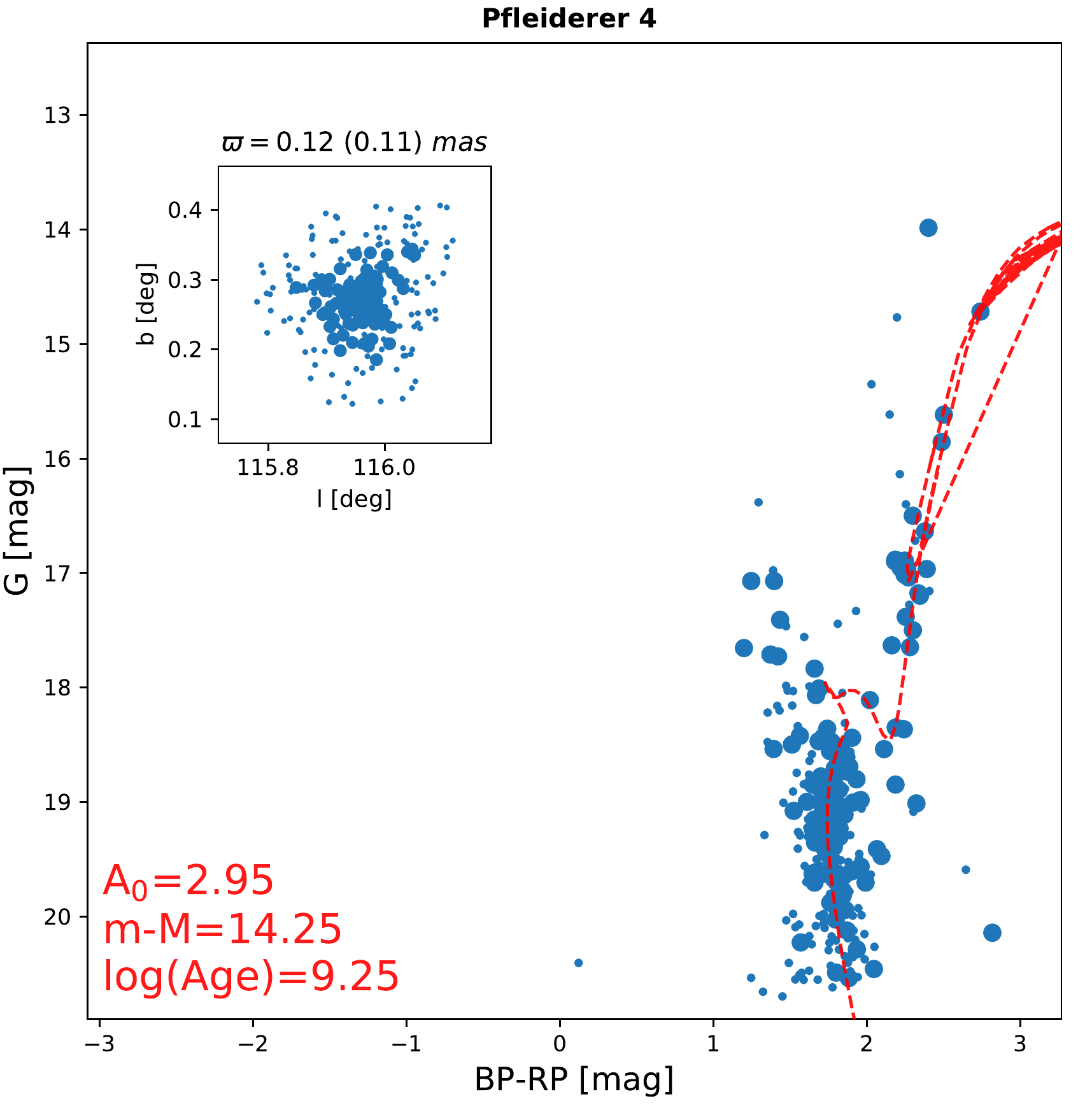}
\includegraphics[width=0.245\linewidth]{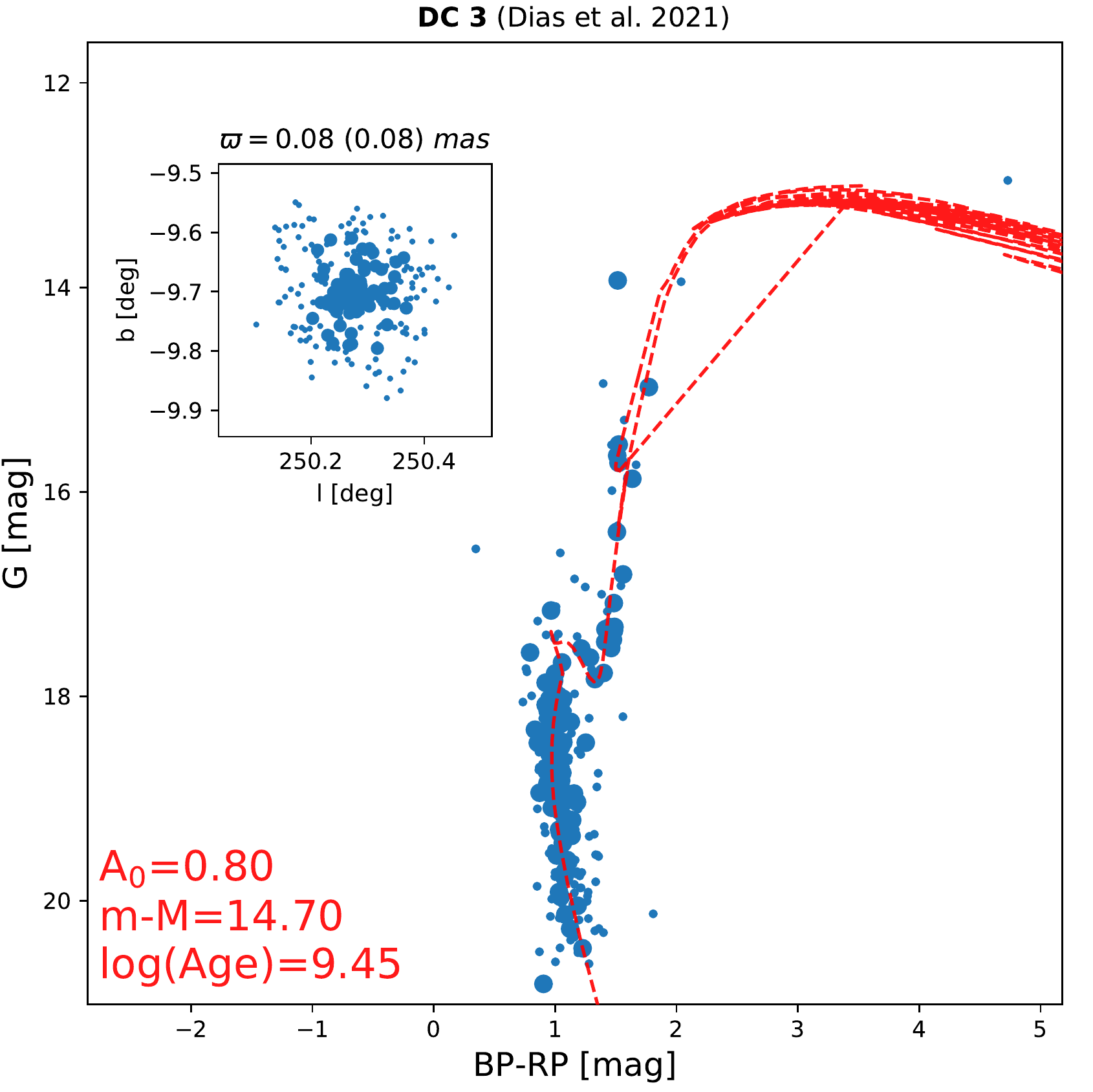}	
\includegraphics[width=0.245\linewidth]{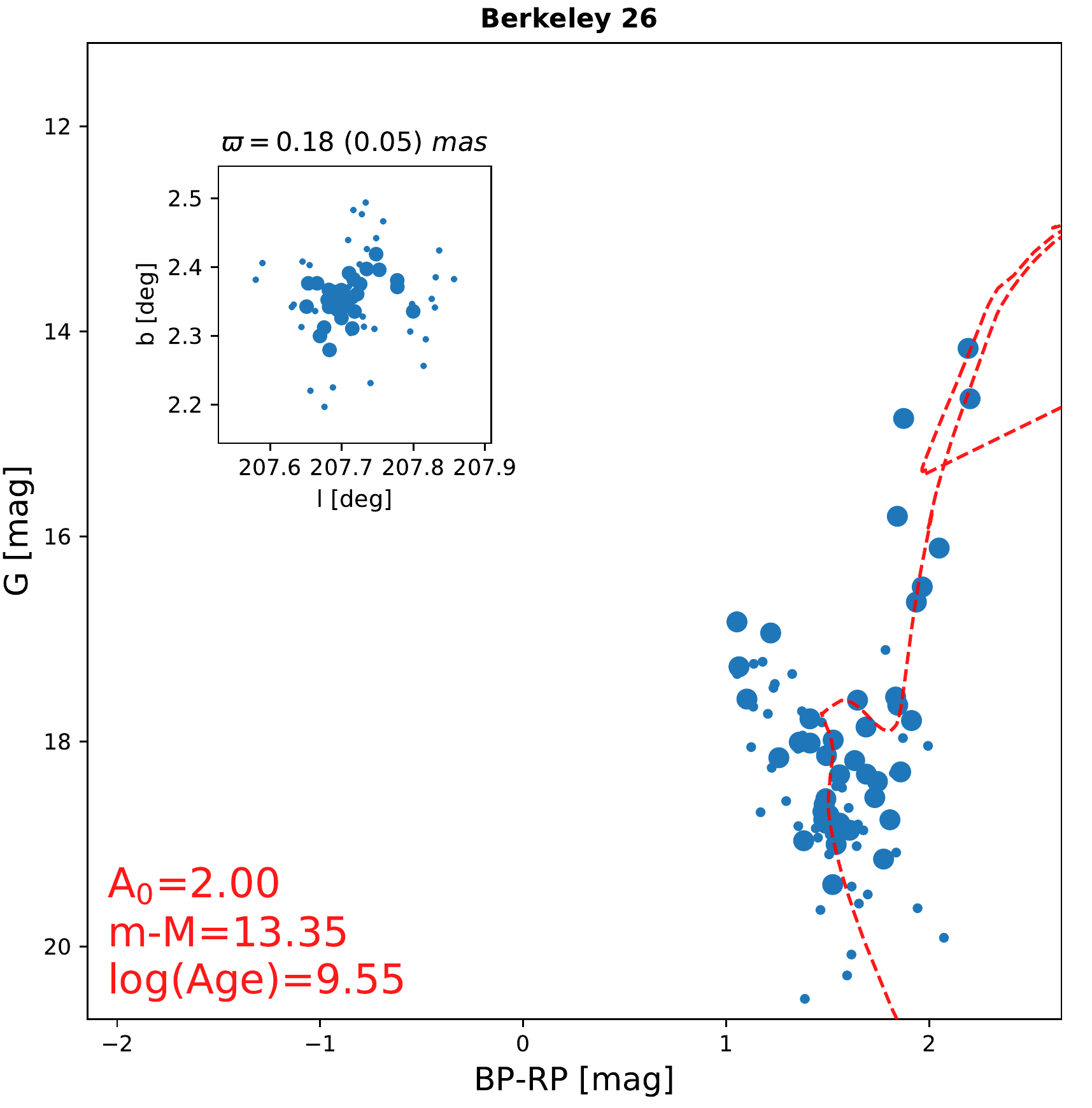}
\includegraphics[width=0.245\linewidth]{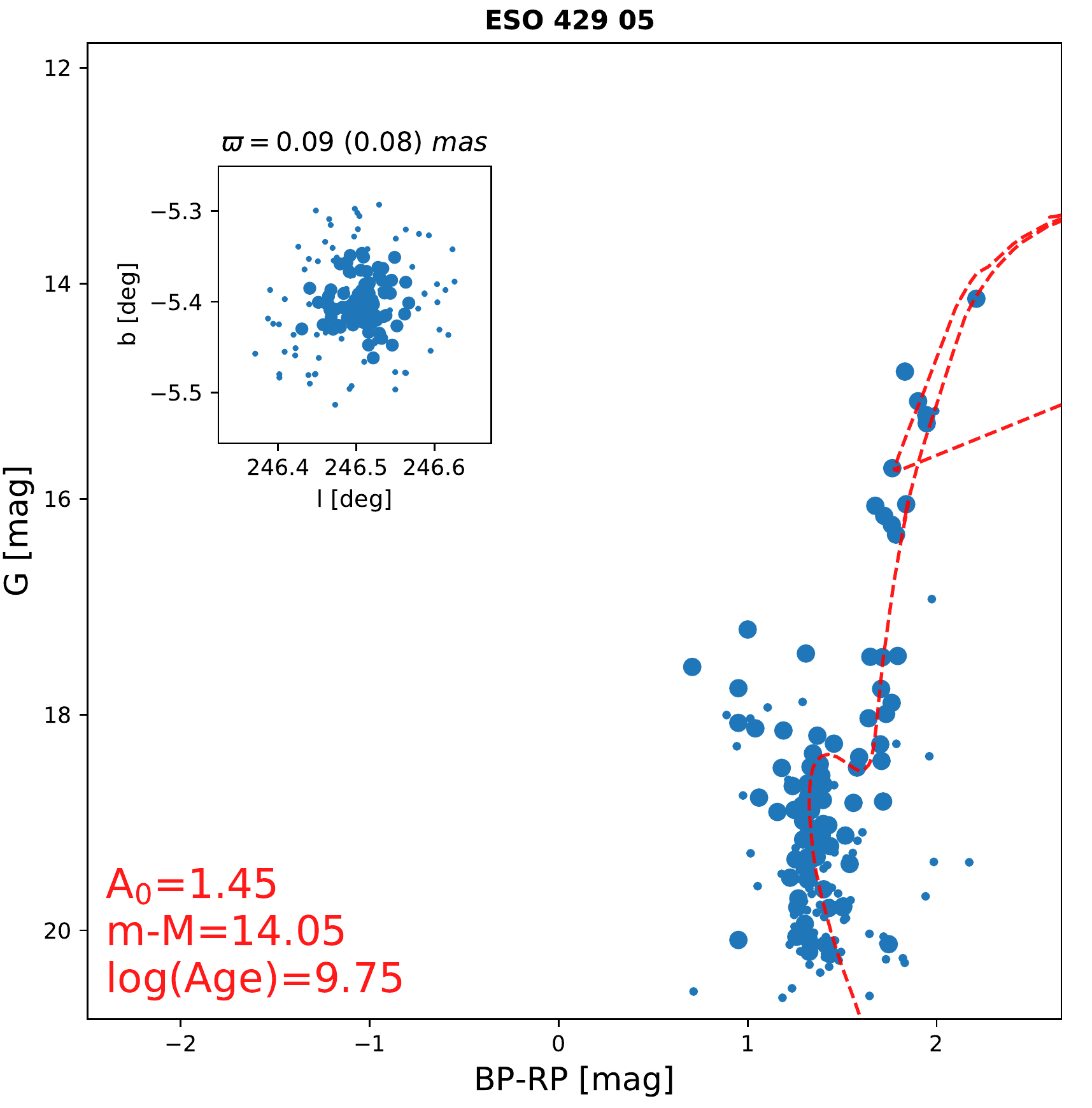}				
\caption{Examples of color-magnitude diagrams with overlaid isochrones for reliable pre-Gaia OCs are shown in this figure, depicting their increasing ages. The subplot represents the position of the clusters, displaying the core and outer member of the cluster as big and small dots, respectively. Additionally, the median and dispersion values of the parallax for the member stars are provided.}
\label{fig:pre_gaia_ocs}
\end{center}
\end{figure*}

\begin{figure*}
\begin{center}	
\includegraphics[width=0.245\linewidth]{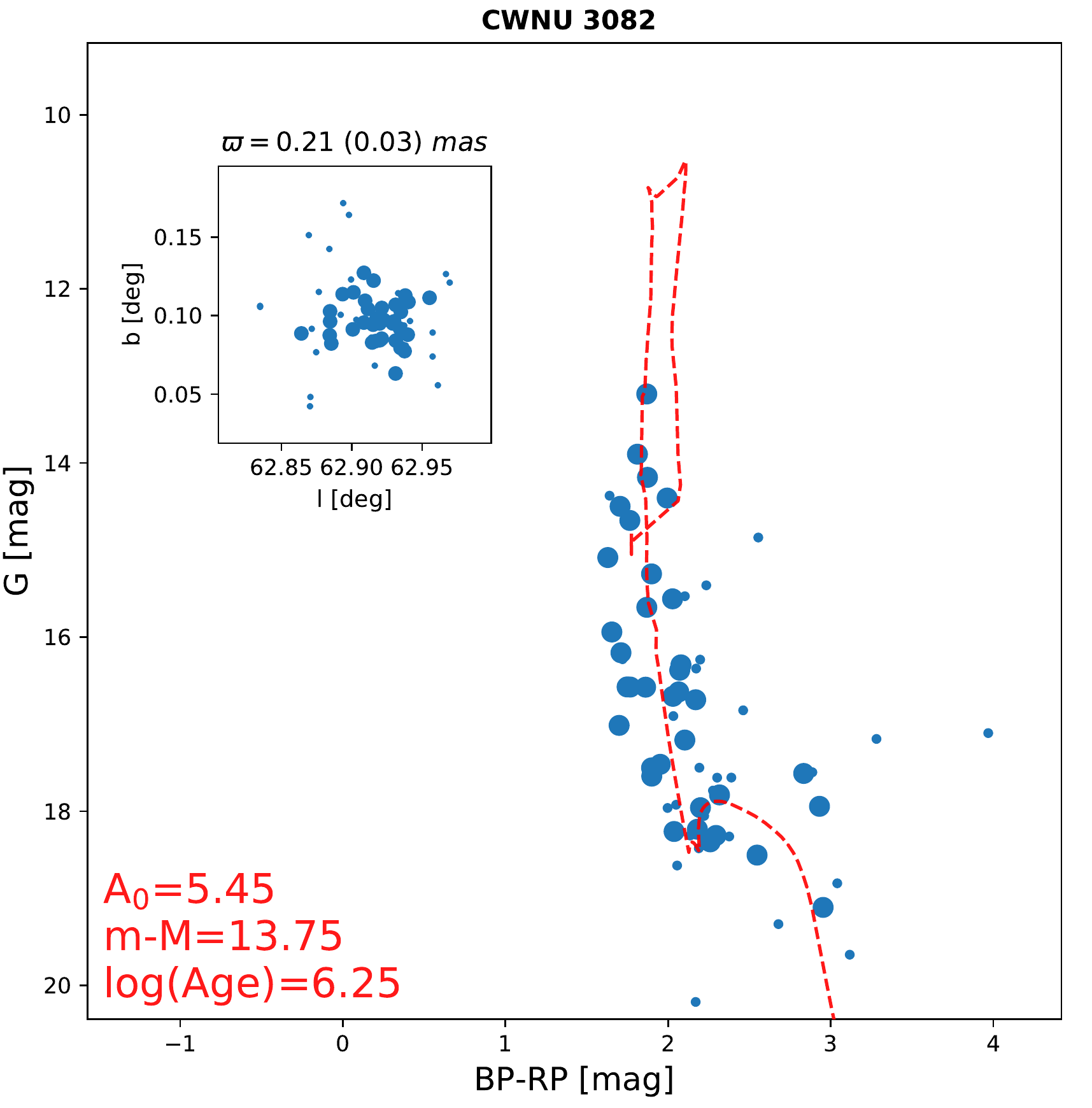}	
\includegraphics[width=0.245\linewidth]{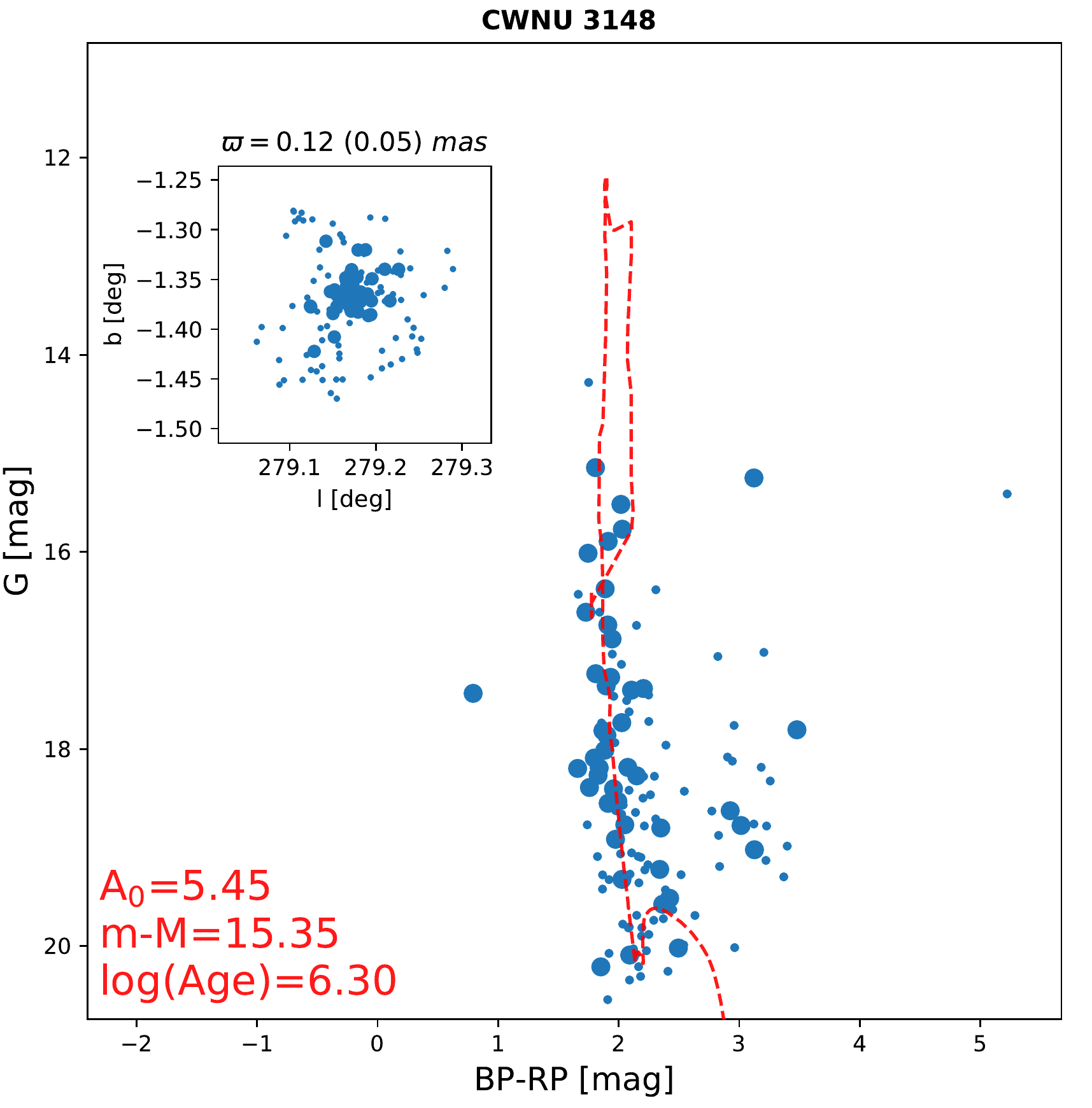}
\includegraphics[width=0.245\linewidth]{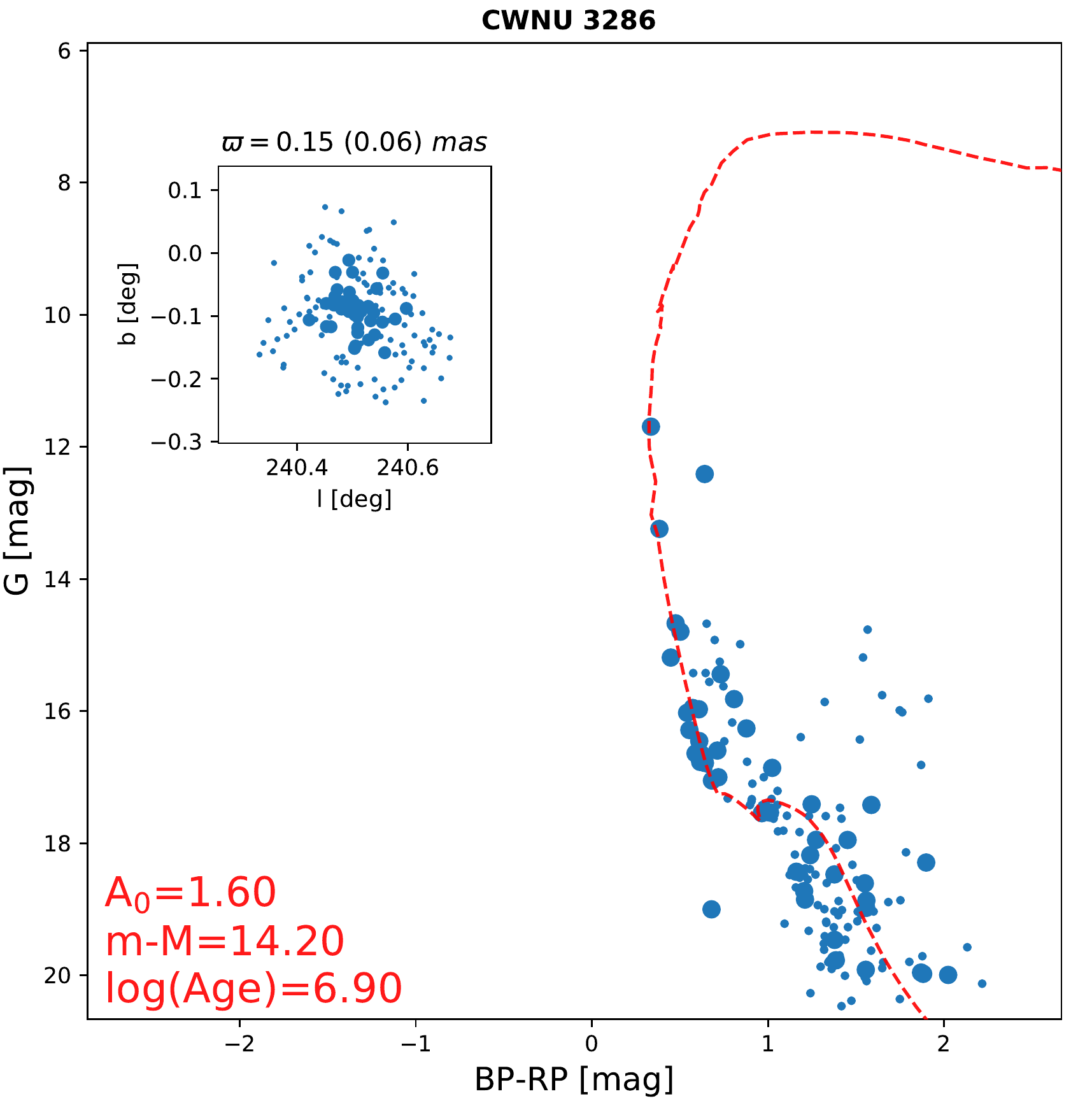}
\includegraphics[width=0.245\linewidth]{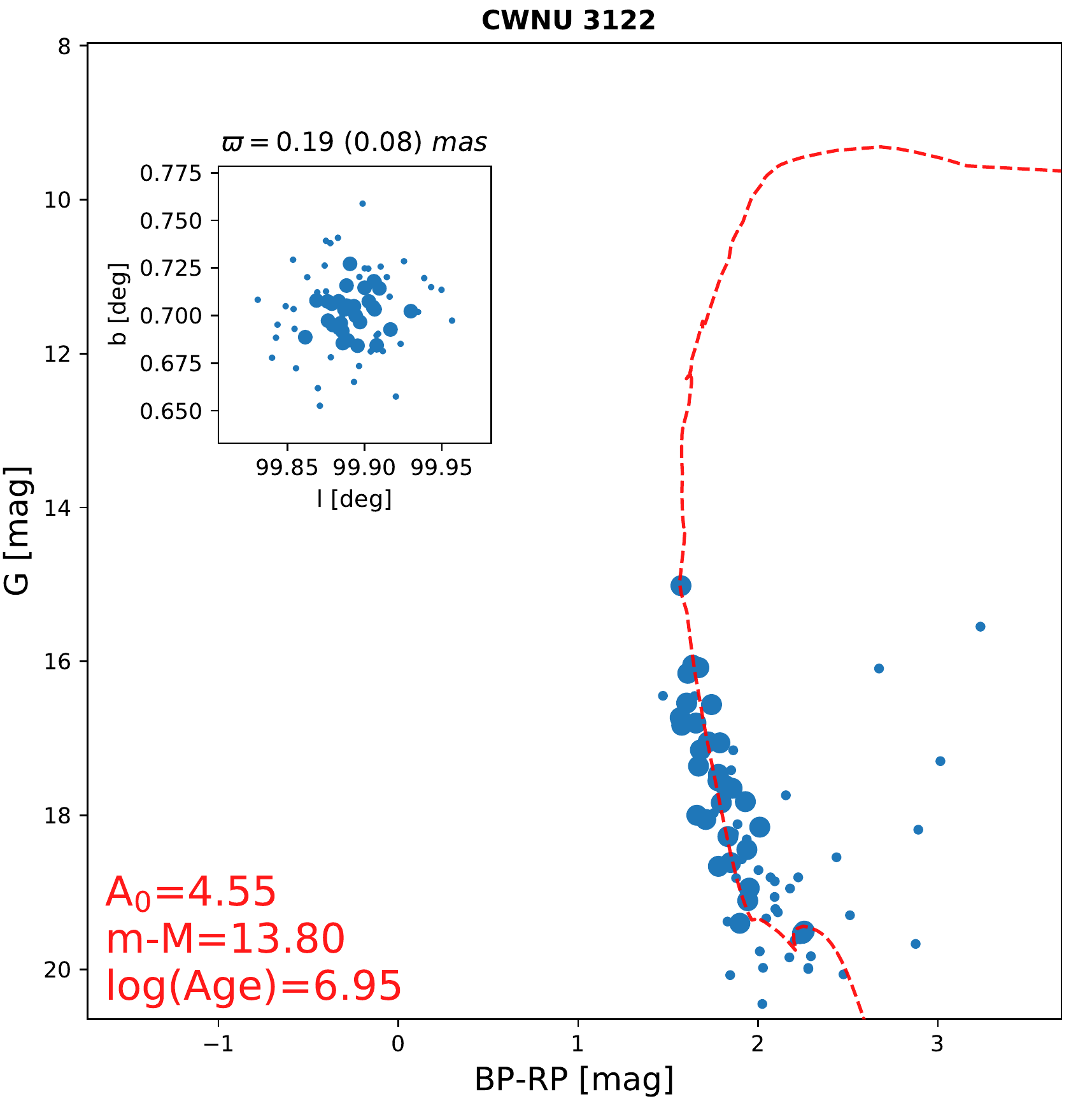}
\includegraphics[width=0.245\linewidth]{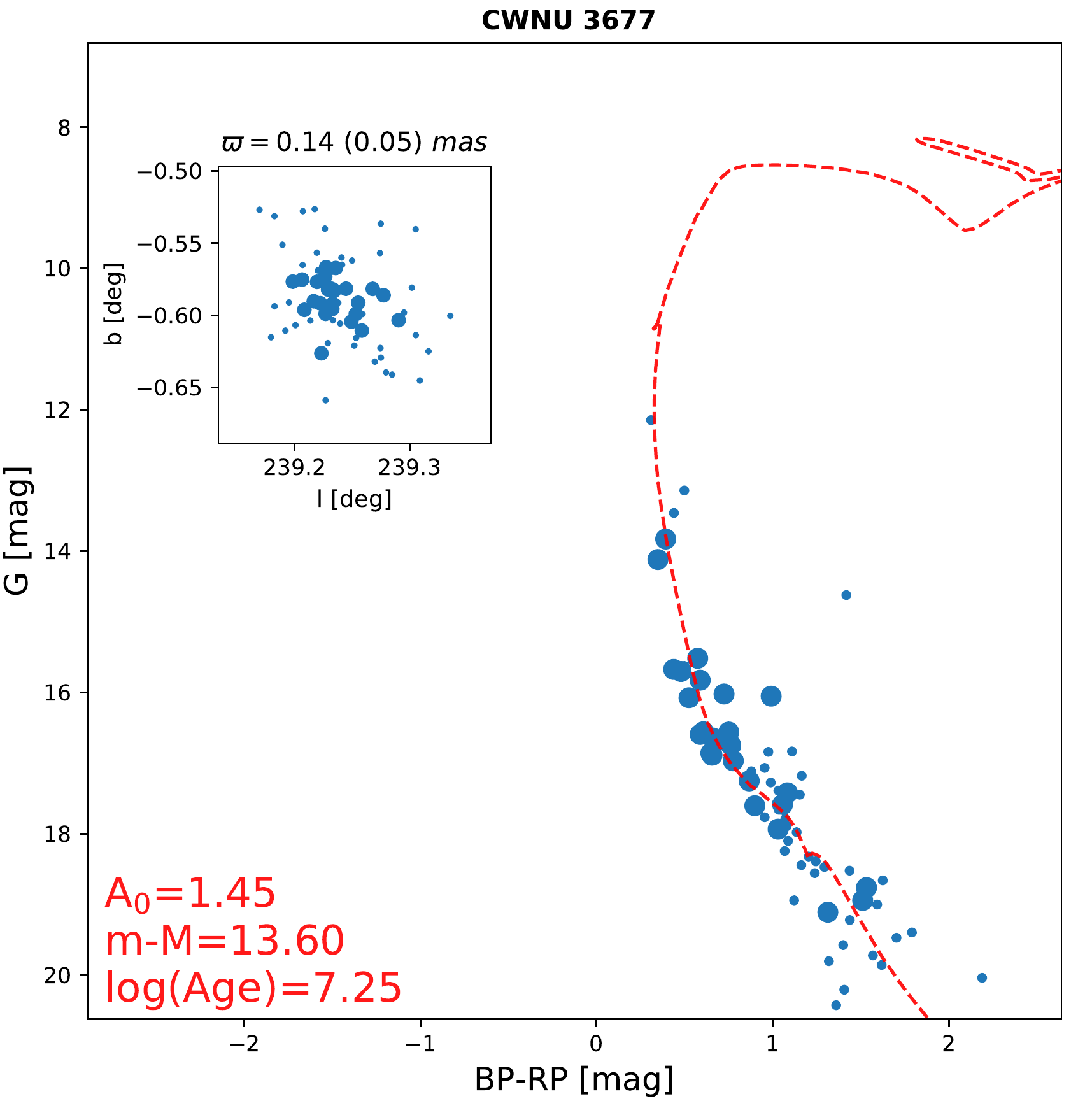}
\includegraphics[width=0.245\linewidth]{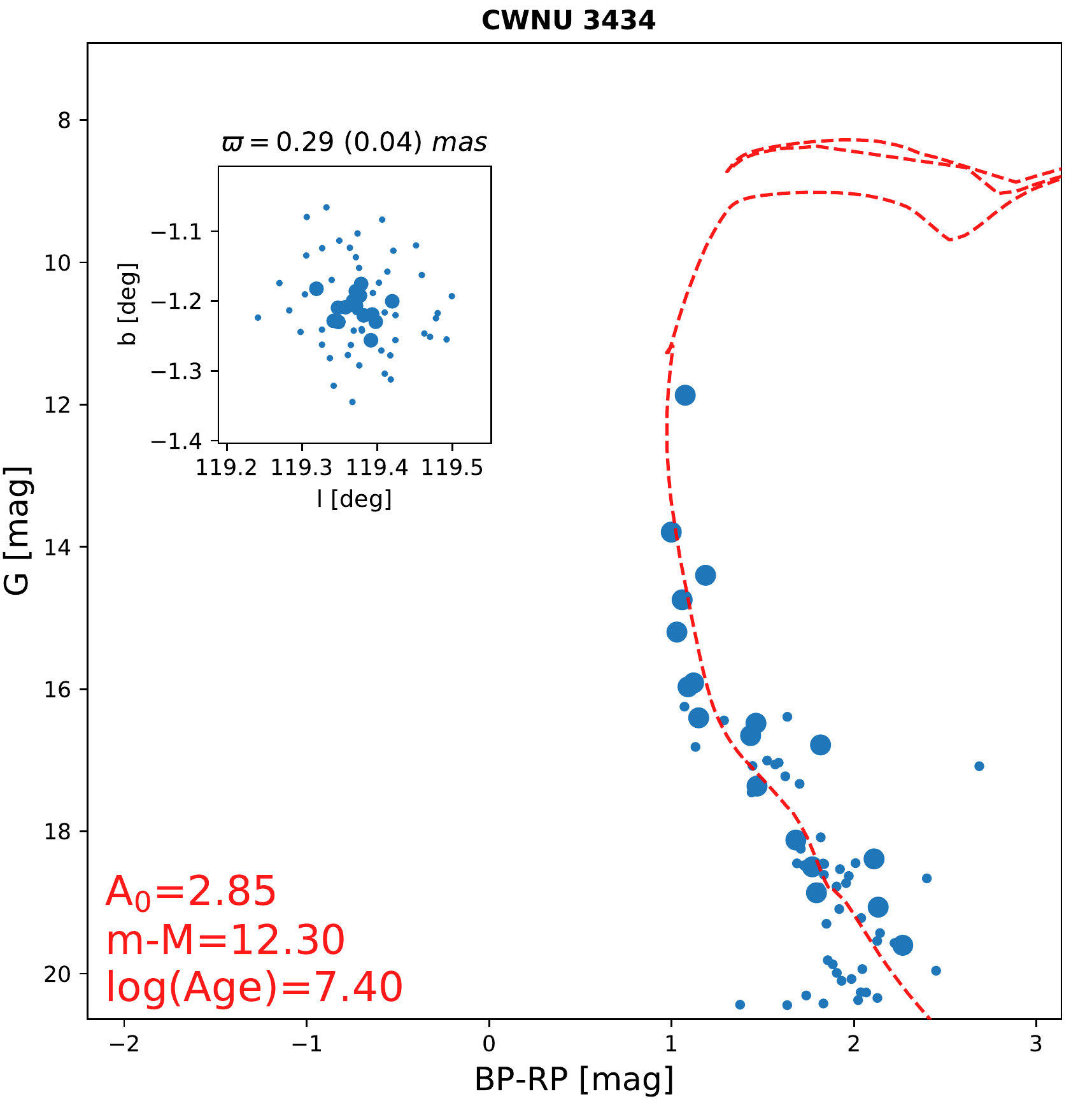}
\includegraphics[width=0.245\linewidth]{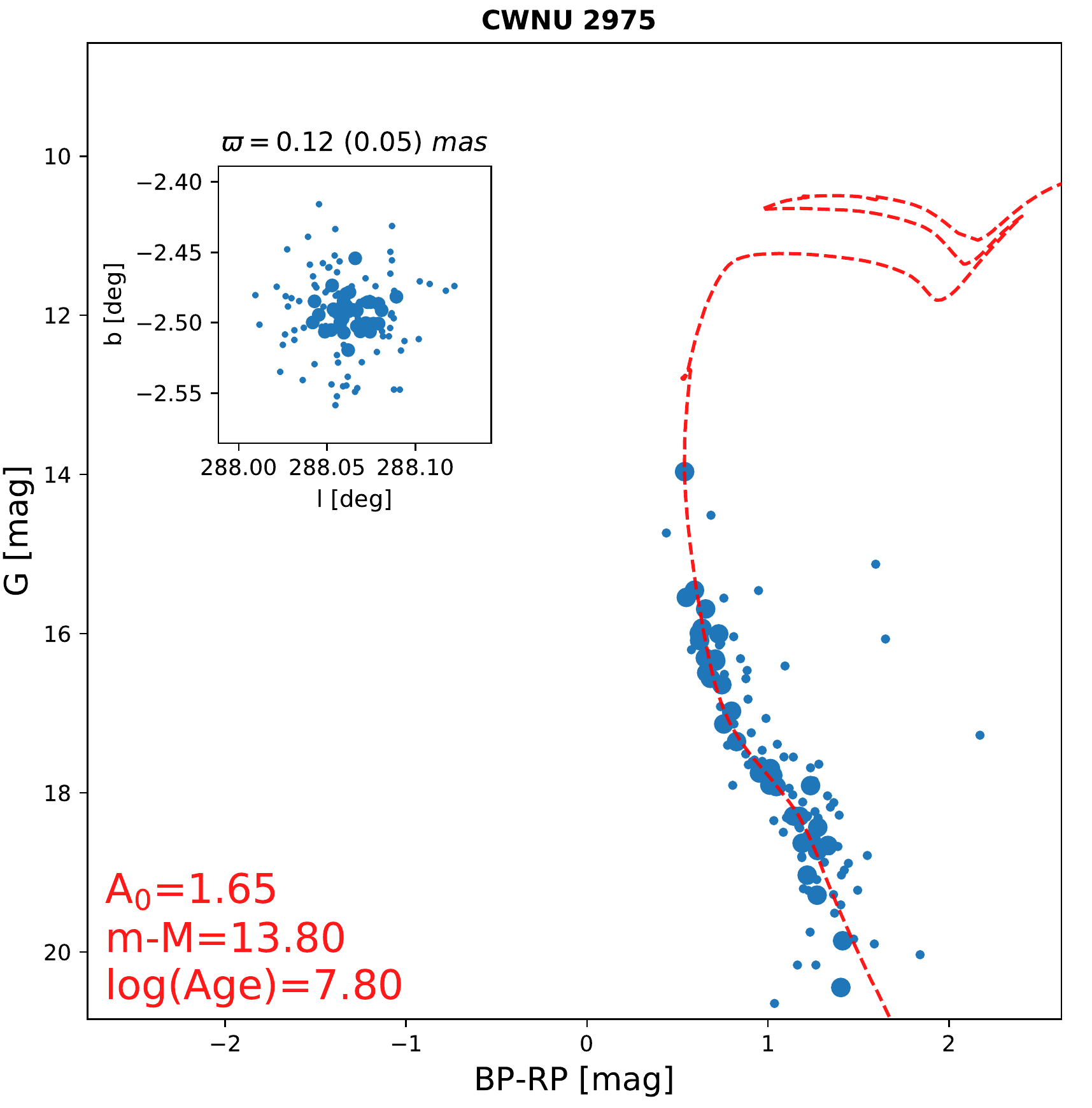}
\includegraphics[width=0.245\linewidth]{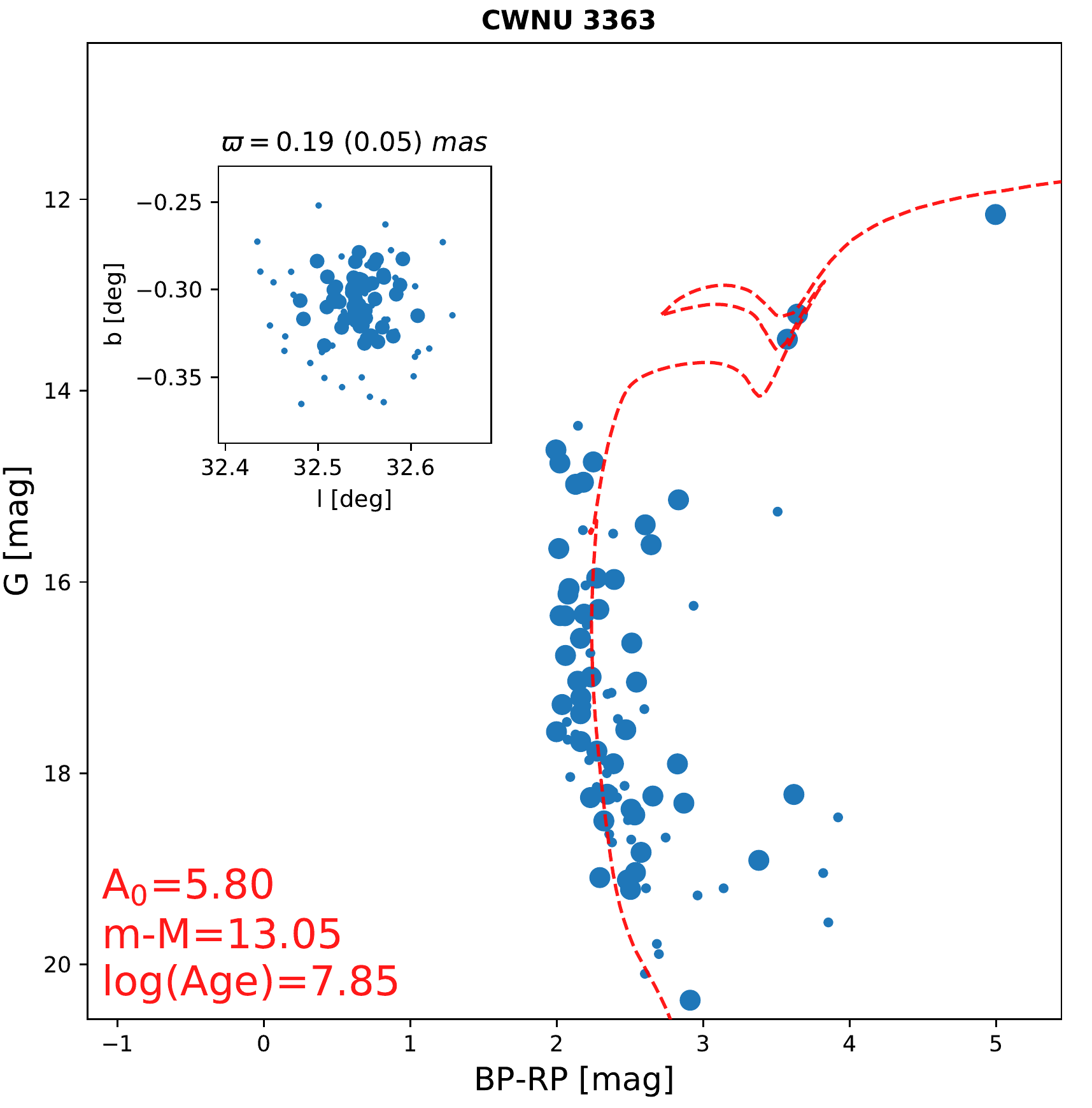}
\includegraphics[width=0.245\linewidth]{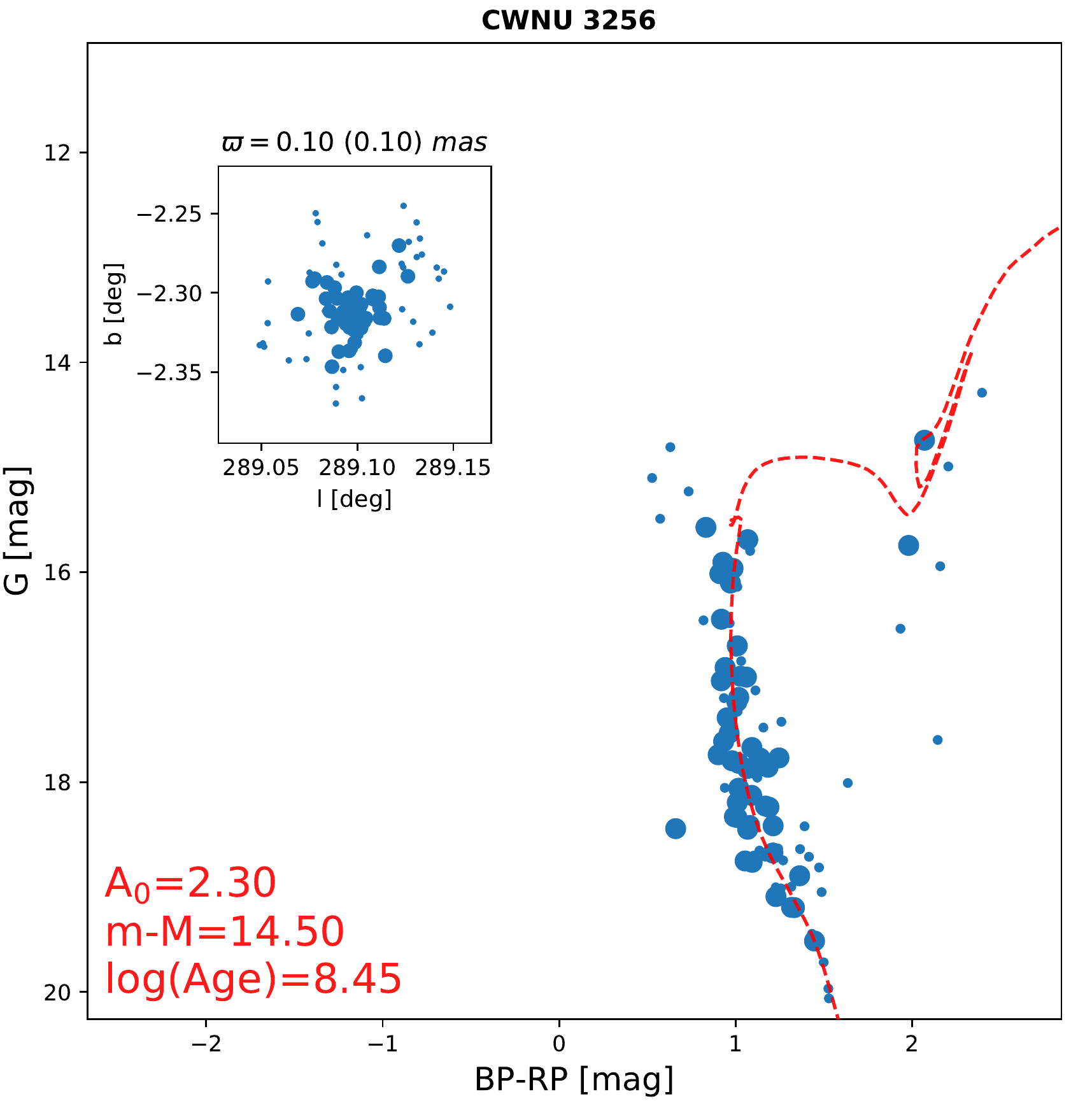}	
\includegraphics[width=0.245\linewidth]{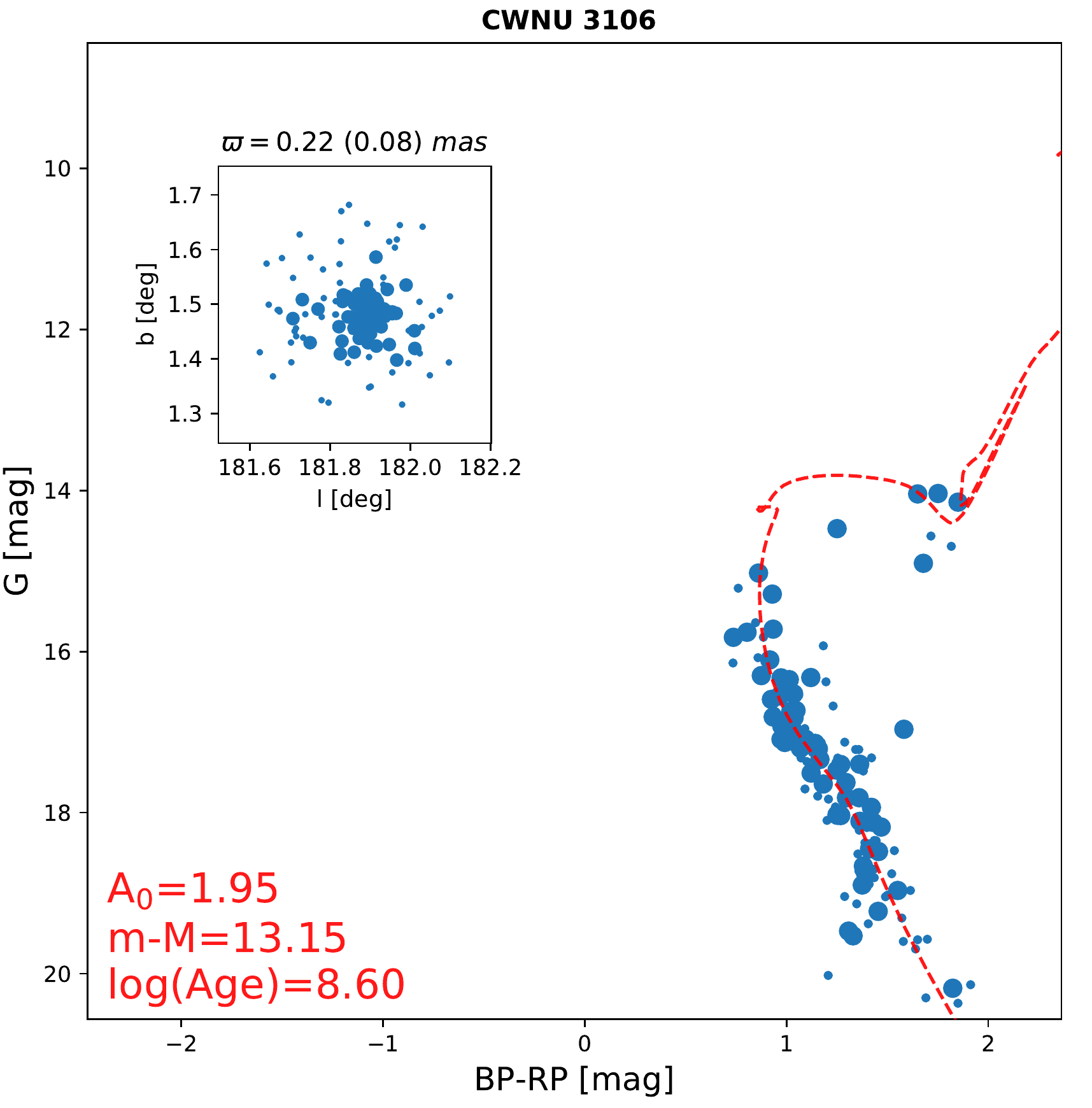}		
\includegraphics[width=0.245\linewidth]{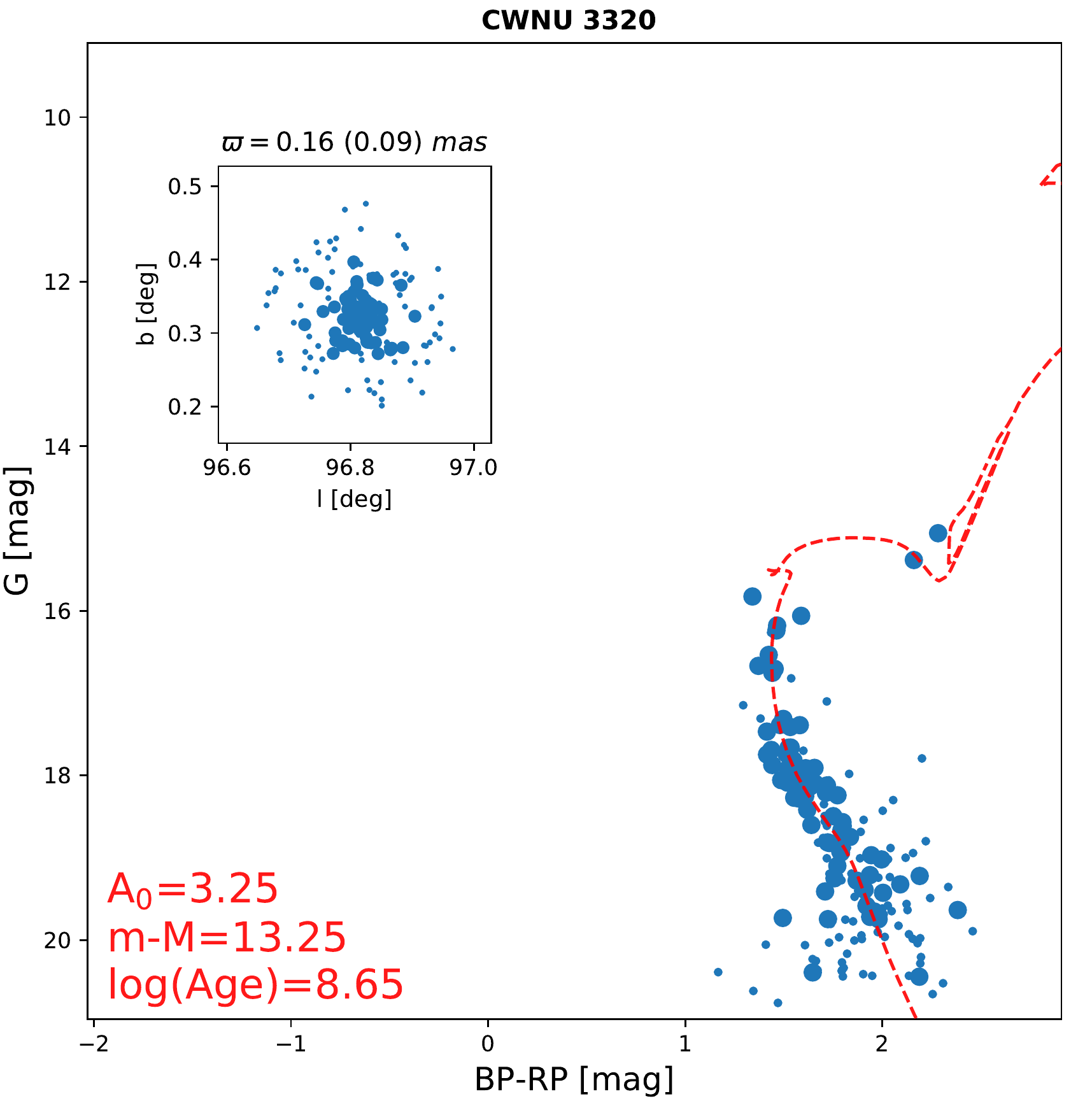}
\includegraphics[width=0.245\linewidth]{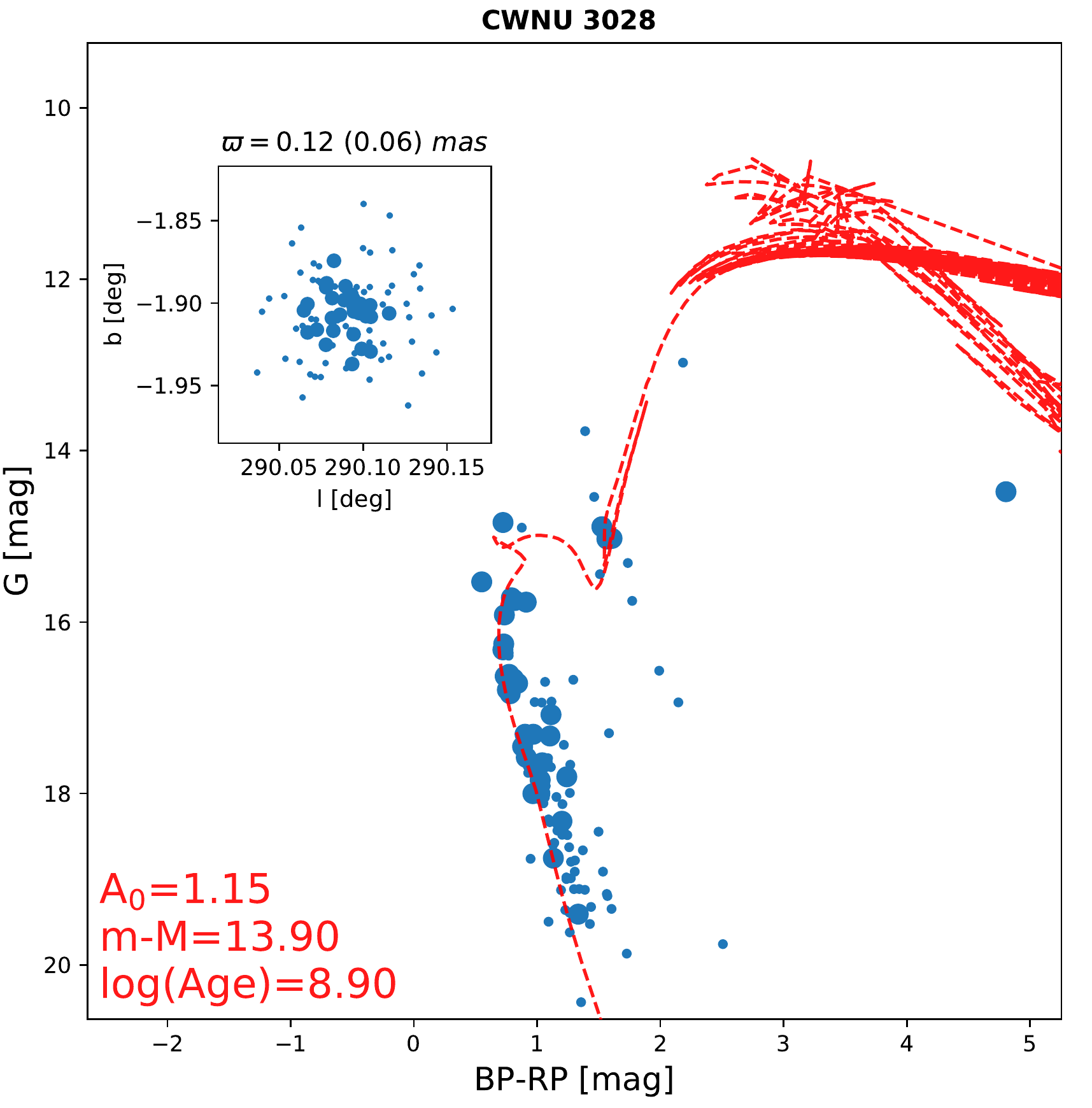}
\includegraphics[width=0.245\linewidth]{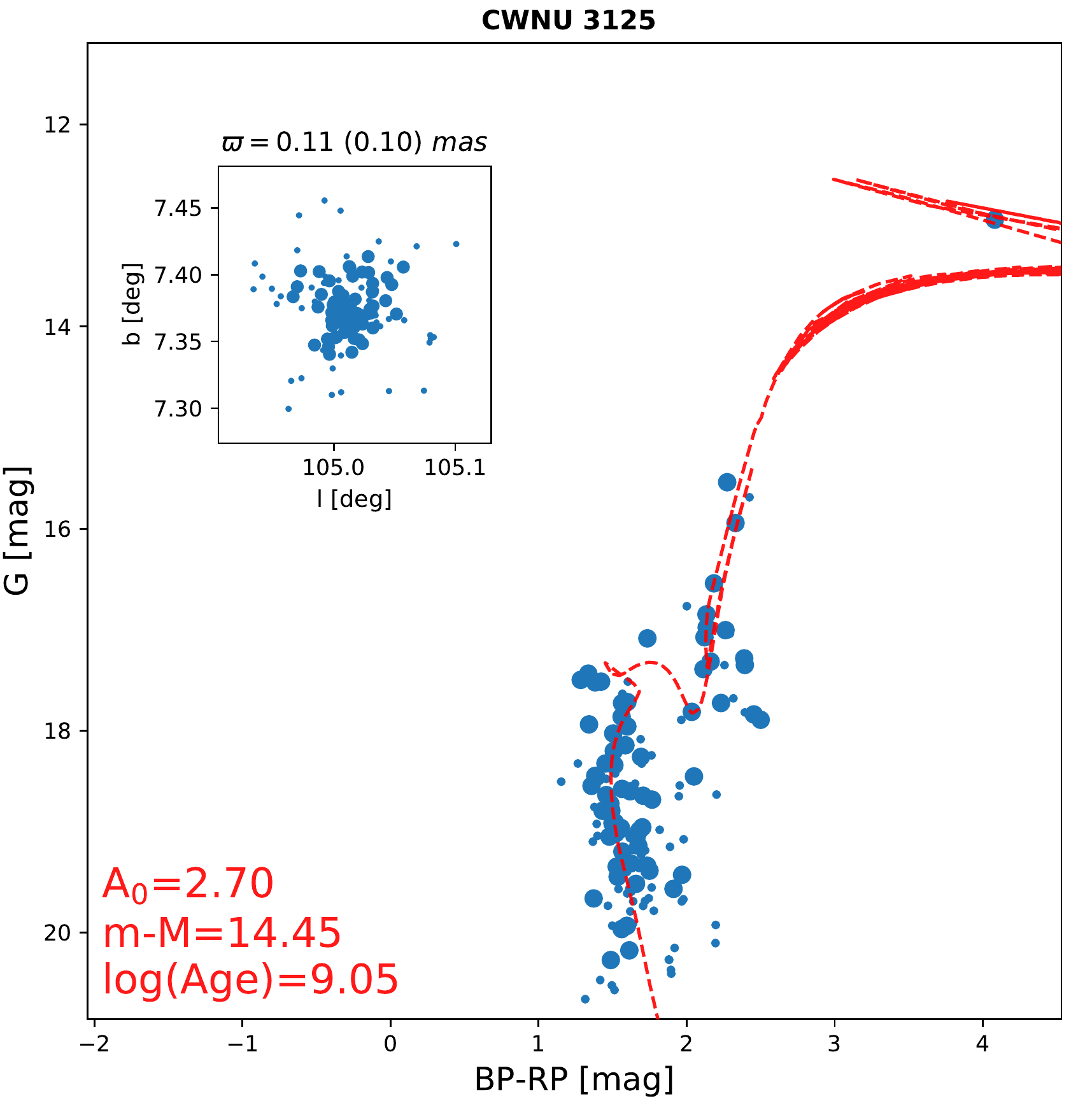}
\includegraphics[width=0.245\linewidth]{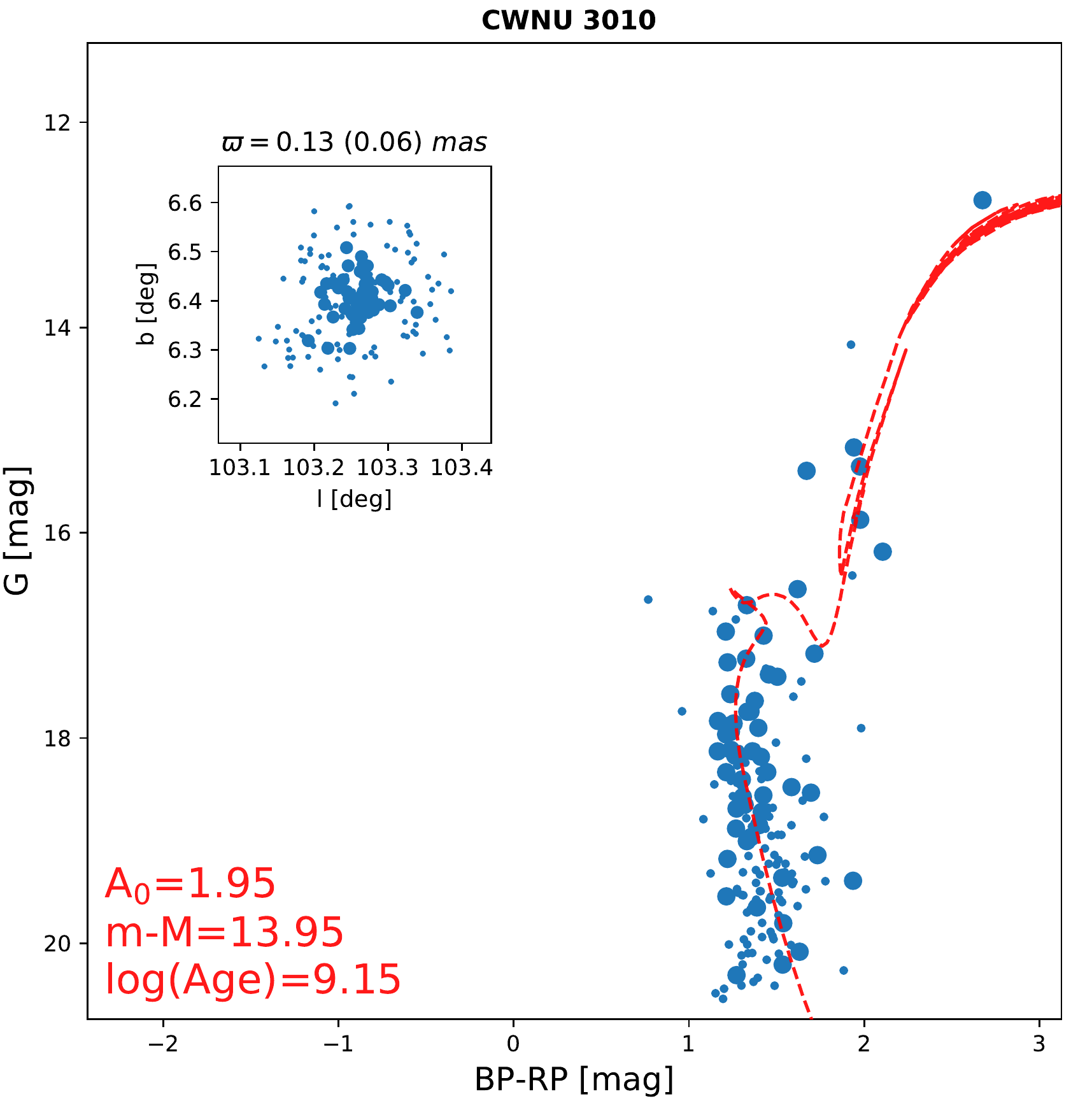}
\includegraphics[width=0.245\linewidth]{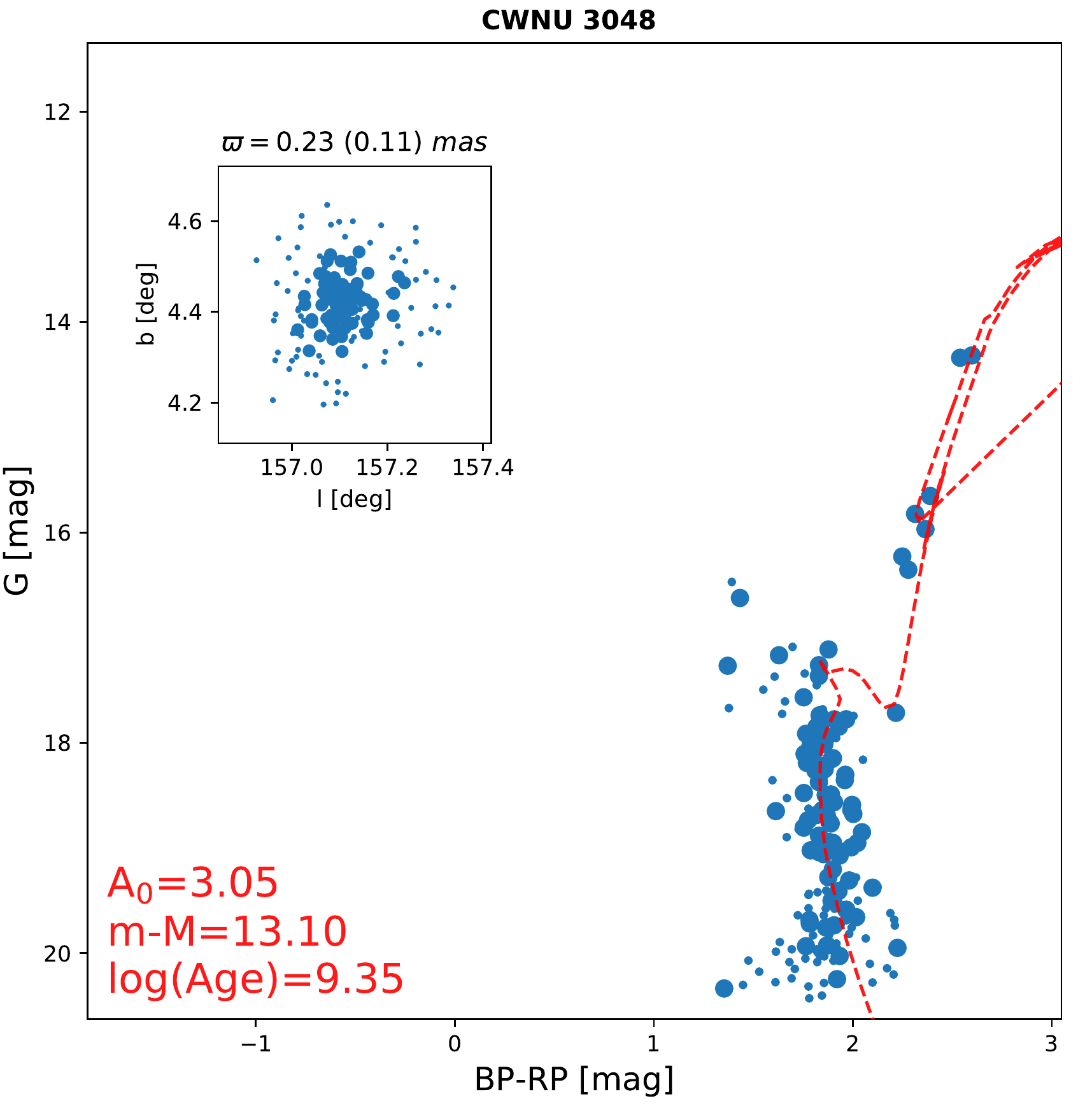}
\includegraphics[width=0.245\linewidth]{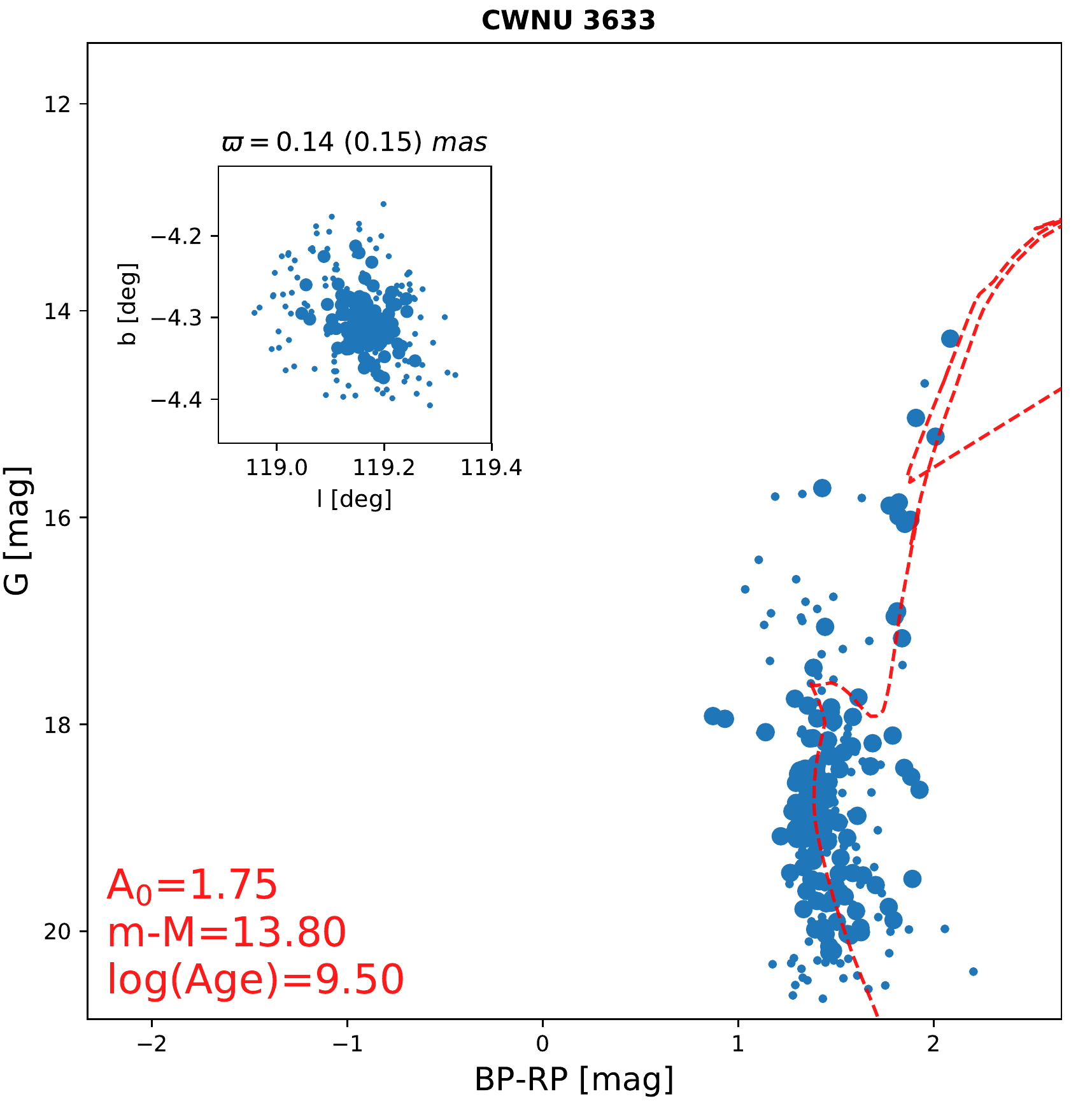}
\caption{Same as Figure~\ref{fig:pre_gaia_ocs}, but for the newly detected reliable clusters (Type~1), which follow a distinct evolutionary sequence as depicted by the increasing age of the clusters.}
\label{fig:Type1}
\end{center}
\end{figure*}

\begin{figure*}
\begin{center}
\includegraphics[width=0.245\linewidth]{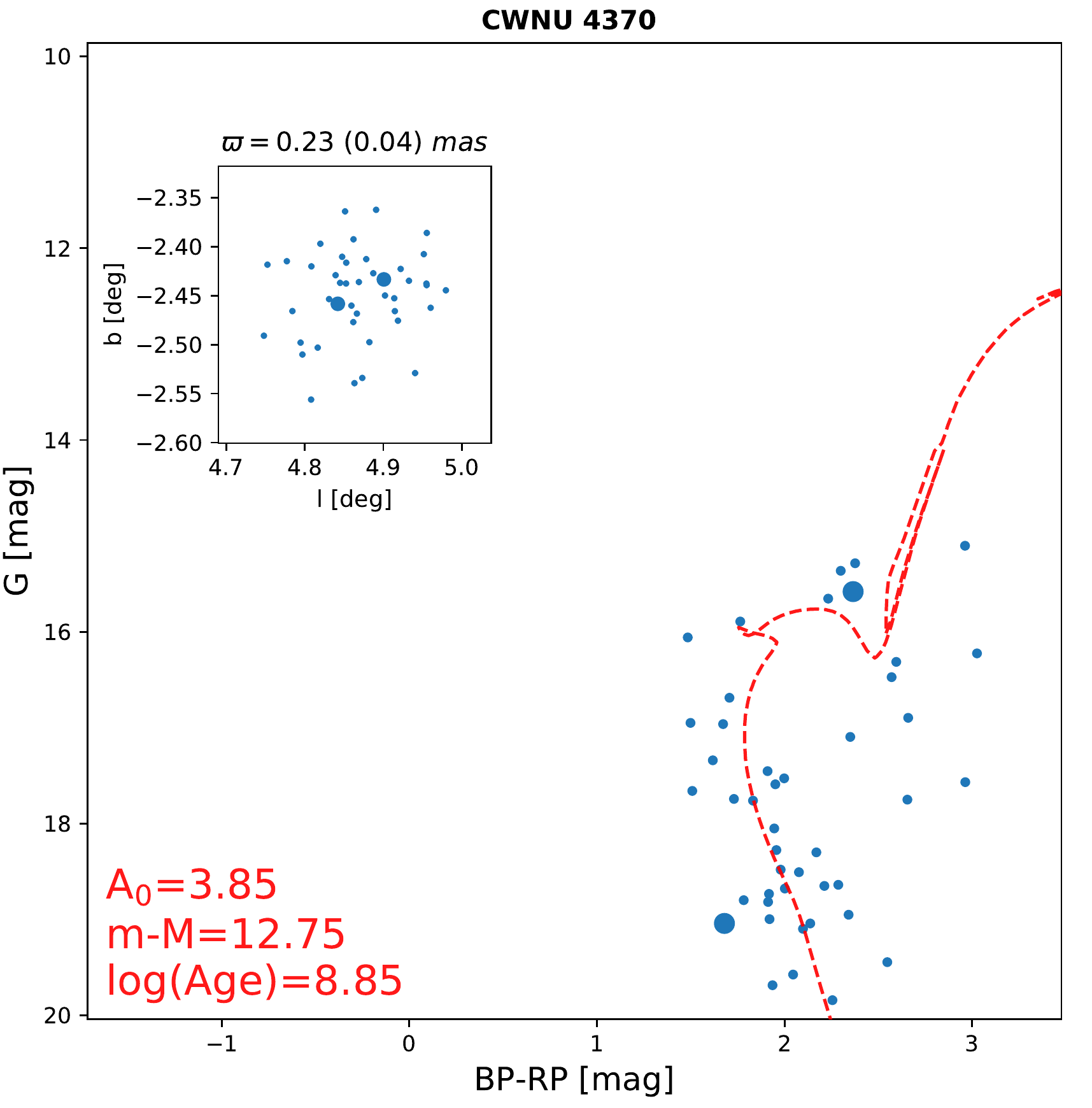}
\includegraphics[width=0.245\linewidth]{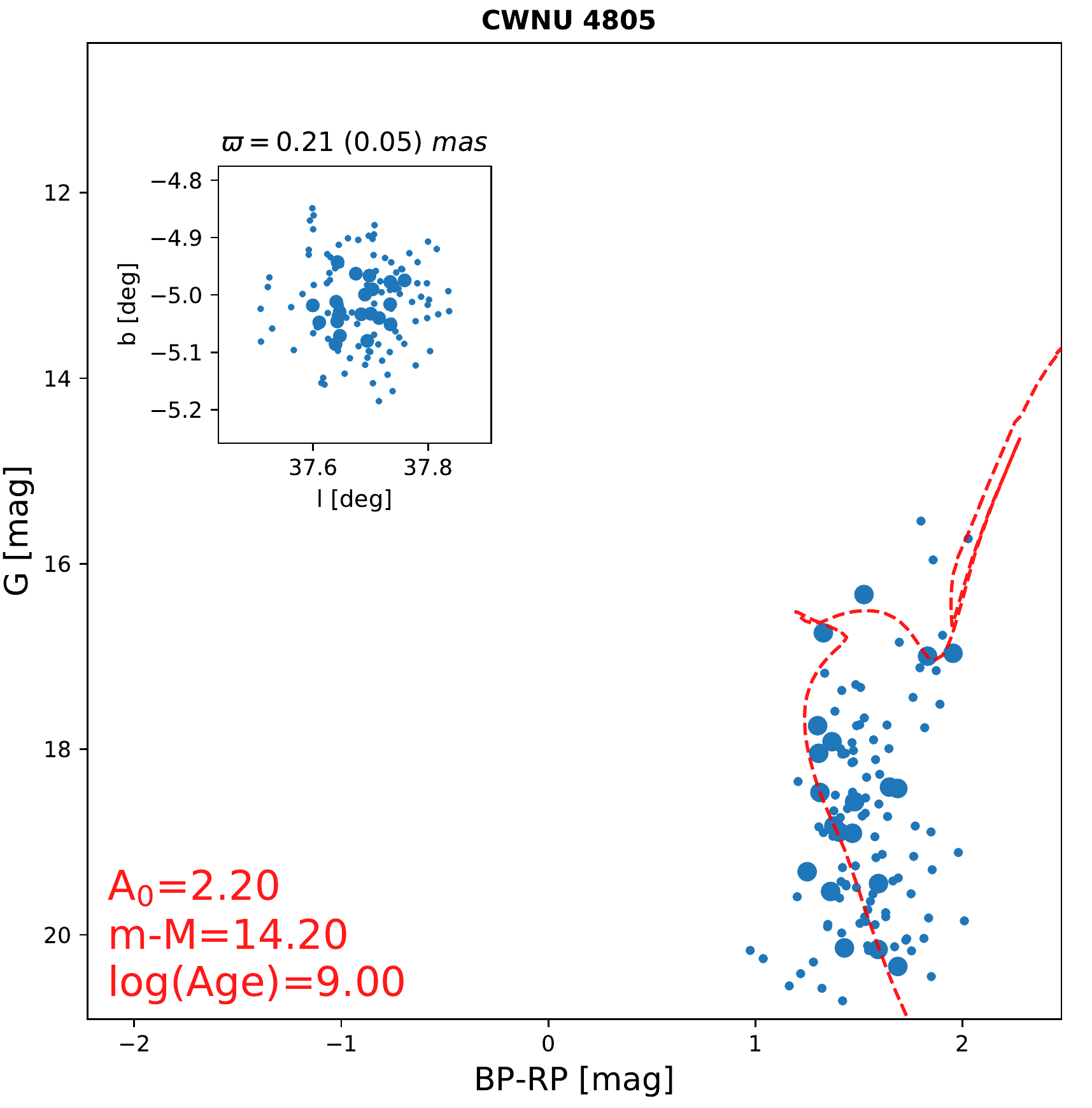}	
\includegraphics[width=0.245\linewidth]{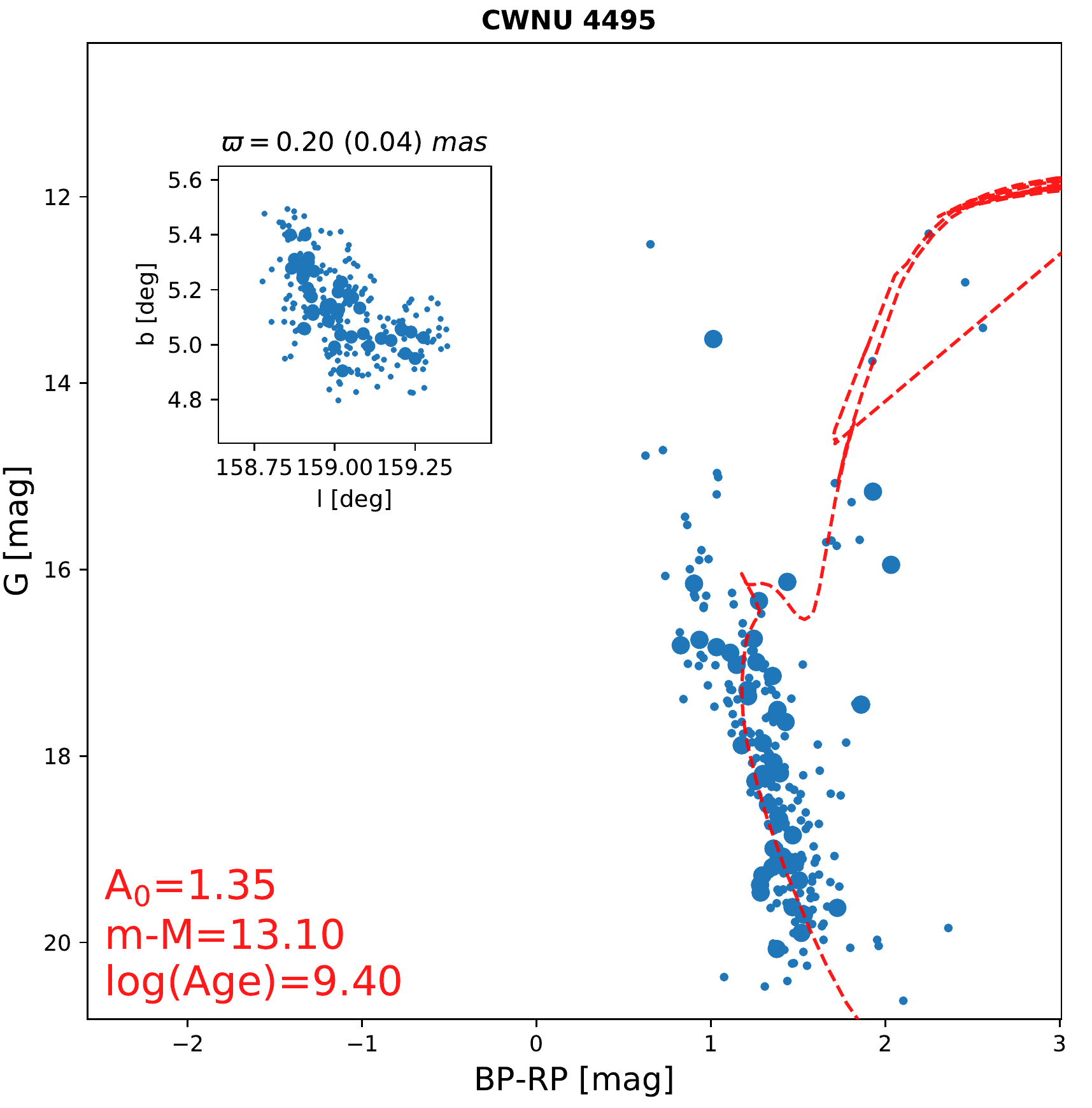}
\includegraphics[width=0.245\linewidth]{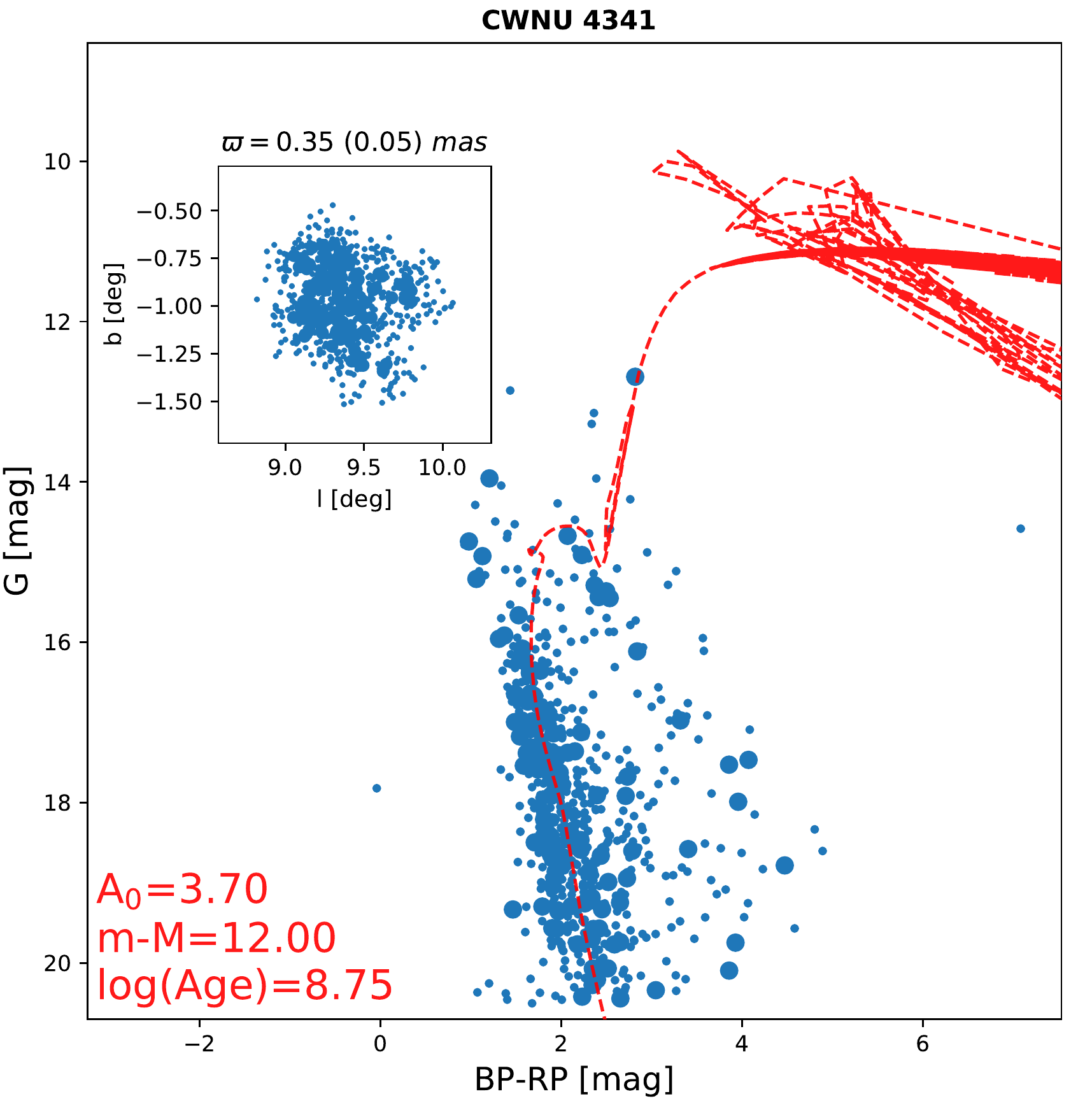}
\caption{Same as Figure~\ref{fig:pre_gaia_ocs}, but for the newly detected candidate clusters (Type~2)}
\label{fig:type2}
\end{center}
\end{figure*}

After following the steps outlined above, we have compiled a catalog of 2057 identified objects similar to OCs, 28 of them were cataloged in ~\citet{dias21}. As illustrated in Figure~\ref{fig:lb}, the majority of these objects are situated in low latitude regions (|$b$| < 5~deg), while some are located in higher latitudes. We believe higher latitude detections originate from the Galactic bulge or warp structure, and we have conducted further analysis on them in another study (He 2023, in preparation). Out of these, 1902 objects could be distinguished twice using our TGFIG method, while 155 responded only once. By visually analyzing the results described above, we have classified them into 1488 Type~1 clusters (Figure~\ref{fig:pre_gaia_ocs} and Figure~\ref{fig:Type1}) and 569 Type~2 clusters (Figure~\ref{fig:type2}). Type~1 cluster candidates have more tightly bound distributions in Galactic coordinates, or have better CMDs, making them the most dependable open clusters detected in this study when compared to Type 2 cluster candidates.

We have fitted all photometry (G, BP-RP) from the OCs and OC candidates with theoretical isochrones, and statistical parameters have also been included in Table~1. For all parameters in Table~1, we have used the core member and median values, with standard deviation displaying dispersion. It is worth noting that since not all clusters have radial velocity members more than 1, for those clusters with only one stellar radial velocity value, we have used the radial velocity error instead. Furthermore, we have provided researchers with member star information for all results, including astrometric data, uncertainties, $ruwe$ values, photometry, and radial velocity, all of which come from the Gaia DR3 catalog. In Table~\ref{tab:table_mem}, we have included an $ifcore$ value to designate core members ($ifcore$=1) and outer members ($ifcore$=0).

\subsection{Newly detected OCs and candidates}\label{sec:newoc}

Regarding the newly detected OCs and candidates, we have cross-matched them with pre-Gaia OCs using MWSC~\citep{Kharchenko13} and DIAS02~\citep{Dias02}. First, we cross-matched them in a radius r, where the r$_1$ value for MWSC catalog was used. To eliminate the clusters that superposition in a foreground cluster, we crossed the clusters in distance in (1/3 $\times$ distance, 3 $\times$ distance), and radius in (0.3 $\times$ r, 3$\times$r). As a result, we identified a total of 85 pre-Gaia clusters (Figure~\ref{fig:pre_gaia_ocs}), 78 of which were Type~1 and 7 of which were Type~2. Figure~\ref{fig:pre_gaia_ocs} shows examples of the pre-Gaia OCs we identified as reliable OCs (Type~1), with increasing ages, including some high extinguished (A${_0}$ > 5~mag) and old clusters (logarithmic age > 9). The BP band overestimation affects the main sequence, making it appear unreliable in the fainter end for high extinguished clusters. However, they remain bound in the space/kinematic vector, and the CMD turn-off point is also visible. For some old clusters, the presence of BSSs sets them apart from other clusters, these stars were identified in the new clusters as well (Section~\ref{sec:comparison}).

The 1972 OCs and OC candidates were labeled 1410 reliable OCs (Type~1) and 562 candidates (Type~2), respectively. Figure~\ref{fig:Type1} and Figure~\ref{fig:type2} depict some Type~1 and Type~2 clusters, respectively, with increasing ages. Type~1 clusters have more tightly bound structures and a visible CMD (with low contamination), while Type~2 clusters are unbound and/or heavily contaminated, which requires further research for candidates. Although the method may not fully detect the OCs and member stars in Gaia data, the large number of newly detected reliable OCs (Type~1) and OC candidates (Type~2) shows the high efficiency of the OC search under TGFIG method. The isochrone fits are also valuable for cluster studies. All CMDs, isochrone fits, and Galactic coordinates distribution diagrams can be viewed online to facilitate their usage.

\subsection{Globular cluster candidates and dwarf galaxy}\label{sec:gc}
\begin{figure*}
\begin{center}
\includegraphics[width=0.245\linewidth]{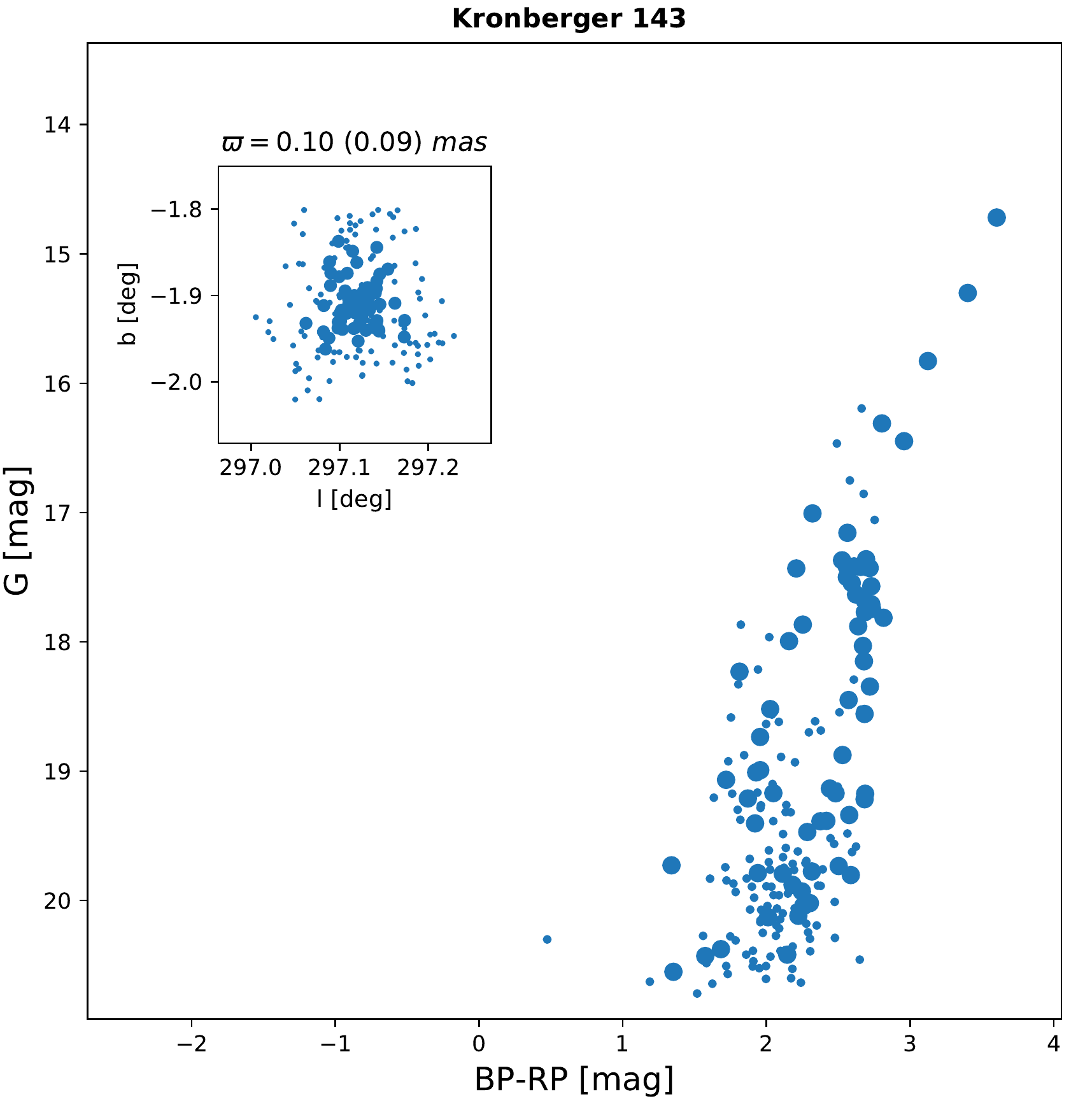}	
\includegraphics[width=0.245\linewidth]{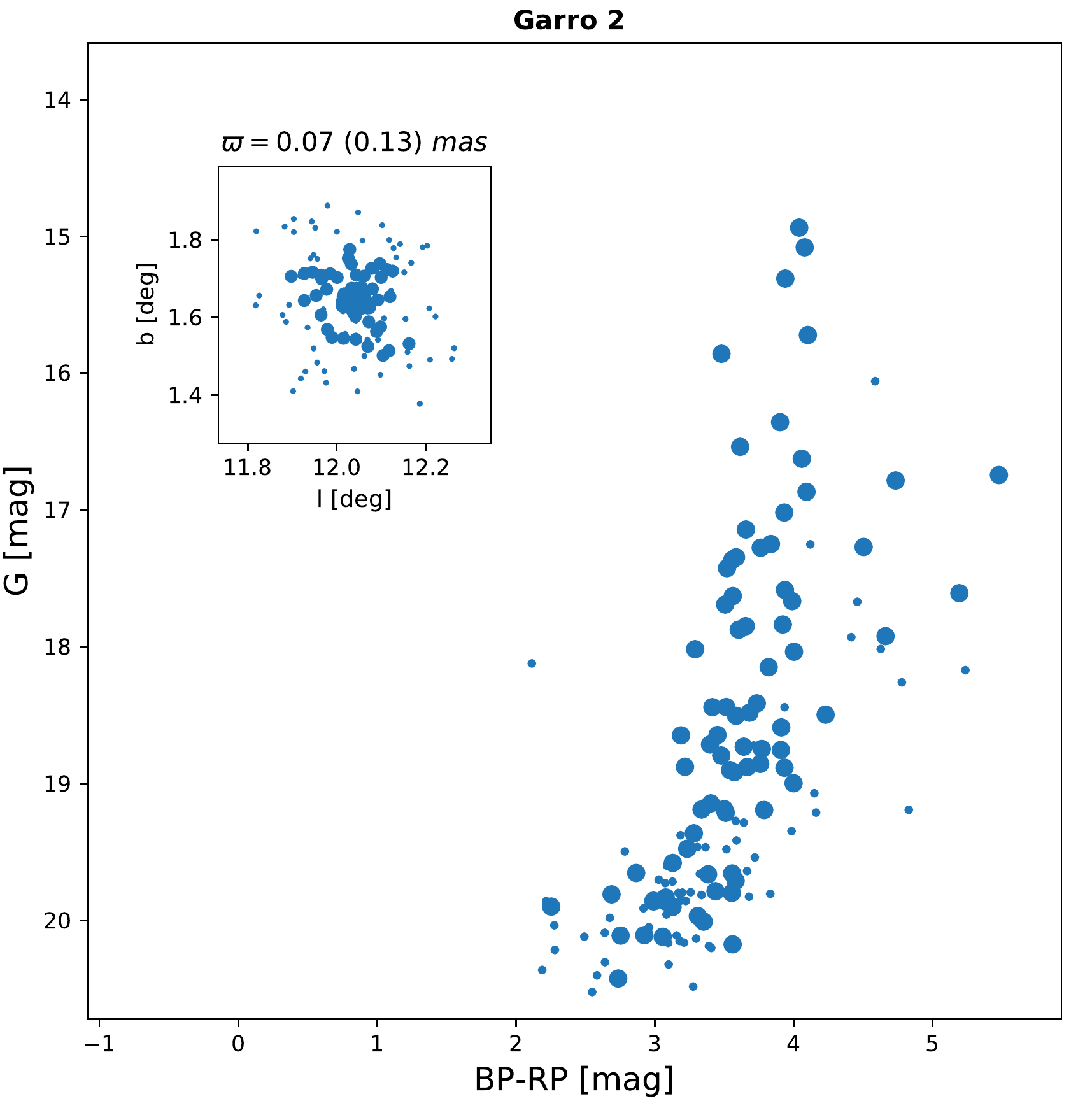}
\includegraphics[width=0.245\linewidth]{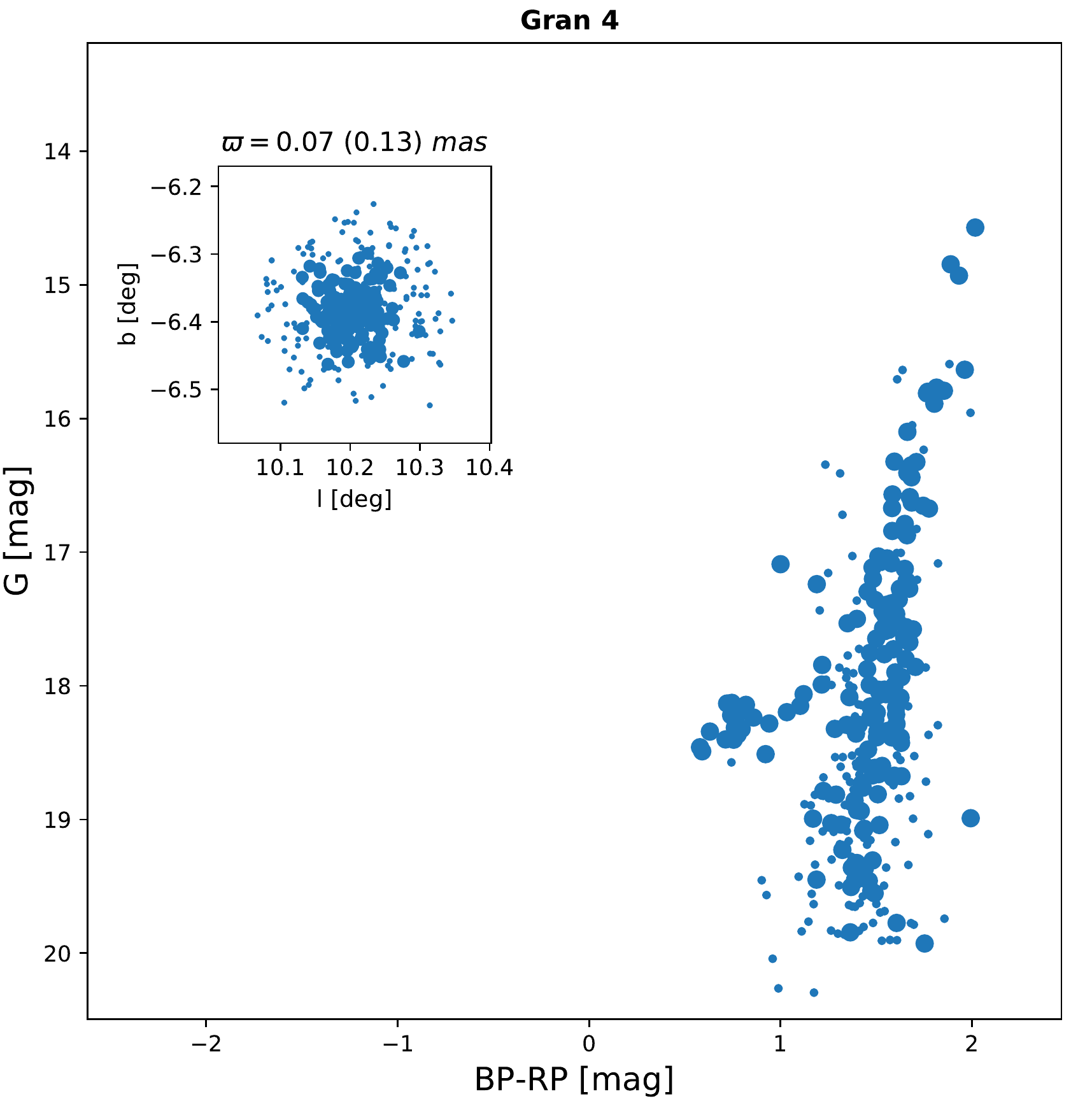}
\includegraphics[width=0.245\linewidth]{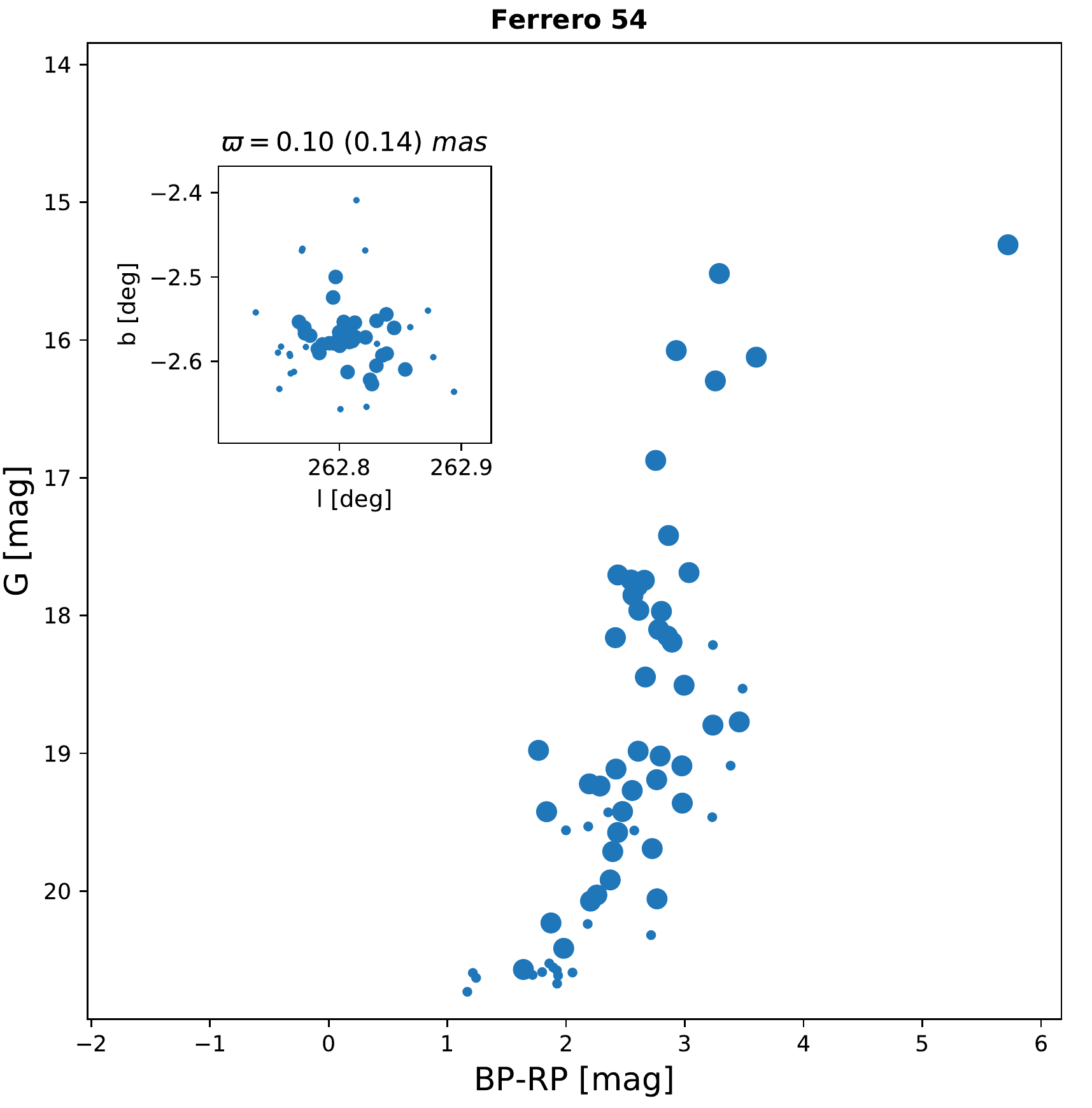}
\includegraphics[width=0.245\linewidth]{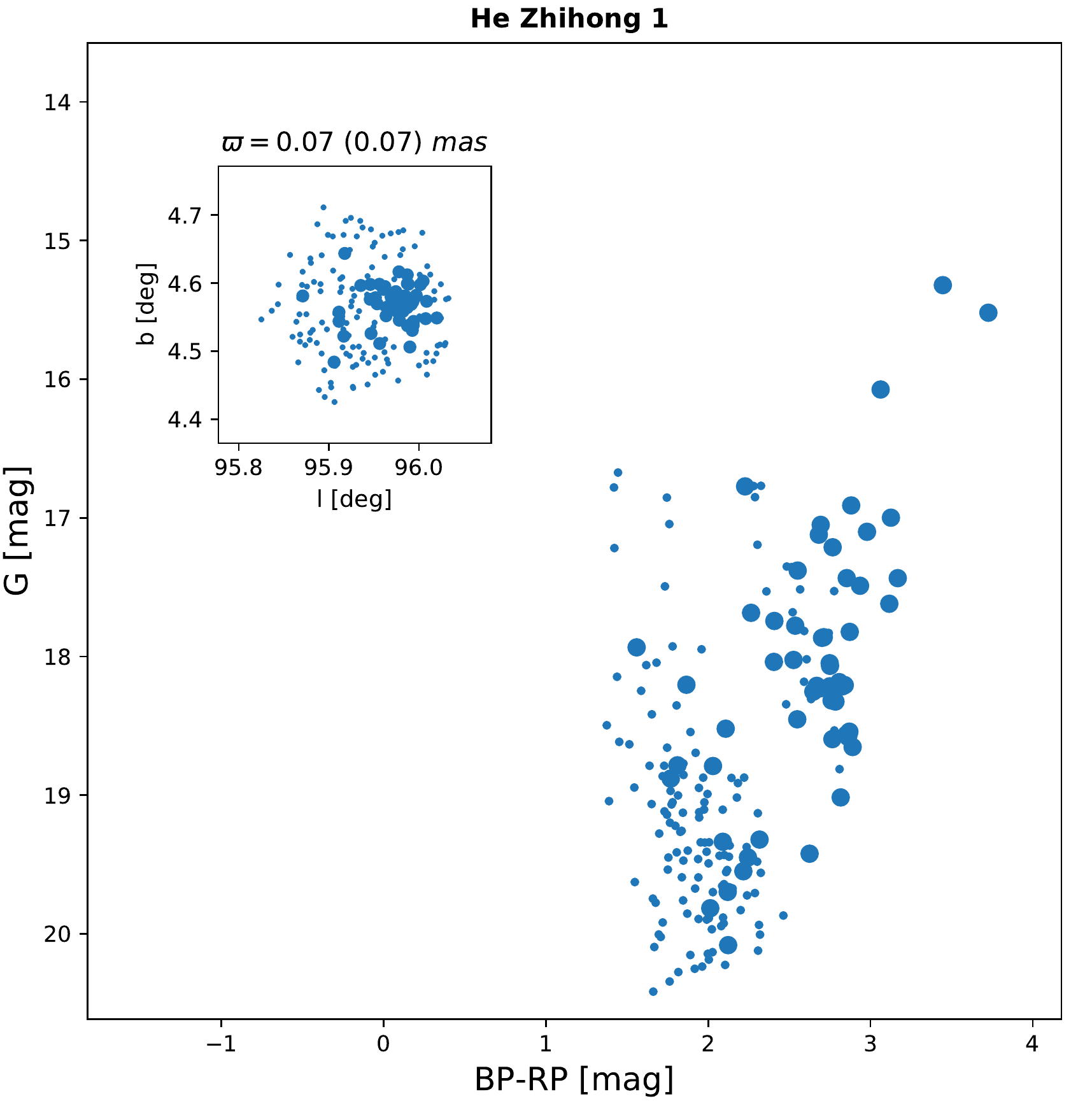}
\includegraphics[width=0.245\linewidth]{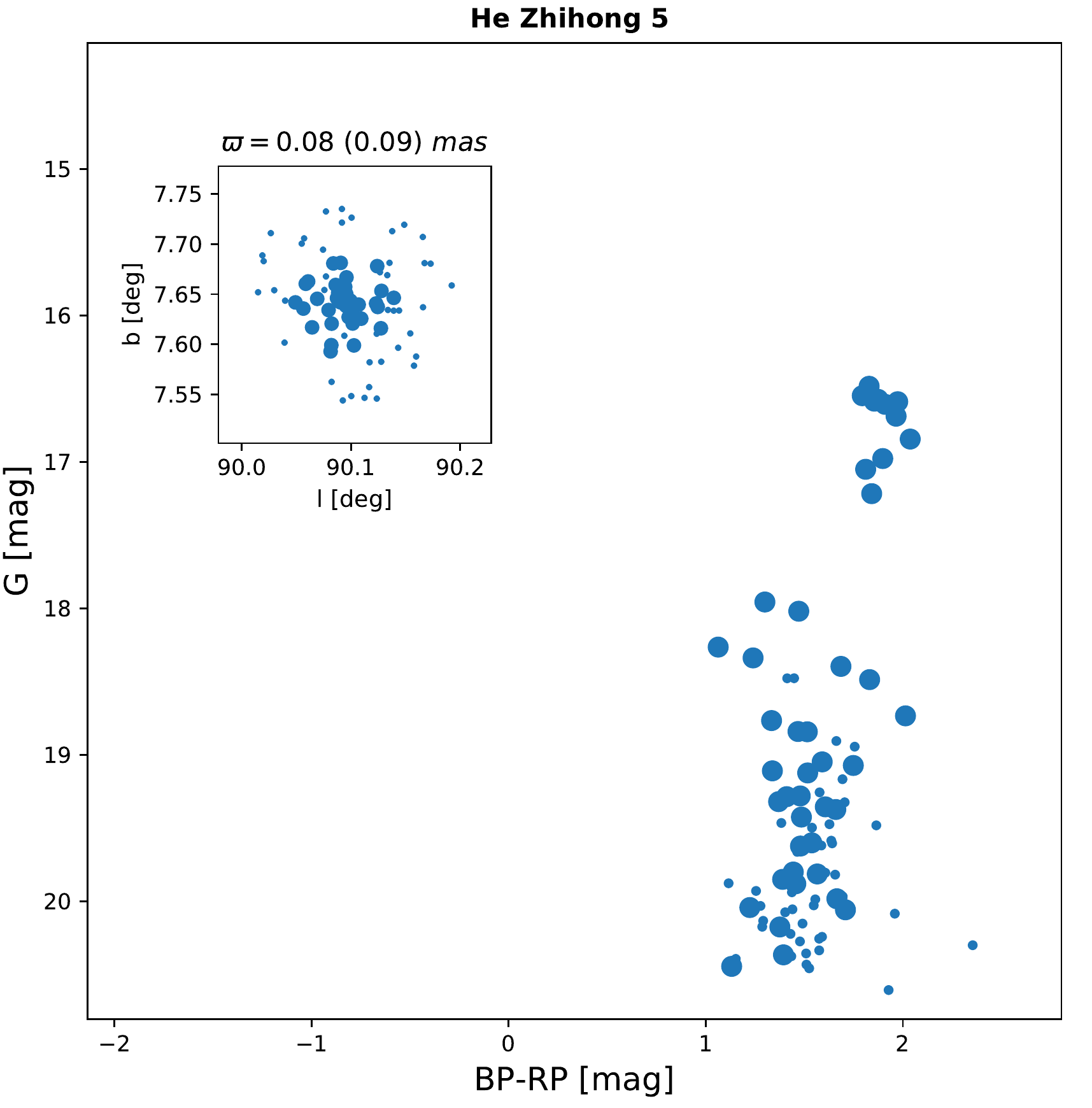}
\includegraphics[width=0.245\linewidth]{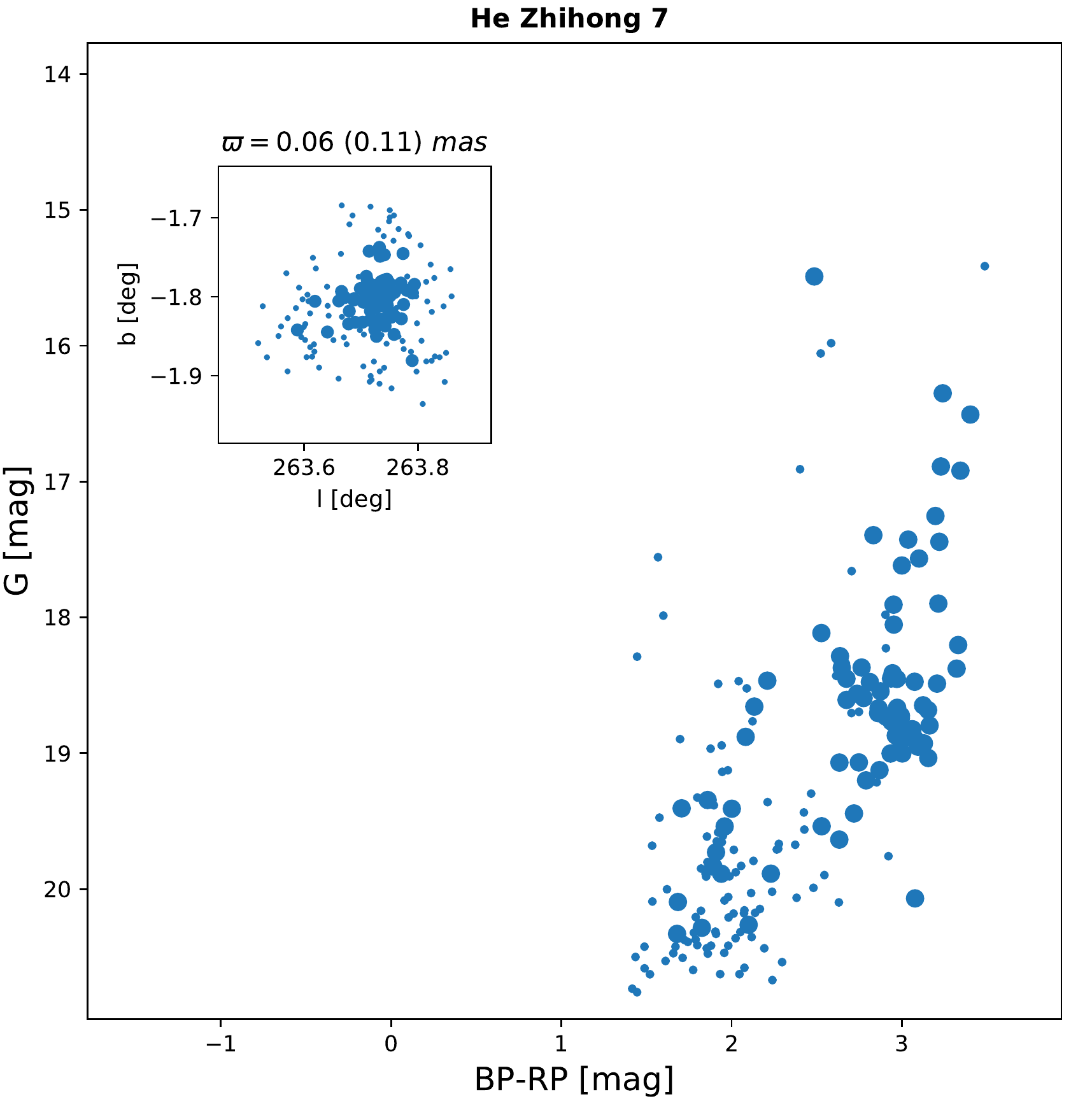}
\includegraphics[width=0.245\linewidth]{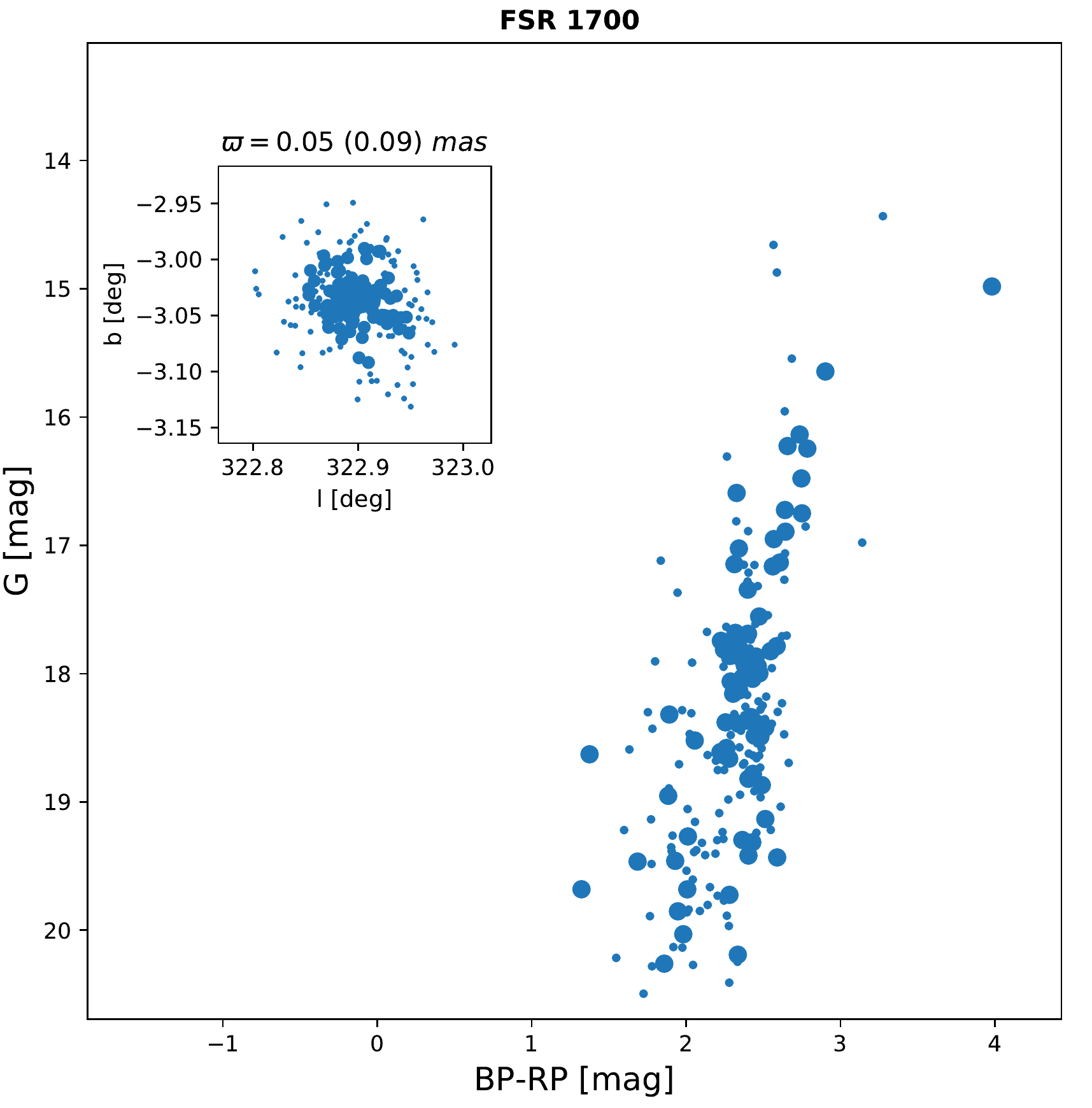}
\caption{CMDs and spatial distribution of the GCCs re-detected in this work (upper panels), and newly identified (lower panels). Gran~4 is already exhibiting a distinct horizontal branch, confirming its classification as a highly reliable globular cluster. Nevertheless, the other clusters included in the dataset only show dense spatial distributions from great distances, as well as red clump stars that are clustered with possible horizontal branch, although this cannot be conclusively determined due to the enormous distance and extinction. Additional research should be conducted to ascertain the source of these potential clusters, particularly the newly identified candidates that rely solely on Gaia data. }
\label{fig:gcc}
\end{center}
\end{figure*}

In addition to the OC samples, we have compiled a list of 28 GC candidates, of which 16 are cataloged in ~\citet{Palma2019,Minniti2017,Minniti2021,Gran2022,Garro2020,Garro2022}, and 12 are newly detected in Gaia data. For eleven of them, we have identified them as He Zhihong~1 to 11, and the remaining one, FSR~2700, is cataloged as an OC in MWSC. Although the matched GCCs were detected before, in our work, we have solely used Gaia data, which shows that the TGFIG method is also applicable in detecting other stellar aggregates in Milky Way. The positions and astrometric statistic parameters of the GCCs can be found in Table~1. Figure~\ref{fig:gcc} presents some examples of the GCCs, and we note that some of those objects located in low-latitude regions are highly extinguished, making them challenging to detect in Gaia data. However, since the sequence of most of the candidates is not distinct, further research is needed, particularly for the new findings that only based on Gaia data.

\begin{figure*}
\begin{center}
\includegraphics[width=0.8\linewidth]{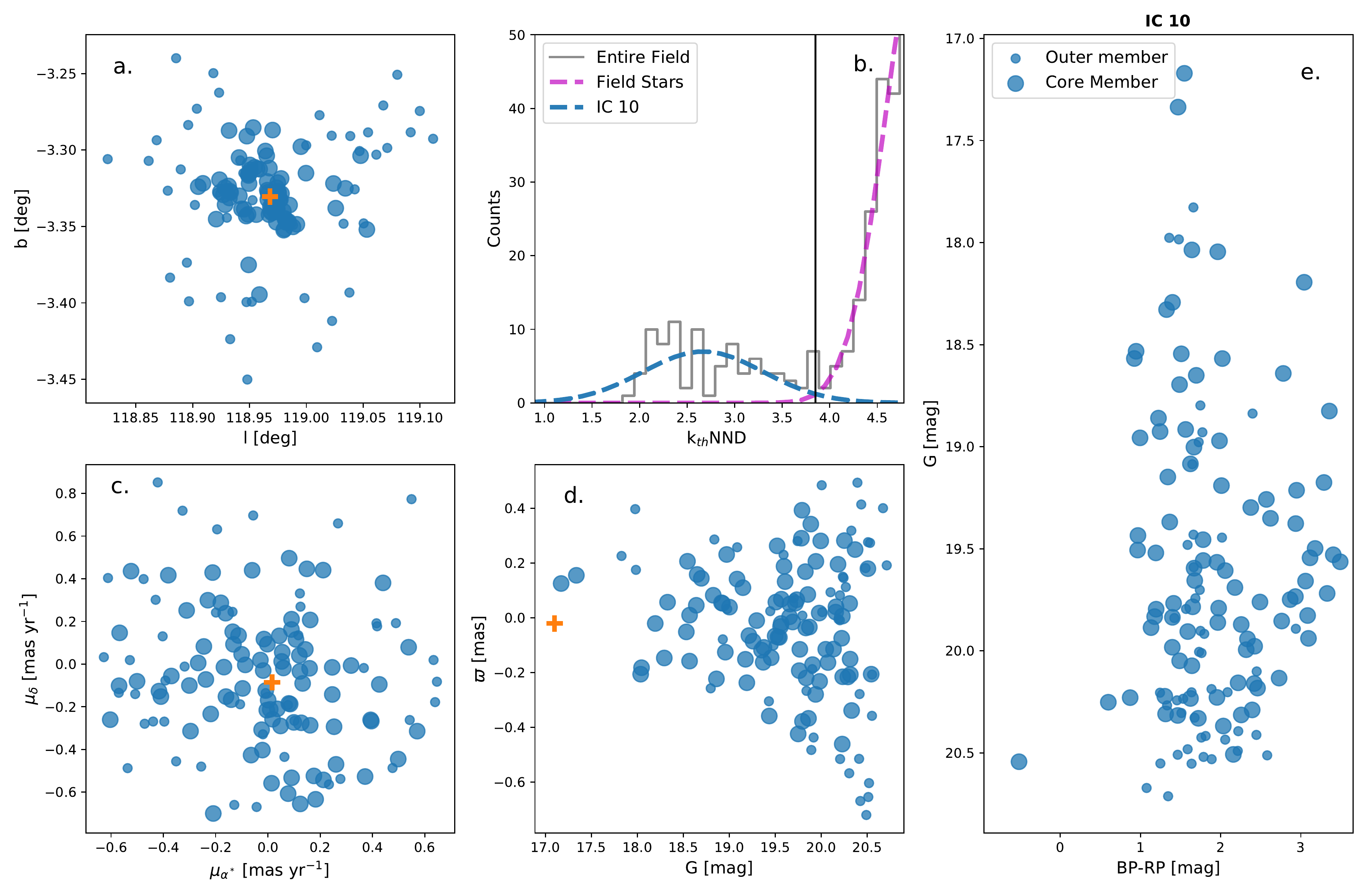}
\caption{Same as Figure~\ref{fig:kds}, but for the dwarf galaxy IC~10. The $k_{th}$NND histogram (b.) shows a significant difference between the field stars and member stars. Despite its low Galactic latitude (a.) and distance of approximately 660~kpc~\citep{Huchra99,Borissova2000}, IC~10 remains detectable in Gaia data up to a magnitude of 17 (e.).}
\label{fig:dwarf}
\end{center}
\end{figure*}

Meanwhile, when we analyzed the results, we noticed a dwarf galaxy located at ($l$ = 118.97~deg, $b$ = -3.33~deg) (Figure~\ref{fig:dwarf} a), with a radius of approximately 0.05~deg (core member) to 0.1~deg (outer member). When we cross-matched the simbad database~\footnote{\url{http://simbad.cds.unistra.fr/simbad/sim-fid}}, we found that the object is the dense part of IC~10, a local dwarf galaxy located at a distance of approximately 660~kpc~\citep{Huchra99,Borissova2000}, which has been discovered 135 years ago~\citep{Swift1888}. As a result, it also displays a unique distribution in the k$_{th}$NND histogram (Figure~\ref{fig:dwarf}b). Although the object have a median parallax of - 0.02 mas ($\sigma_{\varpi}$ = 0.13~mas), which is not advantageous for distance measurement, the proper motion values of IC~10 are still useful. The median proper motion of $\mu_\alpha^\ast$ = 0.02~mas yr$^{-1}$ ($\sigma_{\mu_\alpha^\ast}$ = 0.15~mas~yr$^{-1}$), $\mu_\delta$ = - 0.09~mas yr$^{-1}$ ($\sigma_{\mu_\delta}$ = 0.18~mas~yr$^{-1}$), it shows a tangent velocity of approximately 280~\kms, which is also comparable to its radial velocity~\citep[RV = -348~\kms][]{McConnachie12}.

Although previous observations of the proper motion of IC~10 through VLBA showed a smaller value~\citep[$\mu_\alpha^\ast$ = -0.039 $\pm$ 0.009~mas yr$^{-1}$, $\mu_\delta$ = 0.031$\pm$ 0.008~mas yr$^{-1}$][]{Brunthaler06}. However, the difference between the two values is not significant. In the future, acquiring a more precise measurement from Gaia DR4/DR5 would be of great interest.
This single sample demonstrates that the real error of parallax~\citep[including the parallax zero point -0.017~mas of Gaia data,][]{Lindegren21,Fabricius21} in this work is about 0.02~mas. 
This is helpful for researchers to verify the validity of Gaia parallax. Additionally, after high-Galactic latitude dwarf galaxies found in ~\citet{CG18}, this is the first time that a low latitude dwarf galaxy has been reported in a blindly star cluster search work. This may inspire us to use TGFIG to identify local satellites (even in low-Galactic latitude regions) and search for star clusters in LMC/SMC based on Gaia data. 

\subsection{Comparison with previous OC works}\label{sec:comparison}

\begin{figure*}
\begin{center}
\includegraphics[width=1.0\linewidth]{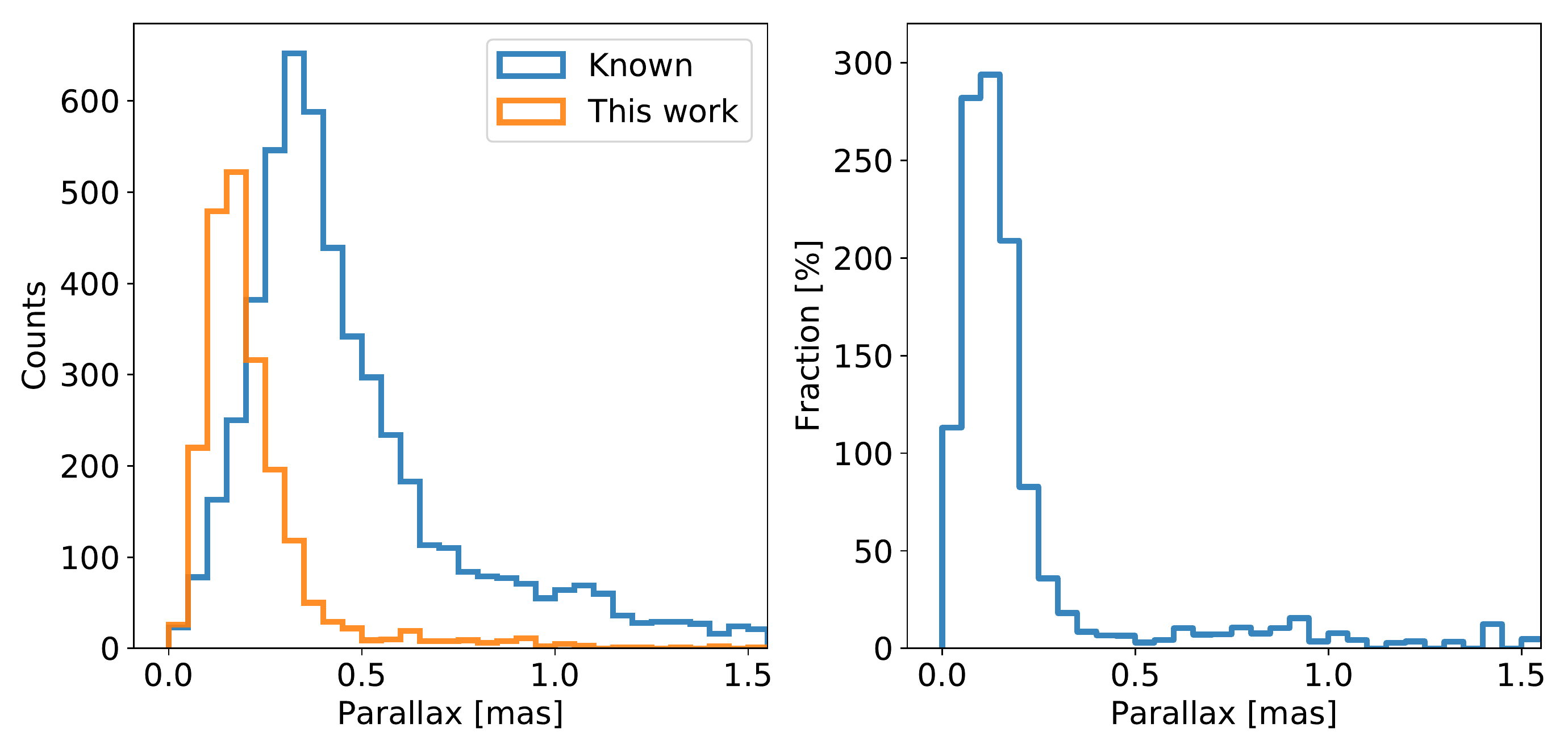}
\caption{Statistics on parallax. The left panel displays a histogram that depicts the distribution of parallaxes. The blue steps indicate a $\sim$5200 Gaia star cluster sample (including GCs) within |$b$| < 10~deg, while the orange steps represent the sample that this work has discovered. On the right panel, the steps exhibits the proportion of increase in the number of known star clusters across varying parallaxes.}
\label{fig:plx}
\end{center}
\end{figure*}

This study focuses on distant OCs and is a follow-up to our previous searches in Gaia DR2 and EDR3~\citep{he21,he22a,he22b,he23a}. Our previous searches discovered 615 new clusters in Gaia DR2, 886 known and new clusters under 1.2~kpc, and 1665 new candidates in 1.2 to $\sim$5~kpc. In this work, as shown in Figure~\ref{fig:plx}, we increased the previously discovered OC sample size by up to three times beyond a distance of 5~kpc, detecting 1249 objects that $\varpi$ < 0.2~mas, with 846 of them being reliable Type~1 OCs. This increase is significant as it enables us to better study the faraway structures of the Milky Way.

\begin{figure*}
\begin{center}
\includegraphics[width=0.49\linewidth]{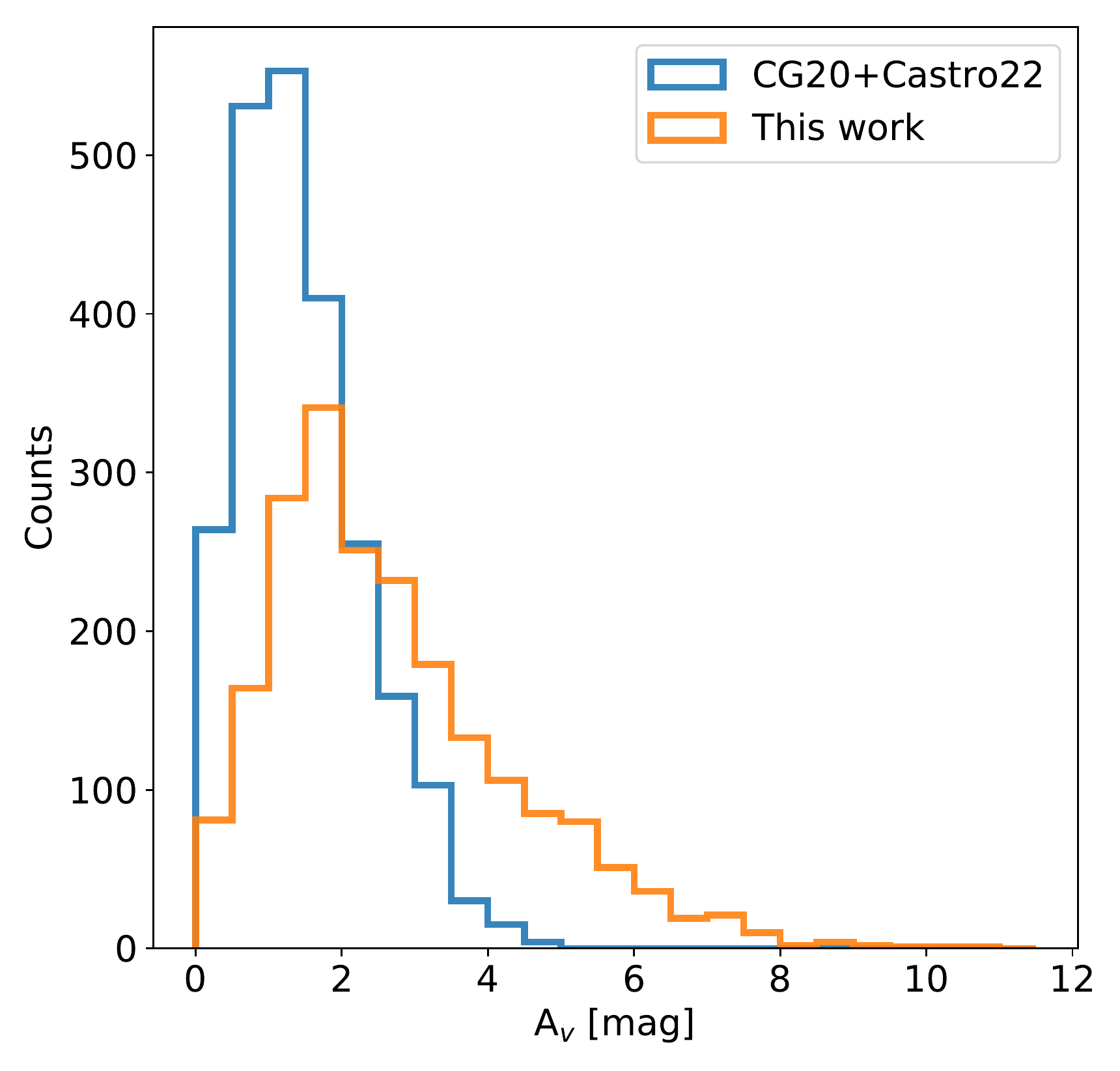}
\includegraphics[width=0.49\linewidth]{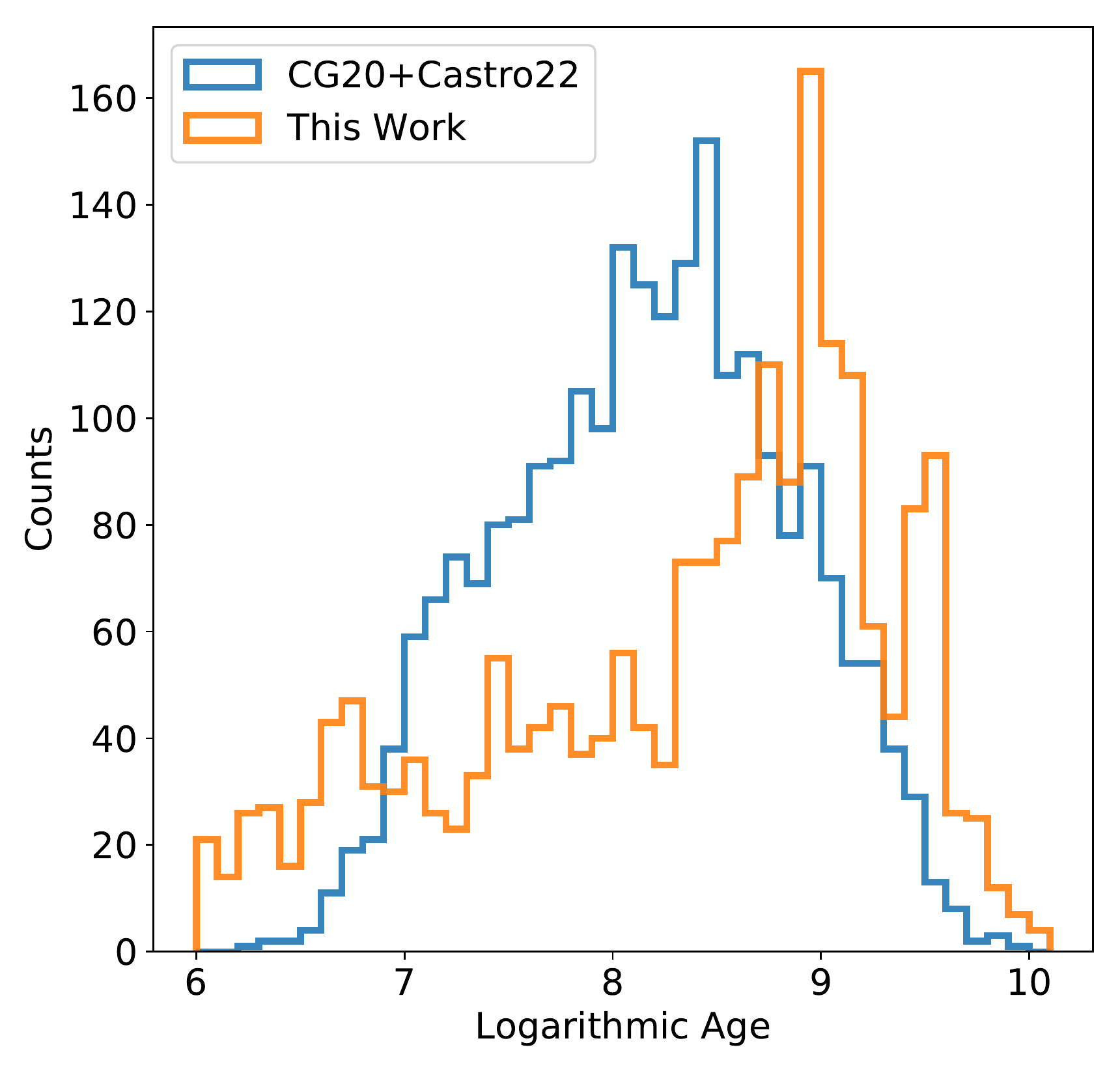}
\caption{Statistics on extinction (left panel) and cluster age (right panel). The newly found clusters have extinction limitation approximately 3 magnitudes greater than those previously detected, suggesting that they are considerably more obscured than the clusters located closer to the sun. This could explain the higher number of detections of old clusters with a logarithmic age greater than 9 (right panel), since the member stars in the main-sequence of these billion-years-old clusters are fainter compared to younger OCs, making them more difficult to detect due to observational limitations. }
\label{fig:av_age}
\end{center}
\end{figure*}

\begin{figure*}
\begin{center}
\includegraphics[width=0.245\linewidth]{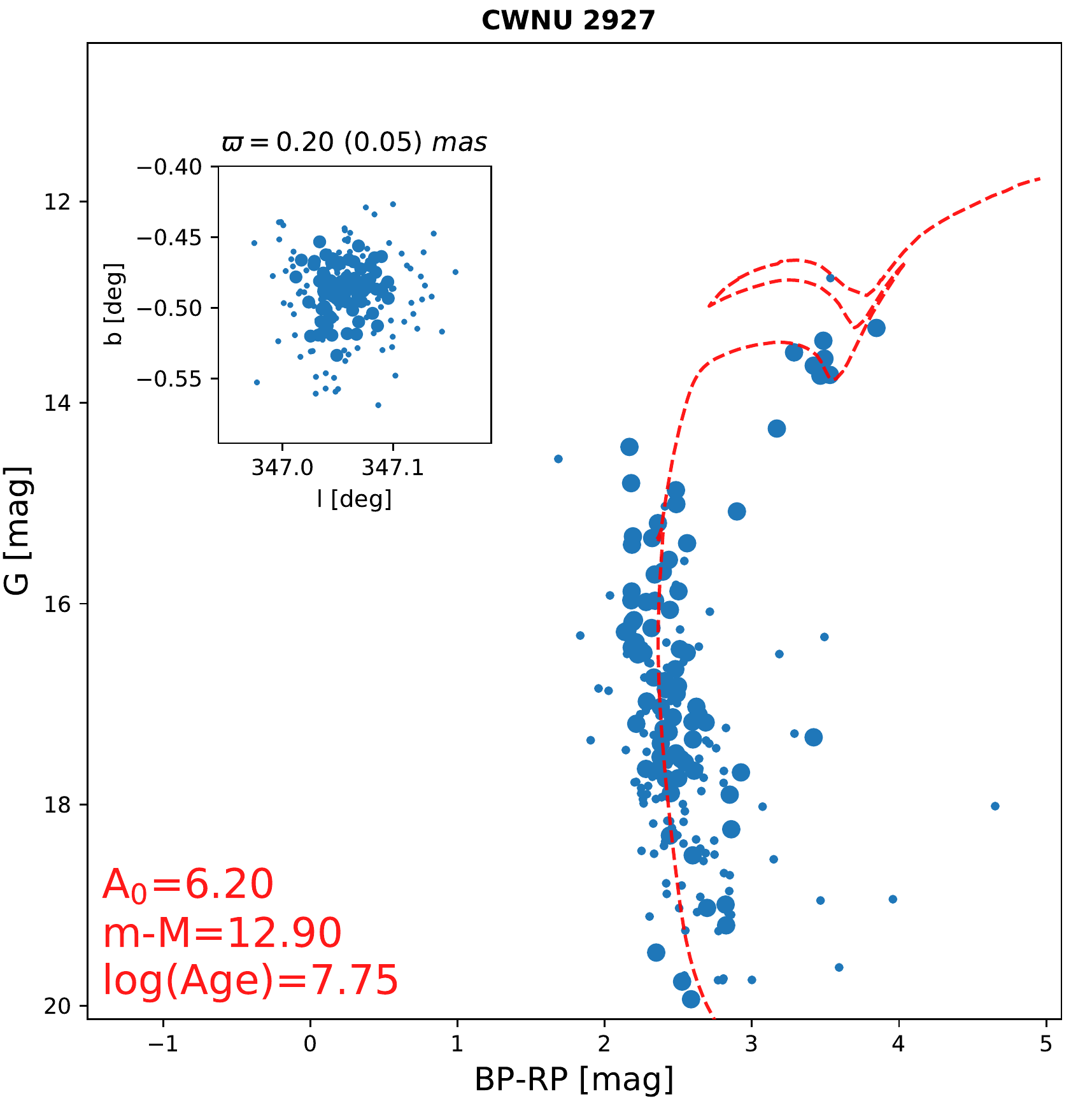}
\includegraphics[width=0.245\linewidth]{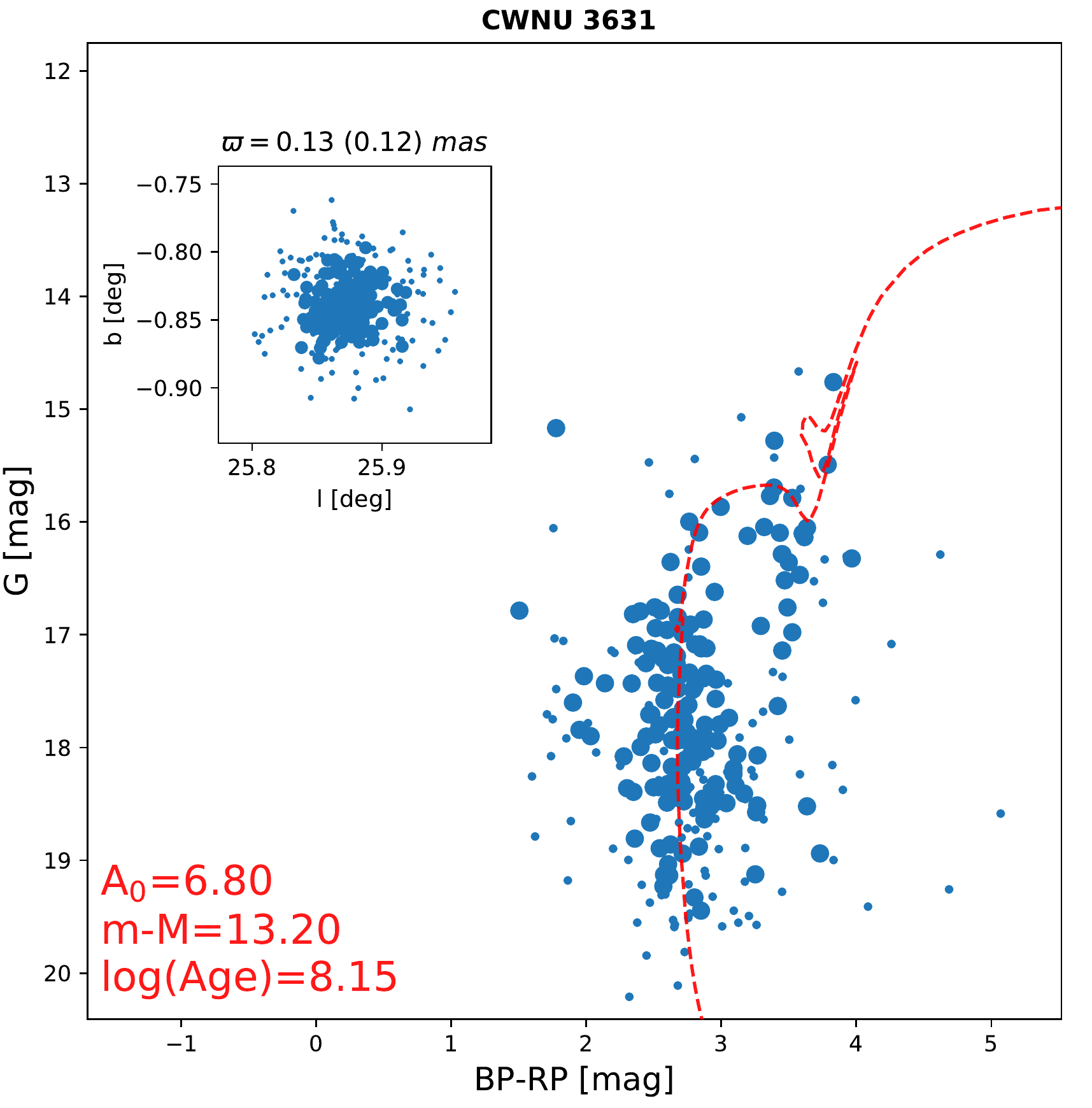}
\includegraphics[width=0.245\linewidth]{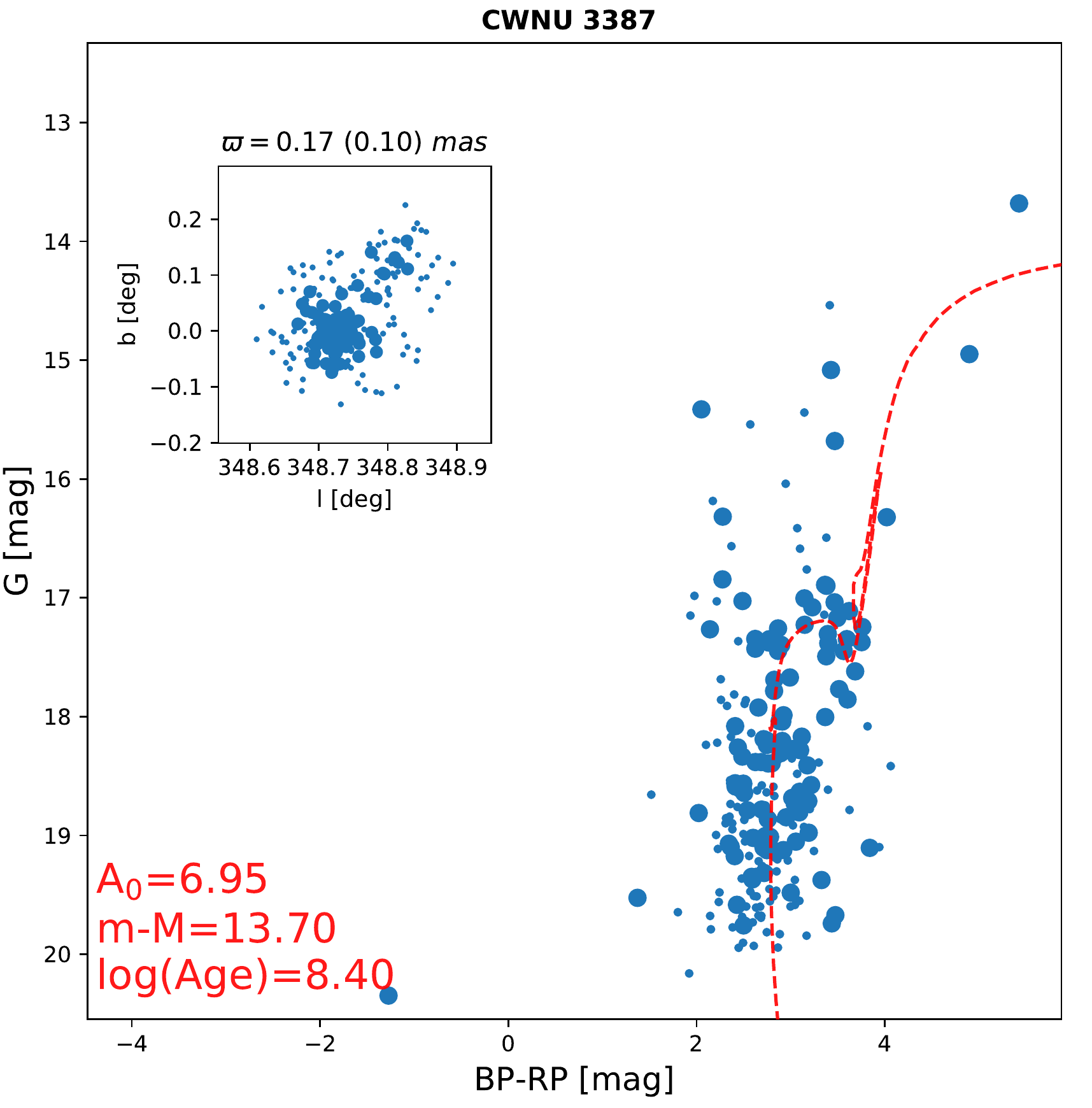}	
\includegraphics[width=0.245\linewidth]{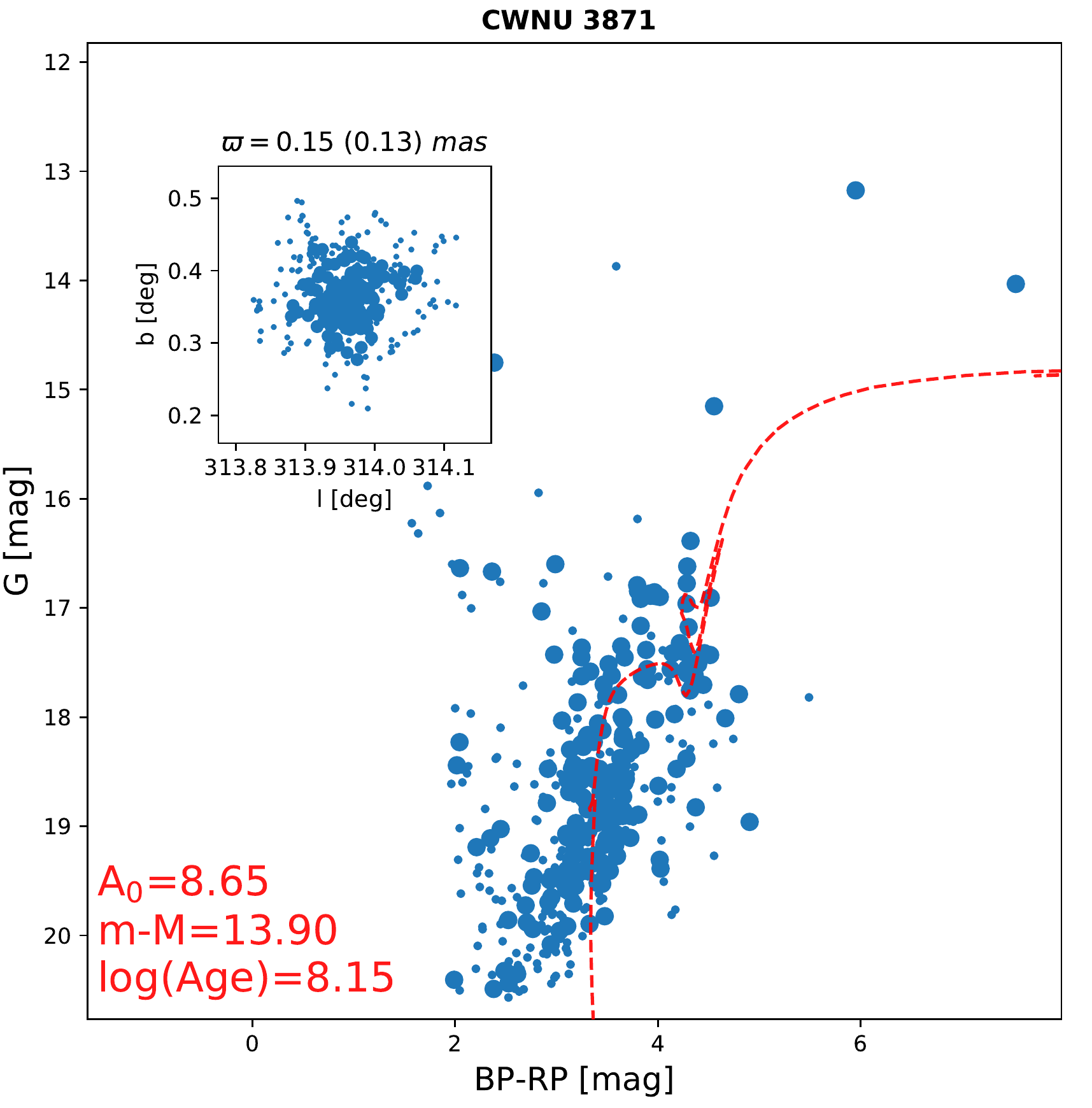}
\includegraphics[width=0.245\linewidth]{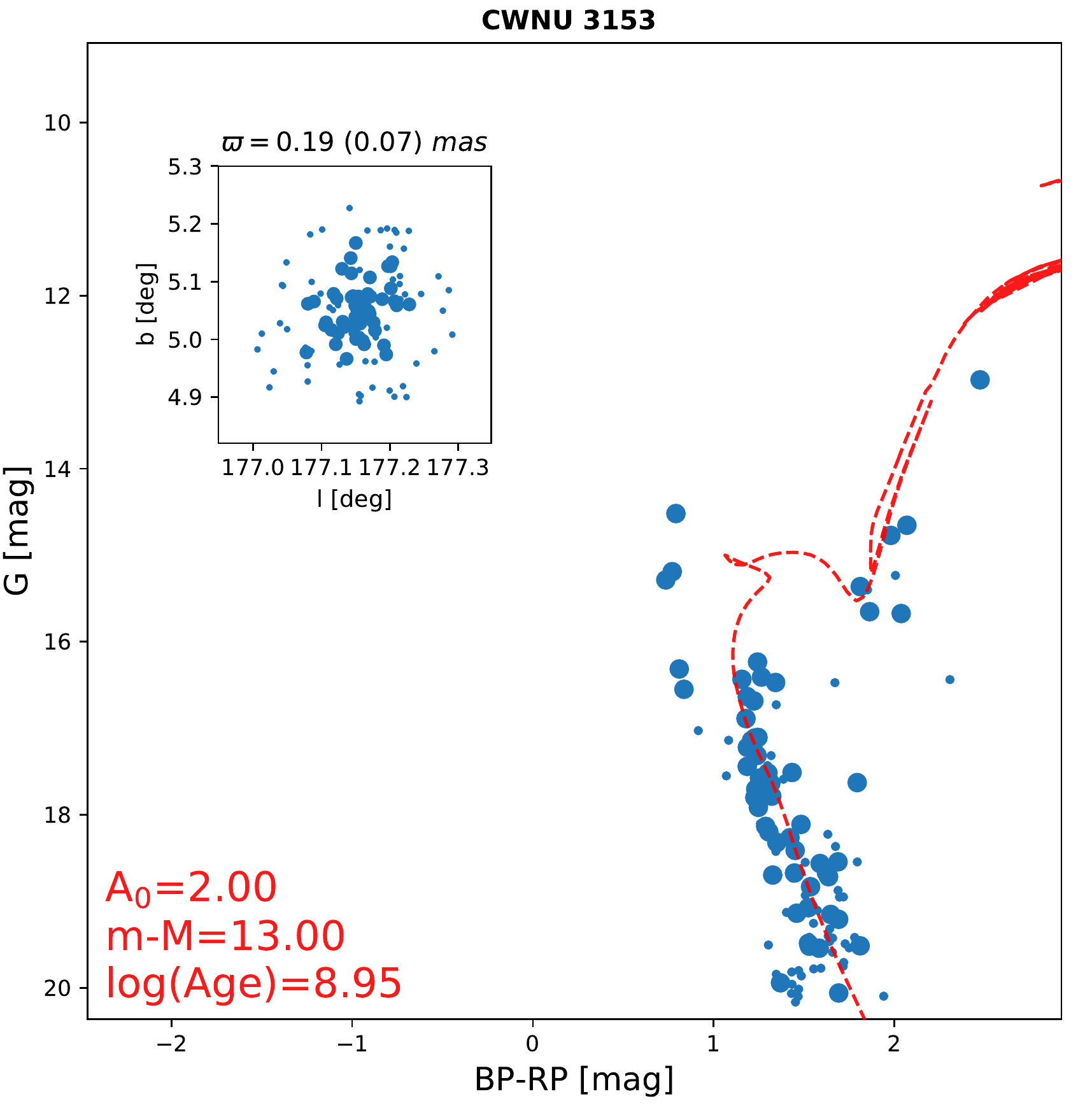}
\includegraphics[width=0.245\linewidth]{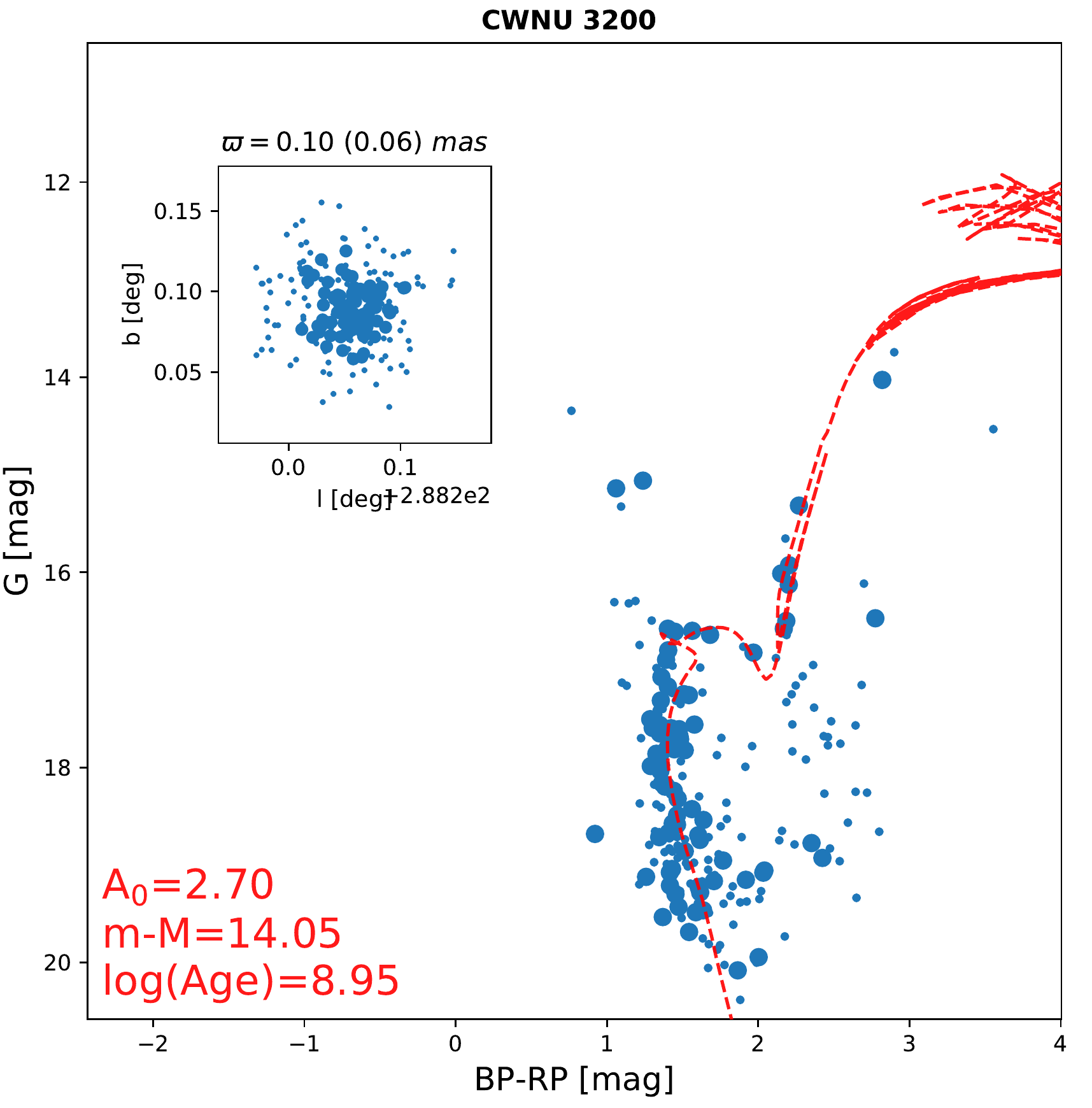}
\includegraphics[width=0.245\linewidth]{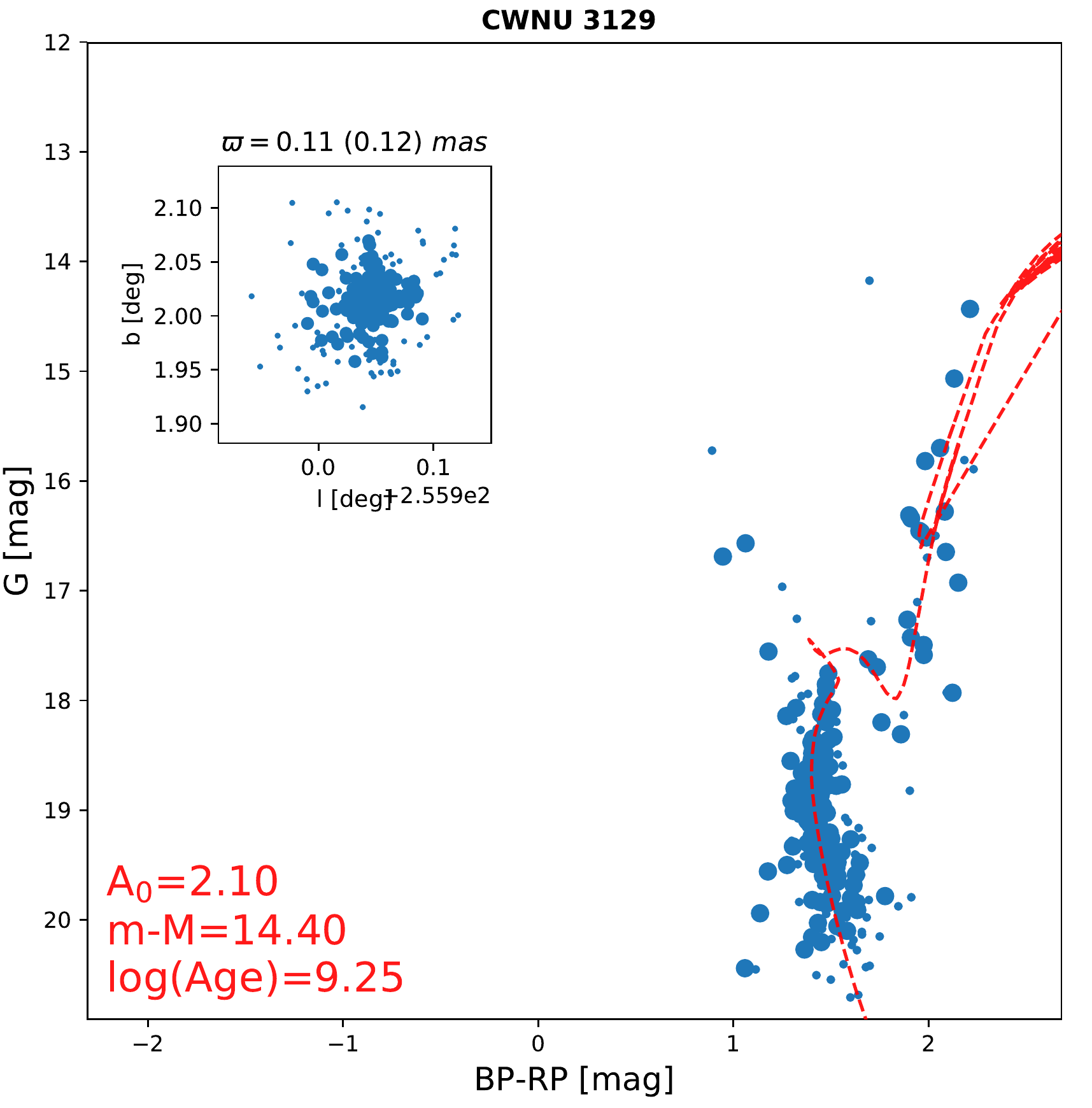}
\includegraphics[width=0.245\linewidth]{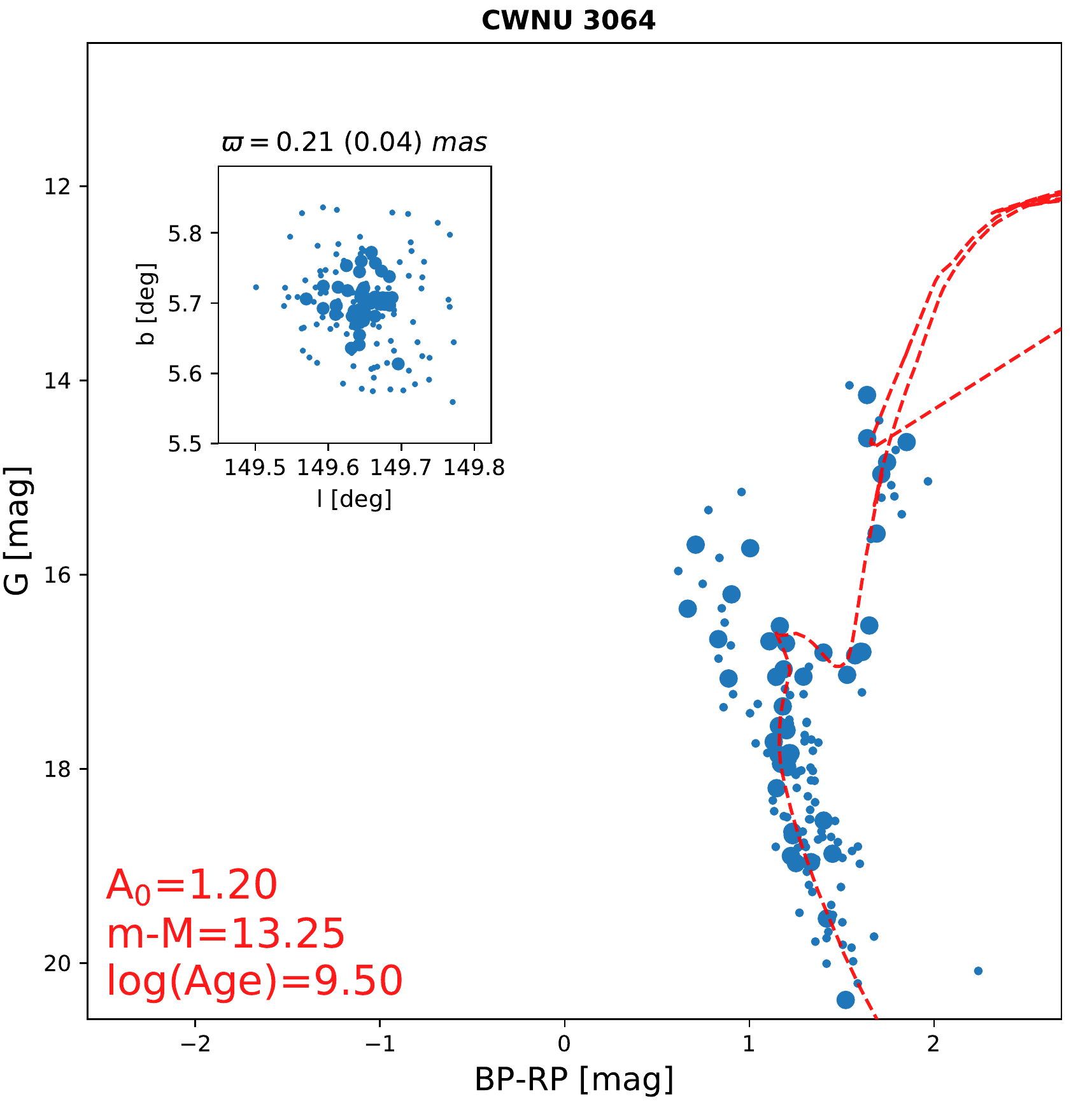}		
\caption{Example of highly extinguished new clusters (upper panels) and old clusters (lower panels) . The old OCs are likely to contain blue straggler stars that are distinctly visible, and lower levels of contamination make them more reliably identifiable.}
\label{fig:a_bss}
\end{center}
\end{figure*}

The new discoveries have larger extinction levels compared to before, with an increase of at least 3~mag (Figure~\ref{fig:av_age}). While we used stars fainter to $\sim$20 mag, comparable to the previous searches~\citep[e.g. 18~mag in CG20,][and our previous searches]{castro22}, most searches show that the Gaia data are limited for the extinction at 5~mag. However, we found that 190 clusters with higher extinction values can still be detected (Figure~\ref{fig:a_bss}), some of them are at logarithmic age $\sim$8.4 (e.g. FSR~0134 in Figure~\ref{fig:pre_gaia_ocs}) or larger, which could useful in studying the Galactic extinction. 

In addition, unlike in our previous discovery of young clusters, we found more old clusters, particularly those with logarithmic age larger than 9. We detected over 618 clusters in this category, with 309 of them being Type~1. Some of the old clusters contain visible BSSs (Figure~\ref{fig:a_bss}), and we listed newly discovered clusters that may have BSSs in the Appendix. These old clusters can be helpful in the study of the evolution of the stars and the Galaxy.

\subsection{Apparent radius and proper motion dispersion}\label{sec:dispersion}
\begin{figure*}
\begin{center}
\includegraphics[width=1.0\linewidth]{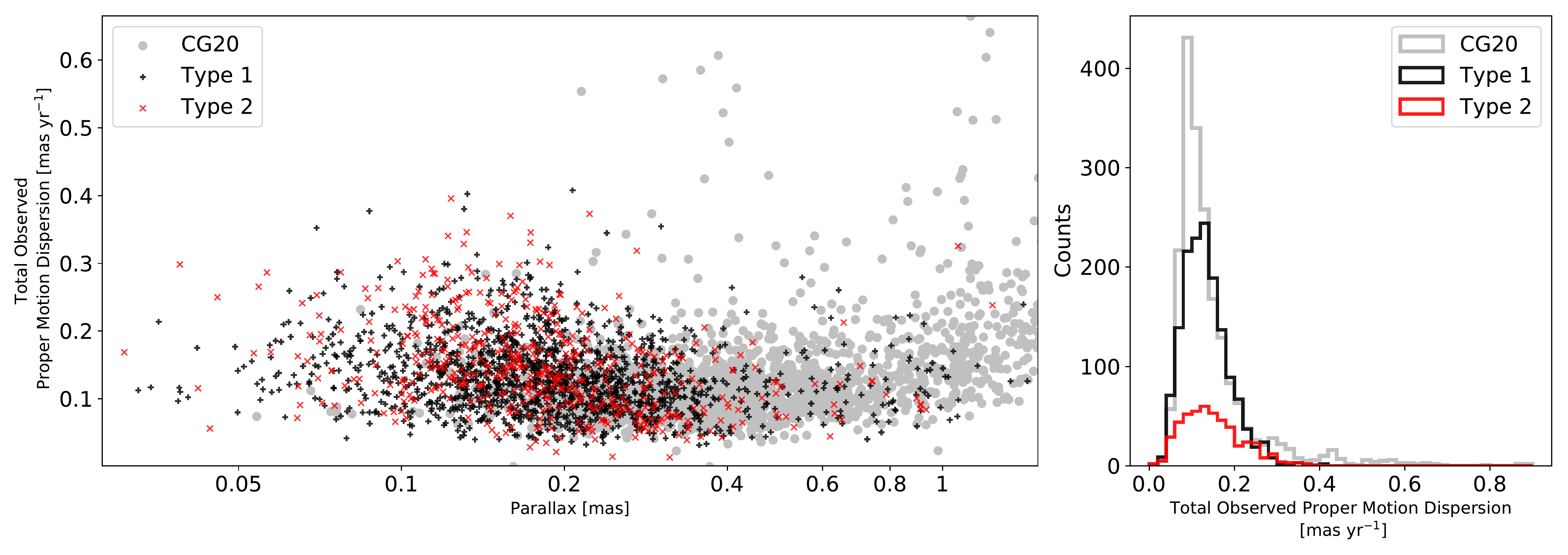}
\caption{A comparison between the total observed proper motion dispersions obtained from CG20, represented by grey dots (member stars were cross-matched in Gaia EDR3), and our work, indicated by black plus signs for Type~1 clusters and red crosses for Type~2 clusters. The x-axis in the left plot is presented in logarithmic scale parallax, while the right plot displays the histograms of the three sets.}
\label{fig:sigpm}
\end{center}
\end{figure*}

To compare our new discoveries with previously reported reliable OCs, we applied the method created by CG20, as done in our previous studies~\citep{he21}, to verify if the new clusters have OC-like apparent radius and proper motion dispersions. Figure~\ref{fig:sigpm} shows the total proper motion dispersion (TPD) of new clusters in Type~1 and Type~2, with all statistical values in CG20 being in a Gaia EDR3 membership, as its dispersions are lower under a smaller astrometric error. The TPD of both sets is comparable to CG20's results, with Type~1 clusters showing slightly higher dispersion since more detections of faint member stars; and Type~2 candidates do not differ significantly from Type~1 clusters. The low-dispersion in proper motion may be due to the proper motion cut in our TGFIG method (Section~\ref{sec:TGFIG}). 

\begin{figure*}
\begin{center}
\includegraphics[width=1.0\linewidth]{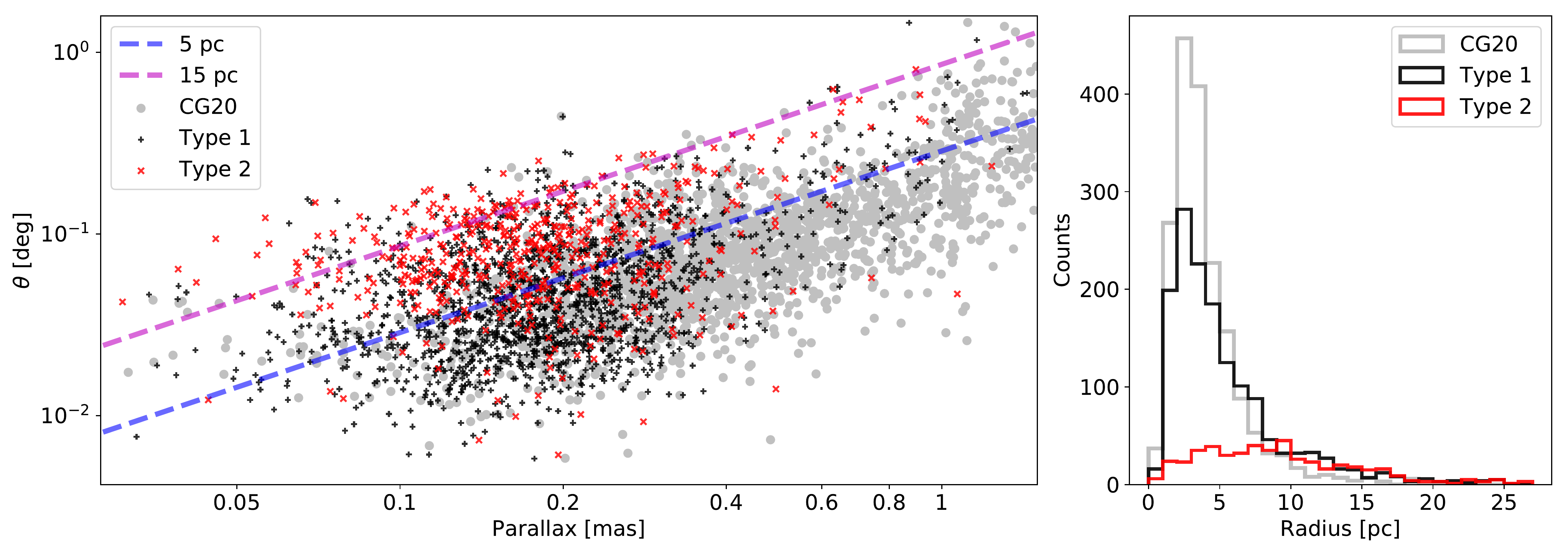}
\caption{Same as Figure~\ref{fig:sigpm}, but for the apparent radius, the blue and magenta dashed lines indicate the angular sizes corresponding to 5 and 15 parsecs, respectively.}
\label{fig:siglb}
\end{center}
\end{figure*}

\begin{figure*}
\begin{center}
\includegraphics[width=0.5\linewidth]{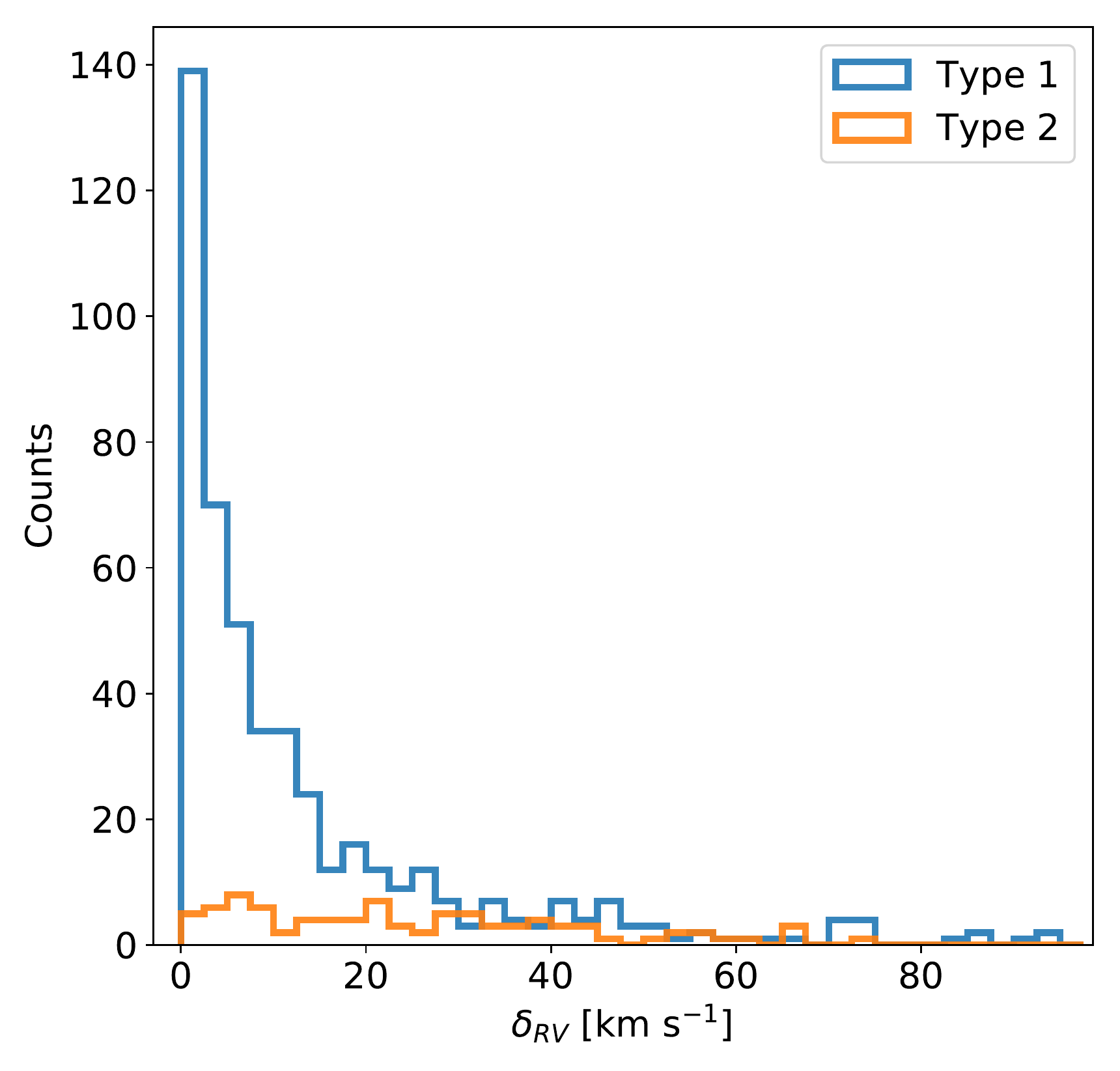}
\caption{The radial velocity dispersion for newly detected clusters, where blue steps represent Type~1 clusters and orange steps represent Type~2 clusters. Only clusters with more than one matched member star in the Gaia DR3 radial velocity database were included.}
\label{fig:rv}
\end{center}
\end{figure*}

The apparent radius (Figure~\ref{fig:siglb}) shows that Type~1 clusters are still comparable to CG20 clusters, but Type~2 clusters are larger than the latter. However, some of the Type~2 clusters have cluster-like CMDs and may be related to moving groups or stellar streams~\citep{Kounkel19,Kounkel20}, requiring further research in future works. We also investigated the radial velocity dispersion (Figure~\ref{fig:rv}), and the Type~1 clusters are more reliable than the Type~2 ones. However, the larger dispersion (especially >~10 \kms) presented in Type~1 clusters shows that field star contamination exists in the new discoveries, necessitating deep membership determinations in future works. These issues highlight the challenges associated with OC searches and member star determinations.

\begin{longrotatetable}
\begin{deluxetable*}{cccccccccccccccccccc}
\tablecaption{Derived astrophysical parameters for the 2085 clusters detected in this work. The positions, parallax, proper motions, and radial velocities of each cluster are calculated as the median value, and the dispersion of each astrometric value are also presented. The typical fitted accuracy for distance modulus, A$_0$, and logarithmic age are 0.05, 0.05~mag, and 0.05~dex, respectively.}
\label{tab:tab_all}
\tablewidth{1.0pt}
\tabletypesize{\scriptsize}
\tablehead{ID  & $l$  & $\sigma_{l}$ & $b$ & $\sigma_{b}$ & N & N$_{core}$ & parallax & $\sigma_{parallax}$ & pmra & $\sigma_{pmra}$ & pmdec & $\sigma_{pmdec}$  & m-M & A$_0$ & Logarithmic Age &RV &$\sigma_{RV}$ & N$_{RV}$ & Type \\ 
& [$^{\circ}$] & [$^{\circ}$] & [$^{\circ}$] & [$^{\circ}$] &  & & [mas] & [mas] & [mas~  yr$^{-1}$] & [mas~  yr$^{-1}$] & [mas~  yr$^{-1}$] & [mas~  yr$^{-1}$] &   & [mag]  & [dex]  & [km~  s$^{-1}$] &[km~  s$^{-1}$] &  & }
\startdata
CWNU  2927 & 347.05 & 0.02 & -0.49 & 0.01 &194&78& 0.20 & 0.05 & -0.79 & 0.11 & -2.27 & 0.07 & 12.90& 6.20& 7.75& -35.2& 1.1 & 8 & 1\\ 
CWNU  2928 & 157.39 & 0.13 & -5.82 & 0.16 &55&26& 0.55 & 0.02 & 0.23 & 0.09 & -2.74 & 0.06 & 11.35& 1.80& 6.85& -31.7& 6.1 & 1 & 1\\ 
CWNU  2929 & 75.09 & 0.07 & -2.71 & 0.04 &84&37& 0.15 & 0.04 & -3.26 & 0.11 & -5.50 & 0.13 & 14.00& 4.60& 9.00& -& - & 0 & 1\\ 
CWNU  2930 & 235.34 & 0.05 & 2.06 & 0.02 &86&35& 0.18 & 0.06 & -1.23 & 0.12 & 1.88 & 0.13 & 13.25& 1.50& 8.80& 63.1& 2.1 & 1 & 1\\ 
CWNU  2931 & 246.10 & 0.01 & -0.26 & 0.02 &56&33& 0.12 & 0.08 & -1.53 & 0.12 & 1.61 & 0.06 & 13.85& 2.35& 8.65& -& - & 0 & 1\\ 
SAI  88 & 258.52 & 0.01 & -4.22 & 0.01 &79&33& 0.13 & 0.07 & -1.13 & 0.10 & 3.32 & 0.17 & 13.65& 3.55& 9.30& -& - & 0 & 1\\ 
CWNU  2932 & 112.28 & 0.11 & -2.98 & 0.06 &131&48& 0.47 & 0.02 & -3.88 & 0.08 & -2.92 & 0.07 & 11.55& 2.60& 8.60& -37.2& 3.8 & 2 & 1\\ 
CWNU  2933 & 278.29 & 0.03 & -4.99 & 0.01 &69&35& 0.06 & 0.02 & -3.12 & 0.11 & 3.14 & 0.08 & 14.55& 1.80& 9.15& 57.3& 0.5 & 1 & 1\\ 
CWNU  2934 & 104.81 & 0.02 & -1.61 & 0.01 &81&17& 0.16 & 0.03 & -2.74 & 0.07 & -1.44 & 0.11 & 13.45& 2.10& 8.50& -52.2& 2.8 & 1 & 1\\ 
CWNU  2935 & 153.38 & 0.03 & 0.52 & 0.04 &120&33& 0.16 & 0.05 & 0.20 & 0.10 & 0.11 & 0.07 & 13.30& 3.50& 8.70& -35.9& 1.1 & 4 & 1\\ 
CWNU  2936 & 182.69 & 0.20 & -7.34 & 0.12 &102&50& 0.75 & 0.06 & -0.45 & 0.14 & 1.08 & 0.10 & 10.85& 2.05& 9.25& 32.8& 9.4 & 4 & 1\\ 
CWNU  2937 & 205.47 & 0.04 & -1.69 & 0.03 &56&18& 0.15 & 0.06 & -0.17 & 0.09 & 0.16 & 0.06 & 13.95& 2.55& 6.80& -& - & 0 & 1\\ 
CWNU  2938 & 239.24 & 0.09 & 8.07 & 0.04 &120&34& 0.19 & 0.05 & -2.18 & 0.10 & 2.48 & 0.09 & 13.45& 0.55& 9.05& 105.1& 12.9 & 2 & 1\\ 
CWNU  2939 & 232.53 & 0.05 & -1.12 & 0.06 &105&40& 0.26 & 0.06 & -0.58 & 0.06 & 0.33 & 0.06 & 12.95& 2.70& 8.15& 57.4& 0.4 & 1 & 1\\ 
CWNU  2940 & 109.56 & 0.01 & 4.69 & 0.02 &101&49& 0.13 & 0.05 & -1.95 & 0.08 & -1.68 & 0.09 & 13.90& 3.50& 8.45& -96.3& 0.6 & 3 & 1\\ 
... & ... & ... & ... & ... & ... &... & ... & ... & ... & ... & ... & ... & ... & ... & ... & ... & ... & ... & ...\\ 
CWNU  4378 & 102.54 & 0.05 & -4.47 & 0.05 &81&20& 0.16 & 0.07 & -2.20 & 0.07 & -1.42 & 0.13 & 13.95& 0.90& 9.50& -32.5& 3.6 & 1 & 2\\ 
Sgr  OB6 & 14.31 & 0.13 & 1.30 & 0.19 &149&26& 0.36 & 0.05 & -0.66 & 0.09 & -2.62 & 0.08 & 12.00& 5.05& 8.60& 50.4& 5.3 & 2 & 2\\ 
CWNU  4379 & 289.58 & 0.10 & 3.05 & 0.03 &71&16& 0.12 & 0.03 & -5.34 & 0.08 & 1.34 & 0.06 & 13.80& 1.25& 9.50& 2.8& 15.2 & 2 & 2\\ 
CWNU  4380 & 291.49 & 0.05 & -4.51 & 0.03 &116&17& 0.09 & 0.04 & -4.43 & 0.13 & 1.89 & 0.13 & 14.60& 1.60& 9.50& 35.4& 3.8 & 1 & 2\\ 
CWNU  4381 & 3.61 & 0.08 & -7.02 & 0.02 &49&20& 0.18 & 0.04 & -5.12 & 0.21 & -4.72 & 0.22 & 14.15& 1.75& 9.05& 48.2& 7.1 & 1 & 2\\ 
Hogg  13 & 291.37 & 0.02 & 0.44 & 0.05 &55&20& 0.16 & 0.03 & -5.59 & 0.05 & 1.97 & 0.07 & 13.35& 2.25& 9.05& -& - & 0 & 2\\ 
... & ... & ... & ... & ... & ... &... & ... & ... & ... & ... & ... & ... & ... & ... & ... & ... & ... & ... & ...\\ 
FSR 1700 & 322.90 & 0.02 & -3.04 & 0.02 &217&86& 0.05 & 0.09 & -4.86 & 0.11 & -4.01 & 0.11 & -& -& -& 10.4& 5.5 & 2 & GCC\\ 
Kronberger 119 & 289.52 & 0.01 & -3.73 & 0.01 &61&28& 0.07 & 0.08 & -3.84 & 0.08 & 1.77 & 0.06 & -& -& -& 97.8& 6.3 & 3 & GCC\\ 
Kronberger 143 & 297.12 & 0.02 & -1.91 & 0.03 &186&65& 0.10 & 0.09 & -7.75 & 0.17 & 0.65 & 0.18 & -& -& -& 17.8& 1.9 & 3 & GCC\\ 
He Zhihong 1 & 95.98 & 0.02 & 4.57 & 0.02 &195&56& 0.07 & 0.07 & -2.40 & 0.13 & -1.98 & 0.11 & -& -& -& -138.2& 2.9 & 2 & GCC\\ 
He Zhihong 2 & 279.41 & 0.03 & 1.78 & 0.03 &80&36& 0.04 & 0.06 & -3.71 & 0.09 & 2.16 & 0.08 & -& -& -& 87.7& 23.6 & 2 & GCC\\ 
Patchick 125 & 349.76 & 0.02 & 3.42 & 0.02 &146&86& 0.10 & 0.19 & -3.77 & 0.19 & 0.68 & 0.22 & -& -& -& 93.8& 6.7 & 5 & GCC\\ 
Patchick 126 & 340.38 & 0.01 & -3.82 & 0.01 &37&21& -0.01 & 0.10 & -4.91 & 0.10 & -6.78 & 0.15 & -& -& -& -124.0& 0.4 & 3 & GCC\\ 
Gran  2 & 359.23 & 0.01 & 8.59 & 0.01 &125&64& 0.03 & 0.15 & 0.20 & 0.18 & -2.57 & 0.16 & -& -& -& 61.2& 6.0 & 1 & GCC\\ 
Gran  4 & 10.20 & 0.03 & -6.39 & 0.03 &314&159& 0.07 & 0.13 & 0.49 & 0.17 & -3.52 & 0.18 & -& -& -& -262.6& 1.6 & 3 & GCC\\ 
\enddata
\end{deluxetable*}
\end{longrotatetable}

\begin{longrotatetable}
\begin{deluxetable*}{cccccccccccccccc}
\tablecaption{The astrometric and photometric parameters of the member stars from Gaia DR3~\citep{gaia2021}. The core member are labeled as $ifcore$ = 1 from DBSCAN.}
\tablewidth{0pt}
\label{tab:table_mem}
\tabletypesize{\scriptsize}
\tablehead{ source$\_$id&	$l	$ & $b$	& ruwe &parallax&	parallax$\_$error	&pmra	&pmra$\_$error&	pmdec&	pmdec$\_$error& radia$\_$velocity &	radial$\_$velocity$\_$error &phot$\_$g$\_$mean$\_$mag	&bp$\_$rp  & $ifcore$ & Cluster ID\\ 
& [$^{\circ}$]  & [$^{\circ}$]  & & [mas] & [mas] & [mas~yr$^{-1}$] & [mas~yr$^{-1}$] & [mas~yr$^{-1}$] & [mas~yr$^{-1}$]   & [km~  s$^{-1}$] &[km~  s$^{-1}$] & [mag]&[mag]&  & }
\startdata
5972163329446185856 & 347.044 & -0.490 & 1.067 & 0.125 & 0.15 & -0.64 & 0.20 & -2.09 & 0.14 & - & - & 17.91 & 2.42 &  0& CWNU 2927 \\ 
5972163333755712384 & 347.041 & -0.493 & 0.988 & 0.206 & 0.29 & -0.95 & 0.40 & -2.56 & 0.24 & - & - & 18.89 & 2.42 &  0& CWNU 2927 \\ 
5972163333763799168 & 347.049 & -0.493 & 0.994 & 0.205 & 0.04 & -0.79 & 0.06 & -2.27 & 0.04 & - & - & 15.71 & 2.34 &  1& CWNU 2927 \\ 
5972163333780135936 & 347.038 & -0.499 & 1.024 & 0.191 & 0.03 & -0.74 & 0.04 & -2.28 & 0.03 & -35.57 & 0.58 & 13.56 & 3.49 &  1& CWNU 2927 \\ 
5972163333780140800 & 347.037 & -0.496 & 1.012 & 0.215 & 0.03 & -0.72 & 0.04 & -2.09 & 0.03 & - & - & 15.03 & 2.41 &  0& CWNU 2927 \\ 
5972163333780154496 & 347.041 & -0.488 & 1.039 & 0.198 & 0.10 & -0.83 & 0.13 & -2.36 & 0.09 & - & - & 17.28 & 2.44 &  1& CWNU 2927 \\ 
5972163333780157312 & 347.051 & -0.488 & 1.334 & 0.206 & 0.04 & -0.83 & 0.05 & -2.43 & 0.03 & -31.07 & 1.93 & 14.26 & 3.17 &  1& CWNU 2927 \\ 
5972163333780413696 & 347.048 & -0.496 & 1.048 & 0.341 & 0.09 & -1.04 & 0.13 & -2.24 & 0.09 & - & - & 17.31 & 2.36 &  0& CWNU 2927 \\ 
5972163333780414208 & 347.046 & -0.492 & 1.014 & 0.170 & 0.03 & -0.79 & 0.04 & -2.21 & 0.02 & -35.91 & 0.79 & 13.73 & 3.47 &  1& CWNU 2927 \\ 
5972163363805079040 & 347.038 & -0.477 & 1.022 & 0.349 & 0.07 & -0.94 & 0.08 & -2.23 & 0.06 & - & - & 16.65 & 2.48 &  1& CWNU 2927 \\ 
5972163363805081088 & 347.039 & -0.475 & 0.987 & 0.168 & 0.05 & -0.76 & 0.05 & -2.48 & 0.04 & - & - & 16.02 & 2.38 &  0& CWNU 2927 \\ 
5972163368139901184 & 347.033 & -0.481 & 0.969 & 0.279 & 0.05 & -0.88 & 0.06 & -2.21 & 0.04 & - & - & 15.88 & 2.50 &  1& CWNU 2927 \\ 
5972163368139905152 & 347.038 & -0.480 & 0.958 & 0.194 & 0.11 & -0.69 & 0.13 & -2.28 & 0.10 & - & - & 17.58 & 2.55 &  1& CWNU 2927 \\ 
5972163398164818944 & 347.045 & -0.484 & 1.015 & 0.200 & 0.03 & -0.86 & 0.03 & -2.27 & 0.02 & -35.07 & 0.46 & 13.38 & 3.49 &  1& CWNU 2927 \\ 
5972163398164822400 & 347.050 & -0.483 & 1.004 & 0.225 & 0.05 & -0.91 & 0.06 & -2.13 & 0.04 & - & - & 16.18 & 2.19 &  1& CWNU 2927 \\ 
5972163398164824960 & 347.055 & -0.483 & 0.945 & 0.271 & 0.05 & -0.77 & 0.06 & -2.18 & 0.04 & - & - & 15.97 & 2.18 &  1& CWNU 2927 \\ 
... & ... & ... & ... & ... & ... & ... & ... & ... & ... & ... & ... & ... & ... & ...& ...\\ 
4054570833932778880 & 355.863 & -0.318 & 1.023 & 0.210 & 0.16 & -0.34 & 0.21 & -1.83 & 0.15 & - & - & 18.31 & 2.78 &  0& Trumpler 28 \\ 
4054572139590192512 & 355.878 & -0.248 & 0.968 & 0.276 & 0.39 & -0.15 & 0.49 & -2.11 & 0.32 & - & - & 19.45 & 3.01 &  0& Trumpler 28 \\ 
4054946145345254528 & 355.890 & -0.343 & 1.093 & 0.261 & 0.07 & -0.36 & 0.08 & -1.54 & 0.06 & - & - & 16.61 & 2.21 &  0& Trumpler 28 \\ 
4054934222516012160 & 355.891 & -0.387 & 1.041 & 0.268 & 0.03 & -0.35 & 0.04 & -1.68 & 0.03 & -20.03 & 3.17 & 14.97 & 3.31 &  0& Trumpler 28 \\ 
4054933153058758528 & 355.891 & -0.442 & 1.109 & 0.270 & 0.09 & -0.25 & 0.11 & -1.80 & 0.08 & - & - & 16.94 & 2.48 &  0& Trumpler 28 \\ 
4054946523292129024 & 355.920 & -0.316 & 1.068 & 0.397 & 0.38 & 0.06 & 0.51 & -1.95 & 0.35 & - & - & 19.59 & 2.94 &  0& Trumpler 28 \\ 
4054946660731011200 & 355.930 & -0.347 & 1.151 & 0.257 & 0.03 & -0.35 & 0.03 & -2.03 & 0.02 & -11.34 & 1.26 & 14.01 & 3.12 &  0& Trumpler 28 \\ 
... & ... & ... & ... & ... & ... & ... & ... & ... & ... & ... & ... & ... & ... & ...& ...\\ 
4112518154715652736 & 359.301 & 8.513 & 1.059 & -0.035 & 0.30 & 0.61 & 0.33 & -2.46 & 0.23 & - & - & 19.07 & 1.80 &  0& Gran 2 \\ 
4112518189075186944 & 359.302 & 8.467 & 1.100 & -0.112 & 0.45 & 0.22 & 0.50 & -2.59 & 0.34 & - & - & 19.60 & 1.93 &  0& Gran 2 \\ 
4112521556329887744 & 359.311 & 8.540 & 1.012 & 0.043 & 0.16 & 0.31 & 0.17 & -2.35 & 0.12 & - & - & 18.26 & 1.78 &  0& Gran 2 \\ 
4112521762488381440 & 359.314 & 8.554 & 0.995 & -0.170 & 0.23 & 0.08 & 0.27 & -3.22 & 0.19 & - & - & 18.88 & 1.87 &  0& Gran 2 \\ 
4112569793124317312 & 359.316 & 8.682 & 1.133 & -0.211 & 0.26 & 0.23 & 0.29 & -3.08 & 0.20 & - & - & 18.87 & 1.51 &  0& Gran 2 \\ 
4112521487616969216 & 359.319 & 8.518 & 1.014 & 0.404 & 0.53 & -0.07 & 0.63 & -2.77 & 0.43 & - & - & 20.00 & 1.95 &  0& Gran 2 \\ 
\enddata
\end{deluxetable*}
\end{longrotatetable}

\section{Summary}\label{sec:summary}
We present the results of a survey aimed at identifying the most distant Galactic star clusters in the Gaia Era.
In the study, we conducted a continuous search for distant star clusters as a follow-up to our previous search in Gaia DR2~\citep{he21,he22a} and nearby~\citep{he22b}, middle-distance~\citep{he23a} clusters. We employed an improved method, TGFIG, to search for the |$b$| < 10~deg sky, cataloging 2085 objects, including 28 re-detected clusters cataloged in ~\citet{dias21}, 1462 newly identified reliable OCs, and 567 OC candidates, 28 GC candidates (16 of which were cataloged before). Importantly, we present a local dwarf galaxy (IC~10) in the study, this may be the first time that the two instruments (Gaia and VLBA) have been used for mutual verification at a distance of hundreds of kpc, confirming the powerful astrometric capabilities of these tools. We conducted isochrone fits for each OC and OCC and provided the membership for each cluster. The cluster members were classified as core members ($ifcore$ = 1) or outer members ($ifcore$ = 0). 

The main idea for the TGFIG method involved clipping other dense parts of the search field, ensuring the result was the densest in the data set. We then conducted two-Gaussian fits to the k$_{th}$NND histogram, calculating the DBSCAN coefficient ($\varepsilon$) as the cross-point. Our TGFIG method proved more efficient than our previous works, and our new findings showed that the method is useful in distant cluster searches, particularly those for highly extinguished and old clusters. Our study demonstrated that the Gaia data allowed for the identification of OCs with A${_v}$ > 5~mag and/or distance more than 5 even to 10~kpc, which are valuable for studying the Galactic structure and evolutions. 

However, the limitations of searching for OCs in Gaia data remain unclear. Our search results have shown that the limitations depend on factors such as extinction, distance, and cluster age. We found that clusters located within 3 to 5~kpc from the sun are often undetectable with A${_v}$ $\geq$ 10~mag and a logarithmic age of around 8.5, indicating that older clusters may be concealed by interstellar extinction and distance. On the other hand, young OCs surrounded by star-forming regions are often embedded clusters, making them difficult to identify. However, despite their challenging nature, their brighter magnitudes may make them easier to detect than older OCs, especially in regions near the sun.

We cross-matched our data with 4114 reliable OCs in HR23 within a 3-sigma range, and identified 232 clusters (matched to 234 reported clusters in HR23), and we present a matched catalog in an online table. Though HR23 also reported an additional 3086 objects that were not classified as reliable clusters, we found 362 of these objects nearby our identified clusters (357 ones). We believe that our matched Type~1 clusters are reliable OCs or GCs detected through the TGFIG method, based on the membership determination and CMD/isochrone data (e.g. CNNU~3451 in Figure~\ref{fig:kds}, pre-Gaia OCs in Figure~\ref{fig:pre_gaia_ocs}, and Gran~4 in Figure~\ref{fig:gcc}). However, more studies are needed to verify the physical origin of those reported objects.

The forthcoming release of Gaia DR4 could greatly improve astrometric data accuracy, and ultimately help to distinguish between physical and random bound clusters in astrometric data, as well as detecting diversity of stellar aggregates, such as OCs, GCs, stellar streams, globular/irregular dwarfs, and extragalactic galaxies.
Although the TGFIG method shows high efficiency in cluster searches, we believe it is not comprehensive and that additional methods should be developed or improved to increase the likelihood of discovering new clusters. With every new discovery of star clusters, astronomers are better able to study the Milky Way and to achieve the primary goals of the Gaia mission.

\section{Acknowledgements}
This work has made use of data from the European Space Agency (ESA) mission GAIA (\url{https://www.cosmos.esa.int/gaia}), processed by the GAIA Data Processing and Analysis Consortium (DPAC,\url{https://www.cosmos.esa.int/web/gaia/dpac/consortium}). Funding for the DPAC has been provided by national institutions, in particular the institutions participating in the GAIA Multilateral Agreement.
This work is supported by "Young Data Scientists" project of the National Astronomical Data Center, CAS; and Fundamental Research Funds of China West Normal University (CWNU, No.21E030), the Innewtion Team Funds of CWNU, and the Sichuan Youth Science and Technology innovation Research Team (21CXTD0038).

\appendix\label{sec:appendix}

\begin{table}[!ht]
\label{tab :bsstab}
\caption{A catalog of newly detected clusters (Type~1 ) exhibits a potential presence of blue straggler stars, with the left portion of the cluster demonstrating a higher likelihood of containing real stragglers, while the right section shows a lower probability.}
\centering
\begin{tabular}{cccccc}
\hline
\hline
\multicolumn{3}{c}{\centering High credibility}&\multicolumn{3}{c}{\centering Low credibility} \\
\hline
\textbf{ID}& \textbf{Logarithmic age} & ~ & \textbf{} & \textbf{ID} &\textbf{Logarithmic age}\\
\hline
        CWNU 3153 & 8.95 & ~ & ~ & CWNU 3114 & 8.70 \\ 
        CWNU 3200 & 8.95 & ~ & ~ & CWNU 3154 & 8.90 \\ 
        Teutsh 48 & 9.00 & ~& ~ & CWNU 3629 & 9.00 \\ 
        CWNU 3096 & 9.10 & ~& ~ & CWNU 3085 & 9.00 \\ 
        CWNU 2987 & 9.15 & ~& ~ & CWNU 3825 & 9.05 \\ 
        Saurer 4 & 9.15 & ~& ~ & FSR 0291 & 9.15 \\ 
        CWNU 3190 & 9.15 & ~ & ~& CWNU 3267 & 9.15 \\ 
        FSR 0687 & 9.15 & ~ & ~& CWNU 4203 & 9.15 \\ 
        CWNU 3067 & 9.15 & ~ & ~& FSR 0508 & 9.25 \\ 
        CWNU 2946 & 9.25 & ~& ~ & CWNU 3989 & 9.25 \\ 
        Pfleiderer 4 & 9.25 & ~& ~ & CWNU 3523 & 9.25 \\ 
        CWNU 3129 & 9.25 & ~& ~ & CWNU 3184 & 9.30 \\ 
        CWNU 3109 & 9.25 & ~& ~ & CWNU 3008 & 9.30 \\ 
        CWNU 3510 & 9.30 & ~& ~ & CWNU 3142 & 9.35 \\ 
        CWNU 3048 & 9.35 & ~& ~ & CWNU 3555 & 9.40 \\ 
        CWNU 3556 & 9.35 & ~& ~ & CWNU 3296 & 9.40 \\ 
        CWNU 3204 & 9.40 & ~ & ~& CWNU 4001 & 9.45 \\ 
        CWNU 4172 & 9.40 & ~& ~ & CWNU 3160 & 9.45 \\ 
        Saurer 1 & 9.45 & ~& ~ & CWNU 3009 & 9.45 \\ 
        CWNU 3102 & 9.45 & ~ & ~& CWNU 2974 & 9.45 \\ 
        CWNU 3863 & 9.45 & ~& ~ & CWNU 3633 & 9.5 \\ 
        CWNU 2961 & 9.50 & ~& ~ & CWNU 4150 & 9.55 \\ 
        CWNU 3252 & 9.50 & ~ & ~& CWNU 3284 & 9.55 \\ 
        CWNU 3064 & 9.50 & ~& ~ & CWNU 3911 & 9.60 \\ 
        CWNU 3282 & 9.55 & ~& ~ & CWNU 3593 & 9.80 \\ 
        Berkeley 26 & 9.55 & ~ & ~ & ~ \\ 
        CWNU 3455 & 9.55 & ~ & ~ & ~ \\ 
        ESO 429 05 & 9.75 & ~ & ~ & ~ \\
\hline
\end{tabular}
\end{table}

\bibliographystyle{aasjournal} 
\bibliography{distant} 

\begin{thebibliography}{}
\expandafter\ifx\csname natexlab\endcsname\relax\def\natexlab#1{#1}\fi

\bibitem[{{Barb{\'a}} {et~al.}(2019{\natexlab{a}}){Barb{\'a}}, {Minniti},
  {Geisler}, {Alonso-Garc{\'\i}a}, {Hempel}, {Monachesi}, {Arias}, \&
  {G{\'o}mez}}]{Barb2019}
{Barb{\'a}}, R.~H., {Minniti}, D., {Geisler}, D., {et~al.} 2019{\natexlab{a}},
  \apjl, 870, L24

\bibitem[{{Barb{\'a}} {et~al.}(2019{\natexlab{b}}){Barb{\'a}}, {Minniti},
  {Geisler}, {Alonso-Garc{\'\i}a}, {Hempel}, {Monachesi}, {Arias}, \&
  {G{\'o}mez}}]{Barba2019}
---. 2019{\natexlab{b}}, \apjl, 870, L24

\bibitem[{{Bellazzini} {et~al.}(2020){Bellazzini}, {Ibata}, {Malhan}, {Martin},
  {Famaey}, \& {Thomas}}]{Bellazzini2020}
{Bellazzini}, M., {Ibata}, R., {Malhan}, K., {et~al.} 2020, \aap, 636, A107

\bibitem[{{Belokurov} {et~al.}(2010){Belokurov}, {Walker}, {Evans}, {Gilmore},
  {Irwin}, {Just}, {Koposov}, {Mateo}, {Olszewski}, {Watkins}, \&
  {Wyrzykowski}}]{Belokurov2010}
{Belokurov}, V., {Walker}, M.~G., {Evans}, N.~W., {et~al.} 2010, \apjl, 712,
  L103

\bibitem[{{Borissova} {et~al.}(2000){Borissova}, {Georgiev}, {Rosado},
  {Kurtev}, {Bullejos}, \& {Valdez-Guti{\'e}rrez}}]{Borissova2000}
{Borissova}, J., {Georgiev}, L., {Rosado}, M., {et~al.} 2000, \aap, 363, 130

\bibitem[{{Bressan} {et~al.}(2012){Bressan}, {Marigo}, {Girardi}, {Salasnich},
  {Dal Cero}, {Rubele}, \& {Nanni}}]{Bressan12}
{Bressan}, A., {Marigo}, P., {Girardi}, L., {et~al.} 2012, \mnras, 427, 127

\bibitem[{{Brunthaler} {et~al.}(2006){Brunthaler}, {Henkel}, {de Blok}, {Reid},
  {Greenhill}, \& {Falcke}}]{Brunthaler06}
{Brunthaler}, A., {Henkel}, C., {de Blok}, W.~J.~G., {et~al.} 2006, \aap, 457,
  109

\bibitem[{{Camargo}(2018)}]{Camargo2018}
{Camargo}, D. 2018, \apjl, 860, L27

\bibitem[{{Camargo} \& {Minniti}(2019)}]{Camargo2019}
{Camargo}, D., \& {Minniti}, D. 2019, \mnras, 484, L90

\bibitem[{{Cantat-Gaudin}(2022)}]{Cantat22}
{Cantat-Gaudin}, T. 2022, Universe, 8, 111

\bibitem[{{Cantat-Gaudin} \& {Anders}(2020)}]{CG20_0}
{Cantat-Gaudin}, T., \& {Anders}, F. 2020, \aap, 633, A99

\bibitem[{{Cantat-Gaudin} {et~al.}(2018){Cantat-Gaudin}, {Jordi}, {Vallenari},
  {Bragaglia}, {Balaguer-N{\'u}{\~n}ez}, {Soubiran}, {Bossini}, {Moitinho},
  {Castro-Ginard}, {Krone-Martins}, {Casamiquela}, {Sordo}, \&
  {Carrera}}]{CG18}
{Cantat-Gaudin}, T., {Jordi}, C., {Vallenari}, A., {et~al.} 2018, \aap, 618,
  A93

\bibitem[{{Cantat-Gaudin} {et~al.}(2019){Cantat-Gaudin}, {Krone-Martins},
  {Sedaghat}, {Farahi}, {de Souza}, {Skalidis}, {Malz}, {Mac{\^e}do}, {Moews},
  {Jordi}, {Moitinho}, {Castro-Ginard}, {Ishida}, {Heneka}, {Boucaud}, \&
  {Trindade}}]{CG19-0}
{Cantat-Gaudin}, T., {Krone-Martins}, A., {Sedaghat}, N., {et~al.} 2019, \aap,
  624, A126

\bibitem[{{Cantat-Gaudin} {et~al.}(2020{\natexlab{a}}){Cantat-Gaudin},
  {Anders}, {Castro-Ginard}, {Jordi}, {Romero-G{\'o}mez}, {Soubiran},
  {Casamiquela}, {Tarricq}, {Moitinho}, {Vallenari}, {Bragaglia},
  {Krone-Martins}, \& {Kounkel}}]{cg20arm}
{Cantat-Gaudin}, T., {Anders}, F., {Castro-Ginard}, A., {et~al.}
  2020{\natexlab{a}}, \aap, 640, A1

\bibitem[{{Cantat-Gaudin} {et~al.}(2020{\natexlab{b}}){Cantat-Gaudin},
  {Anders}, {Castro-Ginard}, {Jordi}, {Romero-G{\'o}mez}, {Soubiran},
  {Casamiquela}, {Tarricq}, {Moitinho}, {Vallenari}, {Bragaglia},
  {Krone-Martins}, \& {Kounkel}}]{CG20}
---. 2020{\natexlab{b}}, \aap, 640, A1

\bibitem[{{Casado}(2021)}]{Casado21}
{Casado}, J. 2021, Research in Astronomy and Astrophysics, 21, 117

\bibitem[{{Casado} \& {Hendy}(2023)}]{casado23}
{Casado}, J., \& {Hendy}, Y. 2023, \mnras, 521, 1399

\bibitem[{{Castro-Ginard} {et~al.}(2019){Castro-Ginard}, {Jordi}, {Luri},
  {Cantat-Gaudin}, \& {Balaguer-N{\'u}{\~n}ez}}]{Castro19}
{Castro-Ginard}, A., {Jordi}, C., {Luri}, X., {Cantat-Gaudin}, T., \&
  {Balaguer-N{\'u}{\~n}ez}, L. 2019, \aap, 627, A35

\bibitem[{{Castro-Ginard} {et~al.}(2018){Castro-Ginard}, {Jordi}, {Luri},
  {Julbe}, {Morvan}, {Balaguer-N{\'u}{\~n}ez}, \& {Cantat-Gaudin}}]{Castro18}
{Castro-Ginard}, A., {Jordi}, C., {Luri}, X., {et~al.} 2018, \aap, 618, A59

\bibitem[{{Castro-Ginard} {et~al.}(2020){Castro-Ginard}, {Jordi}, {Luri},
  {{\'A}lvarez Cid-Fuentes}, {Casamiquela}, {Anders}, {Cantat-Gaudin},
  {Mongui{\'o}}, {Balaguer-N{\'u}{\~n}ez}, {Sol{\`a}}, \& {Badia}}]{Castro20}
---. 2020, \aap, 635, A45

\bibitem[{{Castro-Ginard} {et~al.}(2021){Castro-Ginard}, {McMillan}, {Luri},
  {Jordi}, {Romero-G{\'o}mez}, {Cantat-Gaudin}, {Casamiquela}, {Tarricq},
  {Soubiran}, \& {Anders}}]{Castro21}
{Castro-Ginard}, A., {McMillan}, P.~J., {Luri}, X., {et~al.} 2021, \aap, 652,
  A162

\bibitem[{{Castro-Ginard} {et~al.}(2022){Castro-Ginard}, {Jordi}, {Luri},
  {Cantat-Gaudin}, {Carrasco}, {Casamiquela}, {Anders},
  {Balaguer-N{\'u}{\~n}ez}, \& {Badia}}]{castro22}
{Castro-Ginard}, A., {Jordi}, C., {Luri}, X., {et~al.} 2022, \aap, 661, A118

\bibitem[{{Chi} {et~al.}(2023){Chi}, {Wei}, {Wang}, \& {Li}}]{chi23}
{Chi}, H., {Wei}, S., {Wang}, F., \& {Li}, Z. 2023, \apjs, 265, 20

\bibitem[{{Dias} {et~al.}(2022){Dias}, {Palma}, {Minniti},
  {Fern{\'a}ndez-Trincado}, {Alonso-Garc{\'\i}a}, {Barbuy}, {Clari{\'a}},
  {Gomez}, \& {Saito}}]{Dias2022}
{Dias}, B., {Palma}, T., {Minniti}, D., {et~al.} 2022, \aap, 657, A67

\bibitem[{{Dias} {et~al.}(2002){Dias}, {Alessi}, {Moitinho}, \&
  {L{\'e}pine}}]{Dias02}
{Dias}, W.~S., {Alessi}, B.~S., {Moitinho}, A., \& {L{\'e}pine}, J.~R.~D. 2002,
  \aap, 389, 871

\bibitem[{{Dias} {et~al.}(2021){Dias}, {Monteiro}, {Moitinho}, {L{\'e}pine},
  {Carraro}, {Paunzen}, {Alessi}, \& {Villela}}]{dias21}
{Dias}, W.~S., {Monteiro}, H., {Moitinho}, A., {et~al.} 2021, \mnras, 504, 356

\bibitem[{Ester {et~al.}(1996)Ester, Kriegel, Sander, \& Xu}]{Ester96}
Ester, M., Kriegel, H.-P., Sander, J., \& Xu, X. 1996, in Proc. of 2nd
  International Conference on Knowledge Discovery and Data Mining (KDD-96),
  226--231

\bibitem[{{Fabricius} {et~al.}(2021){Fabricius}, {Luri}, {Arenou}, {Babusiaux},
  {Helmi}, {Muraveva}, {Reyl{\'e}}, {Spoto}, {Vallenari}, {Antoja}, {Balbinot},
  {Barache}, {Bauchet}, {Bragaglia}, {Busonero}, {Cantat-Gaudin}, {Carrasco},
  {Diakit{\'e}}, {Fabrizio}, {Figueras}, {Garcia-Gutierrez}, {Garofalo},
  {Jordi}, {Kervella}, {Khanna}, {Leclerc}, {Licata}, {Lambert}, {Marrese},
  {Masip}, {Ramos}, {Robichon}, {Robin}, {Romero-G{\'o}mez}, {Rubele}, \&
  {Weiler}}]{Fabricius21}
{Fabricius}, C., {Luri}, X., {Arenou}, F., {et~al.} 2021, \aap, 649, A5

\bibitem[{{Ferreira} {et~al.}(2020){Ferreira}, {Corradi}, {Maia}, {Angelo}, \&
  {Santos}}]{Ferreira20}
{Ferreira}, F.~A., {Corradi}, W.~J.~B., {Maia}, F.~F.~S., {Angelo}, M.~S., \&
  {Santos}, J.~F.~C., J. 2020, \mnras, 496, 2021

\bibitem[{{Ferreira} {et~al.}(2021){Ferreira}, {Corradi}, {Maia}, {Angelo}, \&
  {Santos}}]{ferreira21}
---. 2021, \mnras, 502, L90

\bibitem[{{Ferreira} {et~al.}(2019){Ferreira}, {Santos}, {Corradi}, {Maia}, \&
  {Angelo}}]{Ferreira19}
{Ferreira}, F.~A., {Santos}, J.~F.~C., {Corradi}, W.~J.~B., {Maia}, F.~F.~S.,
  \& {Angelo}, M.~S. 2019, \mnras, 483, 5508

\bibitem[{{Gaia Collaboration} {et~al.}(2016){Gaia Collaboration}, {Prusti},
  {de Bruijne}, {Brown}, {Vallenari}, {Babusiaux}, {Bailer-Jones}, {Bastian},
  {Biermann}, {Evans}, {Eyer}, {Jansen}, {Jordi}, {Klioner}, {Lammers},
  {Lindegren}, {Luri}, {Mignard}, {Milligan}, {Panem}, {Poinsignon},
  {Pourbaix}, {Randich}, {Sarri}, {Sartoretti}, {Siddiqui}, {Soubiran},
  {Valette}, {van Leeuwen}, {Walton}, {Aerts}, {Arenou}, {Cropper}, {Drimmel},
  {H{\o}g}, {Katz}, {Lattanzi}, {O'Mullane}, {Grebel}, {Holland}, {Huc},
  {Passot}, {Bramante}, {Cacciari}, {Casta{\~n}eda}, {Chaoul}, {Cheek}, {De
  Angeli}, {Fabricius}, {Guerra}, {Hern{\'a}ndez}, {Jean-Antoine-Piccolo},
  {Masana}, {Messineo}, {Mowlavi}, {Nienartowicz}, {Ord{\'o}{\~n}ez-Blanco},
  {Panuzzo}, {Portell}, {Richards}, {Riello}, {Seabroke}, {Tanga},
  {Th{\'e}venin}, {Torra}, {Els}, {Gracia-Abril}, {Comoretto},
  {Garcia-Reinaldos}, {Lock}, {Mercier}, {Altmann}, {Andrae}, {Astraatmadja},
  {Bellas-Velidis}, {Benson}, {Berthier}, {Blomme}, {Busso}, {Carry},
  {Cellino}, {Clementini}, {Cowell}, {Creevey}, {Cuypers}, {Davidson}, {De
  Ridder}, {de Torres}, {Delchambre}, {Dell'Oro}, {Ducourant}, {Fr{\'e}mat},
  {Garc{\'\i}a-Torres}, {Gosset}, {Halbwachs}, {Hambly}, {Harrison}, {Hauser},
  {Hestroffer}, {Hodgkin}, {Huckle}, {Hutton}, {Jasniewicz}, {Jordan},
  {Kontizas}, {Korn}, {Lanzafame}, {Manteiga}, {Moitinho}, {Muinonen},
  {Osinde}, {Pancino}, {Pauwels}, {Petit}, {Recio-Blanco}, {Robin}, {Sarro},
  {Siopis}, {Smith}, {Smith}, {Sozzetti}, {Thuillot}, {van Reeven}, {Viala},
  {Abbas}, {Abreu Aramburu}, {Accart}, {Aguado}, {Allan}, {Allasia},
  {Altavilla}, {{\'A}lvarez}, {Alves}, {Anderson}, {Andrei}, {Anglada Varela},
  {Antiche}, {Antoja}, {Ant{\'o}n}, {Arcay}, {Atzei}, {Ayache}, {Bach},
  {Baker}, {Balaguer-N{\'u}{\~n}ez}, {Barache}, {Barata}, {Barbier}, {Barblan},
  {Baroni}, {Barrado y Navascu{\'e}s}, {Barros}, {Barstow}, {Becciani},
  {Bellazzini}, {Bellei}, {Bello Garc{\'\i}a}, {Belokurov}, {Bendjoya},
  {Berihuete}, {Bianchi}, {Bienaym{\'e}}, {Billebaud}, {Blagorodnova},
  {Blanco-Cuaresma}, {Boch}, {Bombrun}, {Borrachero}, {Bouquillon}, {Bourda},
  {Bouy}, {Bragaglia}, {Breddels}, {Brouillet}, {Br{\"u}semeister},
  {Bucciarelli}, {Budnik}, {Burgess}, {Burgon}, {Burlacu}, {Busonero}, {Buzzi},
  {Caffau}, {Cambras}, {Campbell}, {Cancelliere}, {Cantat-Gaudin}, {Carlucci},
  {Carrasco}, {Castellani}, {Charlot}, {Charnas}, {Charvet}, {Chassat},
  {Chiavassa}, {Clotet}, {Cocozza}, {Collins}, {Collins}, {Costigan}, {Crifo},
  {Cross}, {Crosta}, {Crowley}, {Dafonte}, {Damerdji}, {Dapergolas}, {David},
  {David}, {De Cat}, {de Felice}, {de Laverny}, {De Luise}, {De March}, {de
  Martino}, {de Souza}, {Debosscher}, {del Pozo}, {Delbo}, {Delgado},
  {Delgado}, {di Marco}, {Di Matteo}, {Diakite}, {Distefano}, {Dolding}, {Dos
  Anjos}, {Drazinos}, {Dur{\'a}n}, {Dzigan}, {Ecale}, {Edvardsson}, {Enke},
  {Erdmann}, {Escolar}, {Espina}, {Evans}, {Eynard Bontemps}, {Fabre},
  {Fabrizio}, {Faigler}, {Falc{\~a}o}, {Farr{\`a}s Casas}, {Faye}, {Federici},
  {Fedorets}, {Fern{\'a}ndez-Hern{\'a}ndez}, {Fernique}, {Fienga}, {Figueras},
  {Filippi}, {Findeisen}, {Fonti}, {Fouesneau}, {Fraile}, {Fraser}, {Fuchs},
  {Furnell}, {Gai}, {Galleti}, {Galluccio}, {Garabato}, {Garc{\'\i}a-Sedano},
  {Gar{\'e}}, {Garofalo}, {Garralda}, {Gavras}, {Gerssen}, {Geyer}, {Gilmore},
  {Girona}, {Giuffrida}, {Gomes}, {Gonz{\'a}lez-Marcos},
  {Gonz{\'a}lez-N{\'u}{\~n}ez}, {Gonz{\'a}lez-Vidal}, {Granvik}, {Guerrier},
  {Guillout}, {Guiraud}, {G{\'u}rpide}, {Guti{\'e}rrez-S{\'a}nchez}, {Guy},
  {Haigron}, {Hatzidimitriou}, {Haywood}, {Heiter}, {Helmi}, {Hobbs},
  {Hofmann}, {Holl}, {Holland }, {Hunt}, {Hypki}, {Icardi}, {Irwin}, {Jevardat
  de Fombelle}, {Jofr{\'e}}, {Jonker}, {Jorissen}, {Julbe}, {Karampelas},
  {Kochoska}, {Kohley}, {Kolenberg}, {Kontizas}, {Koposov}, {Kordopatis},
  {Koubsky}, {Kowalczyk}, {Krone-Martins}, {Kudryashova}, {Kull}, {Bachchan},
  {Lacoste-Seris}, {Lanza}, {Lavigne}, {Le Poncin-Lafitte}, {Lebreton},
  {Lebzelter}, {Leccia}, {Leclerc}, {Lecoeur-Taibi}, {Lemaitre}, {Lenhardt},
  {Leroux}, {Liao}, {Licata}, {Lindstr{\o}m}, {Lister}, {Livanou}, {Lobel},
  {L{\"o}ffler}, {L{\'o}pez}, {Lopez-Lozano}, {Lorenz}, {Loureiro},
  {MacDonald}, {Magalh{\~a}es Fernandes}, {Managau}, {Mann}, {Mantelet},
  {Marchal}, {Marchant}, {Marconi}, {Marie}, {Marinoni}, {Marrese},
  {Marschalk{\'o}}, {Marshall}, {Mart{\'\i}n-Fleitas}, {Martino}, {Mary},
  {Matijevi{\v{c}}}, {Mazeh}, {McMillan}, {Messina}, {Mestre}, {Michalik},
  {Millar}, {Miranda}, {Molina}, {Molinaro}, {Molinaro}, {Moln{\'a}r},
  {Moniez}, {Montegriffo}, {Monteiro}, {Mor}, {Mora}, {Morbidelli}, {Morel},
  {Morgenthaler}, {Morley}, {Morris}, {Mulone}, {Muraveva}, {Musella},
  {Narbonne}, {Nelemans}, {Nicastro}, {Noval}, {Ord{\'e}novic},
  {Ordieres-Mer{\'e}}, {Osborne}, {Pagani}, {Pagano}, {Pailler}, {Palacin},
  {Palaversa}, {Parsons}, {Paulsen}, {Pecoraro}, {Pedrosa}, {Pentik{\"a}inen},
  {Pereira}, {Pichon}, {Piersimoni}, {Pineau}, {Plachy}, {Plum}, {Poujoulet},
  {Pr{\v{s}}a}, {Pulone}, {Ragaini}, {Rago}, {Rambaux}, {Ramos-Lerate},
  {Ranalli}, {Rauw}, {Read}, {Regibo}, {Renk}, {Reyl{\'e}}, {Ribeiro},
  {Rimoldini}, {Ripepi}, {Riva}, {Rixon}, {Roelens}, {Romero-G{\'o}mez},
  {Rowell}, {Royer}, {Rudolph}, {Ruiz-Dern}, {Sadowski}, {Sagrist{\`a}
  Sell{\'e}s}, {Sahlmann}, {Salgado}, {Salguero}, {Sarasso}, {Savietto},
  {Schnorhk}, {Schultheis}, {Sciacca}, {Segol}, {Segovia}, {Segransan},
  {Serpell}, {Shih}, {Smareglia}, {Smart}, {Smith}, {Solano}, {Solitro},
  {Sordo}, {Soria Nieto}, {Souchay}, {Spagna}, {Spoto}, {Stampa}, {Steele},
  {Steidelm{\"u}ller}, {Stephenson}, {Stoev}, {Suess}, {S{\"u}veges}, {Surdej},
  {Szabados}, {Szegedi-Elek}, {Tapiador}, {Taris}, {Tauran}, {Taylor},
  {Teixeira}, {Terrett}, {Tingley}, {Trager}, {Turon}, {Ulla}, {Utrilla},
  {Valentini}, {van Elteren}, {Van Hemelryck}, {van Leeuwen}, {Varadi},
  {Vecchiato}, {Veljanoski}, {Via}, {Vicente}, {Vogt}, {Voss}, {Votruba},
  {Voutsinas}, {Walmsley}, {Weiler}, {Weingrill}, {Werner}, {Wevers},
  {Whitehead}, {Wyrzykowski}, {Yoldas}, {{\v{Z}}erjal}, {Zucker}, {Zurbach},
  {Zwitter}, {Alecu}, {Allen}, {Allende Prieto}, {Amorim},
  {Anglada-Escud{\'e}}, {Arsenijevic}, {Azaz}, {Balm}, {Beck}, {Bernstein},
  {Bigot}, {Bijaoui}, {Blasco}, {Bonfigli}, {Bono}, {Boudreault}, {Bressan},
  {Brown}, {Brunet}, {Bunclark}, {Buonanno}, {Butkevich}, {Carret}, {Carrion},
  {Chemin}, {Ch{\'e}reau}, {Corcione}, {Darmigny}, {de Boer}, {de Teodoro}, {de
  Zeeuw}, {Delle Luche}, {Domingues}, {Dubath}, {Fodor}, {Fr{\'e}zouls},
  {Fries}, {Fustes}, {Fyfe}, {Gallardo}, {Gallegos}, {Gardiol}, {Gebran},
  {Gomboc}, {G{\'o}mez}, {Grux}, {Gueguen}, {Heyrovsky}, {Hoar}, {Iannicola},
  {Isasi Parache}, {Janotto}, {Joliet}, {Jonckheere}, {Keil}, {Kim},
  {Klagyivik}, {Klar}, {Knude}, {Kochukhov}, {Kolka}, {Kos}, {Kutka}, {Lainey},
  {LeBouquin}, {Liu}, {Loreggia}, {Makarov}, {Marseille}, {Martayan},
  {Martinez-Rubi}, {Massart}, {Meynadier}, {Mignot}, {Munari}, {Nguyen},
  {Nordlander}, {Ocvirk}, {O'Flaherty}, {Olias Sanz}, {Ortiz}, {Osorio},
  {Oszkiewicz}, {Ouzounis}, {Palmer}, {Park}, {Pasquato}, {Peltzer}, {Peralta},
  {P{\'e}turaud}, {Pieniluoma}, {Pigozzi}, {Poels}, {Prat}, {Prod'homme},
  {Raison}, {Rebordao}, {Risquez}, {Rocca-Volmerange}, {Rosen}, {Ruiz-Fuertes},
  {Russo}, {Sembay}, {Serraller Vizcaino}, {Short}, {Siebert}, {Silva},
  {Sinachopoulos}, {Slezak}, {Soffel}, {Sosnowska}, {Strai{\v{z}}ys}, {ter
  Linden}, {Terrell}, {Theil}, {Tiede}, {Troisi}, {Tsalmantza}, {Tur},
  {Vaccari}, {Vachier}, {Valles}, {Van Hamme}, {Veltz}, {Virtanen}, {Wallut},
  {Wichmann}, {Wilkinson}, {Ziaeepour}, \& {Zschocke}}]{Gaia16}
{Gaia Collaboration}, {Prusti}, T., {de Bruijne}, J.~H.~J., {et~al.} 2016,
  \aap, 595, A1

\bibitem[{{Gaia Collaboration} {et~al.}(2018){Gaia Collaboration}, {Brown},
  {Vallenari}, {Prusti}, {de Bruijne}, {Babusiaux}, {Bailer-Jones}, {Biermann},
  {Evans}, {Eyer}, {Jansen}, {Jordi}, {Klioner}, {Lammers}, {Lindegren},
  {Luri}, {Mignard}, {Panem}, {Pourbaix}, {Randich}, {Sartoretti}, {Siddiqui},
  {Soubiran}, {van Leeuwen}, {Walton}, {Arenou}, {Bastian}, {Cropper},
  {Drimmel}, {Katz}, {Lattanzi}, {Bakker}, {Cacciari}, {Casta{\~n}eda},
  {Chaoul}, {Cheek}, {De Angeli}, {Fabricius}, {Guerra}, {Holl}, {Masana},
  {Messineo}, {Mowlavi}, {Nienartowicz}, {Panuzzo}, {Portell}, {Riello},
  {Seabroke}, {Tanga}, {Th{\'e}venin}, {Gracia-Abril}, {Comoretto},
  {Garcia-Reinaldos}, {Teyssier}, {Altmann}, {Andrae}, {Audard},
  {Bellas-Velidis}, {Benson}, {Berthier}, {Blomme}, {Burgess}, {Busso},
  {Carry}, {Cellino}, {Clementini}, {Clotet}, {Creevey}, {Davidson}, {De
  Ridder}, {Delchambre}, {Dell'Oro}, {Ducourant},
  {Fern{\'a}ndez-Hern{\'a}ndez}, {Fouesneau}, {Fr{\'e}mat}, {Galluccio},
  {Garc{\'\i}a-Torres}, {Gonz{\'a}lez-N{\'u}{\~n}ez}, {Gonz{\'a}lez-Vidal},
  {Gosset}, {Guy}, {Halbwachs}, {Hambly}, {Harrison}, {Hern{\'a}ndez},
  {Hestroffer}, {Hodgkin}, {Hutton}, {Jasniewicz}, {Jean-Antoine-Piccolo},
  {Jordan}, {Korn}, {Krone-Martins}, {Lanzafame}, {Lebzelter}, {L{\"o}ffler},
  {Manteiga}, {Marrese}, {Mart{\'\i}n-Fleitas}, {Moitinho}, {Mora}, {Muinonen},
  {Osinde}, {Pancino}, {Pauwels}, {Petit}, {Recio-Blanco}, {Richards},
  {Rimoldini}, {Robin}, {Sarro}, {Siopis}, {Smith}, {Sozzetti}, {S{\"u}veges},
  {Torra}, {van Reeven}, {Abbas}, {Abreu Aramburu}, {Accart}, {Aerts},
  {Altavilla}, {{\'A}lvarez}, {Alvarez}, {Alves}, {Anderson}, {Andrei},
  {Anglada Varela}, {Antiche}, {Antoja}, {Arcay}, {Astraatmadja}, {Bach},
  {Baker}, {Balaguer-N{\'u}{\~n}ez}, {Balm}, {Barache}, {Barata}, {Barbato},
  {Barblan}, {Barklem}, {Barrado}, {Barros}, {Barstow}, {Bartholom{\'e}
  Mu{\~n}oz}, {Bassilana}, {Becciani}, {Bellazzini}, {Berihuete}, {Bertone},
  {Bianchi}, {Bienaym{\'e}}, {Blanco-Cuaresma}, {Boch}, {Boeche}, {Bombrun},
  {Borrachero}, {Bossini}, {Bouquillon}, {Bourda}, {Bragaglia}, {Bramante},
  {Breddels}, {Bressan}, {Brouillet}, {Br{\"u}semeister}, {Brugaletta},
  {Bucciarelli}, {Burlacu}, {Busonero}, {Butkevich}, {Buzzi}, {Caffau},
  {Cancelliere}, {Cannizzaro}, {Cantat-Gaudin}, {Carballo}, {Carlucci},
  {Carrasco}, {Casamiquela}, {Castellani}, {Castro-Ginard}, {Charlot},
  {Chemin}, {Chiavassa}, {Cocozza}, {Costigan}, {Cowell}, {Crifo}, {Crosta},
  {Crowley}, {Cuypers}, {Dafonte}, {Damerdji}, {Dapergolas}, {David}, {David},
  {de Laverny}, {De Luise}, {De March}, {de Martino}, {de Souza}, {de Torres},
  {Debosscher}, {del Pozo}, {Delbo}, {Delgado}, {Delgado}, {Di Matteo},
  {Diakite}, {Diener}, {Distefano}, {Dolding}, {Drazinos}, {Dur{\'a}n},
  {Edvardsson}, {Enke}, {Eriksson}, {Esquej}, {Eynard Bontemps}, {Fabre},
  {Fabrizio}, {Faigler}, {Falc{\~a}o}, {Farr{\`a}s Casas}, {Federici},
  {Fedorets}, {Fernique}, {Figueras}, {Filippi}, {Findeisen}, {Fonti},
  {Fraile}, {Fraser}, {Fr{\'e}zouls}, {Gai}, {Galleti}, {Garabato},
  {Garc{\'\i}a-Sedano}, {Garofalo}, {Garralda}, {Gavel}, {Gavras}, {Gerssen},
  {Geyer}, {Giacobbe}, {Gilmore}, {Girona}, {Giuffrida}, {Glass}, {Gomes},
  {Granvik}, {Gueguen}, {Guerrier}, {Guiraud}, {Guti{\'e}rrez-S{\'a}nchez},
  {Haigron}, {Hatzidimitriou}, {Hauser}, {Haywood}, {Heiter}, {Helmi}, {Heu},
  {Hilger}, {Hobbs}, {Hofmann}, {Holland}, {Huckle}, {Hypki}, {Icardi},
  {Jan{\ss}en}, {Jevardat de Fombelle}, {Jonker}, {Juh{\'a}sz}, {Julbe},
  {Karampelas}, {Kewley}, {Klar}, {Kochoska}, {Kohley}, {Kolenberg},
  {Kontizas}, {Kontizas}, {Koposov}, {Kordopatis}, {Kostrzewa-Rutkowska},
  {Koubsky}, {Lambert}, {Lanza}, {Lasne}, {Lavigne}, {Le Fustec}, {Le
  Poncin-Lafitte}, {Lebreton}, {Leccia}, {Leclerc}, {Lecoeur-Taibi},
  {Lenhardt}, {Leroux}, {Liao}, {Licata}, {Lindstr{\o}m}, {Lister}, {Livanou},
  {Lobel}, {L{\'o}pez}, {Managau}, {Mann}, {Mantelet}, {Marchal}, {Marchant},
  {Marconi}, {Marinoni}, {Marschalk{\'o}}, {Marshall}, {Martino}, {Marton},
  {Mary}, {Massari}, {Matijevi{\v{c}}}, {Mazeh}, {McMillan}, {Messina},
  {Michalik}, {Millar}, {Molina}, {Molinaro}, {Moln{\'a}r}, {Montegriffo},
  {Mor}, {Morbidelli}, {Morel}, {Morris}, {Mulone}, {Muraveva}, {Musella},
  {Nelemans}, {Nicastro}, {Noval}, {O'Mullane}, {Ord{\'e}novic},
  {Ord{\'o}{\~n}ez-Blanco}, {Osborne}, {Pagani}, {Pagano}, {Pailler},
  {Palacin}, {Palaversa}, {Panahi}, {Pawlak}, {Piersimoni}, {Pineau}, {Plachy},
  {Plum}, {Poggio}, {Poujoulet}, {Pr{\v{s}}a}, {Pulone}, {Racero}, {Ragaini},
  {Rambaux}, {Ramos-Lerate}, {Regibo}, {Reyl{\'e}}, {Riclet}, {Ripepi}, {Riva},
  {Rivard}, {Rixon}, {Roegiers}, {Roelens}, {Romero-G{\'o}mez}, {Rowell},
  {Royer}, {Ruiz-Dern}, {Sadowski}, {Sagrist{\`a} Sell{\'e}s}, {Sahlmann},
  {Salgado}, {Salguero}, {Sanna}, {Santana-Ros}, {Sarasso}, {Savietto},
  {Schultheis}, {Sciacca}, {Segol}, {Segovia}, {S{\'e}gransan}, {Shih},
  {Siltala}, {Silva}, {Smart}, {Smith}, {Solano}, {Solitro}, {Sordo}, {Soria
  Nieto}, {Souchay}, {Spagna}, {Spoto}, {Stampa}, {Steele},
  {Steidelm{\"u}ller}, {Stephenson}, {Stoev}, {Suess}, {Surdej}, {Szabados},
  {Szegedi-Elek}, {Tapiador}, {Taris}, {Tauran}, {Taylor}, {Teixeira},
  {Terrett}, {Teyssand ier}, {Thuillot}, {Titarenko}, {Torra Clotet}, {Turon},
  {Ulla}, {Utrilla}, {Uzzi}, {Vaillant}, {Valentini}, {Valette}, {van Elteren},
  {Van Hemelryck}, {van Leeuwen}, {Vaschetto}, {Vecchiato}, {Veljanoski},
  {Viala}, {Vicente}, {Vogt}, {von Essen}, {Voss}, {Votruba}, {Voutsinas},
  {Walmsley}, {Weiler}, {Wertz}, {Wevers}, {Wyrzykowski}, {Yoldas},
  {{\v{Z}}erjal}, {Ziaeepour}, {Zorec}, {Zschocke}, {Zucker}, {Zurbach}, \&
  {Zwitter}}]{Gaia18-Brown}
{Gaia Collaboration}, {Brown}, A.~G.~A., {Vallenari}, A., {et~al.} 2018, \aap,
  616, A1

\bibitem[{{Gaia Collaboration} {et~al.}(2021){Gaia Collaboration}, {Brown},
  {Vallenari}, {Prusti}, {de Bruijne}, {Babusiaux}, {Biermann}, {Creevey},
  {Evans}, {Eyer}, {Hutton}, {Jansen}, {Jordi}, {Klioner}, {Lammers},
  {Lindegren}, {Luri}, {Mignard}, {Panem}, {Pourbaix}, {Randich}, {Sartoretti},
  {Soubiran}, {Walton}, {Arenou}, {Bailer-Jones}, {Bastian}, {Cropper},
  {Drimmel}, {Katz}, {Lattanzi}, {van Leeuwen}, {Bakker}, {Cacciari},
  {Casta{\~n}eda}, {De Angeli}, {Ducourant}, {Fabricius}, {Fouesneau},
  {Fr{\'e}mat}, {Guerra}, {Guerrier}, {Guiraud}, {Jean-Antoine Piccolo},
  {Masana}, {Messineo}, {Mowlavi}, {Nicolas}, {Nienartowicz}, {Pailler},
  {Panuzzo}, {Riclet}, {Roux}, {Seabroke}, {Sordo}, {Tanga}, {Th{\'e}venin},
  {Gracia-Abril}, {Portell}, {Teyssier}, {Altmann}, {Andrae}, {Bellas-Velidis},
  {Benson}, {Berthier}, {Blomme}, {Brugaletta}, {Burgess}, {Busso}, {Carry},
  {Cellino}, {Cheek}, {Clementini}, {Damerdji}, {Davidson}, {Delchambre},
  {Dell'Oro}, {Fern{\'a}ndez-Hern{\'a}ndez}, {Galluccio}, {Garc{\'\i}a-Lario},
  {Garcia-Reinaldos}, {Gonz{\'a}lez-N{\'u}{\~n}ez}, {Gosset}, {Haigron},
  {Halbwachs}, {Hambly}, {Harrison}, {Hatzidimitriou}, {Heiter},
  {Hern{\'a}ndez}, {Hestroffer}, {Hodgkin}, {Holl}, {Jan{\ss}en}, {Jevardat de
  Fombelle}, {Jordan}, {Krone-Martins}, {Lanzafame}, {L{\"o}ffler}, {Lorca},
  {Manteiga}, {Marchal}, {Marrese}, {Moitinho}, {Mora}, {Muinonen}, {Osborne},
  {Pancino}, {Pauwels}, {Petit}, {Recio-Blanco}, {Richards}, {Riello},
  {Rimoldini}, {Robin}, {Roegiers}, {Rybizki}, {Sarro}, {Siopis}, {Smith},
  {Sozzetti}, {Ulla}, {Utrilla}, {van Leeuwen}, {van Reeven}, {Abbas}, {Abreu
  Aramburu}, {Accart}, {Aerts}, {Aguado}, {Ajaj}, {Altavilla}, {{\'A}lvarez},
  {{\'A}lvarez Cid-Fuentes}, {Alves}, {Anderson}, {Anglada Varela}, {Antoja},
  {Audard}, {Baines}, {Baker}, {Balaguer-N{\'u}{\~n}ez}, {Balbinot}, {Balog},
  {Barache}, {Barbato}, {Barros}, {Barstow}, {Bartolom{\'e}}, {Bassilana},
  {Bauchet}, {Baudesson-Stella}, {Becciani}, {Bellazzini}, {Bernet}, {Bertone},
  {Bianchi}, {Blanco-Cuaresma}, {Boch}, {Bombrun}, {Bossini}, {Bouquillon},
  {Bragaglia}, {Bramante}, {Breedt}, {Bressan}, {Brouillet}, {Bucciarelli},
  {Burlacu}, {Busonero}, {Butkevich}, {Buzzi}, {Caffau}, {Cancelliere},
  {C{\'a}novas}, {Cantat-Gaudin}, {Carballo}, {Carlucci}, {Carnerero},
  {Carrasco}, {Casamiquela}, {Castellani}, {Castro-Ginard}, {Castro Sampol},
  {Chaoul}, {Charlot}, {Chemin}, {Chiavassa}, {Cioni}, {Comoretto}, {Cooper},
  {Cornez}, {Cowell}, {Crifo}, {Crosta}, {Crowley}, {Dafonte}, {Dapergolas},
  {David}, {David}, {de Laverny}, {De Luise}, {De March}, {De Ridder}, {de
  Souza}, {de Teodoro}, {de Torres}, {del Peloso}, {del Pozo}, {Delbo},
  {Delgado}, {Delgado}, {Delisle}, {Di Matteo}, {Diakite}, {Diener},
  {Distefano}, {Dolding}, {Eappachen}, {Edvardsson}, {Enke}, {Esquej}, {Fabre},
  {Fabrizio}, {Faigler}, {Fedorets}, {Fernique}, {Fienga}, {Figueras},
  {Fouron}, {Fragkoudi}, {Fraile}, {Franke}, {Gai}, {Garabato},
  {Garcia-Gutierrez}, {Garc{\'\i}a-Torres}, {Garofalo}, {Gavras}, {Gerlach},
  {Geyer}, {Giacobbe}, {Gilmore}, {Girona}, {Giuffrida}, {Gomel}, {Gomez},
  {Gonzalez-Santamaria}, {Gonz{\'a}lez-Vidal}, {Granvik},
  {Guti{\'e}rrez-S{\'a}nchez}, {Guy}, {Hauser}, {Haywood}, {Helmi}, {Hidalgo},
  {Hilger}, {H{\l}adczuk}, {Hobbs}, {Holland}, {Huckle}, {Jasniewicz},
  {Jonker}, {Juaristi Campillo}, {Julbe}, {Karbevska}, {Kervella}, {Khanna},
  {Kochoska}, {Kontizas}, {Kordopatis}, {Korn}, {Kostrzewa-Rutkowska},
  {Kruszy{\'n}ska}, {Lambert}, {Lanza}, {Lasne}, {Le Campion}, {Le Fustec},
  {Lebreton}, {Lebzelter}, {Leccia}, {Leclerc}, {Lecoeur-Taibi}, {Liao},
  {Licata}, {Lindstr{\o}m}, {Lister}, {Livanou}, {Lobel}, {Madrero Pardo},
  {Managau}, {Mann}, {Marchant}, {Marconi}, {Marcos Santos}, {Marinoni},
  {Marocco}, {Marshall}, {Martin Polo}, {Mart{\'\i}n-Fleitas}, {Masip},
  {Massari}, {Mastrobuono-Battisti}, {Mazeh}, {McMillan}, {Messina},
  {Michalik}, {Millar}, {Mints}, {Molina}, {Molinaro}, {Moln{\'a}r},
  {Montegriffo}, {Mor}, {Morbidelli}, {Morel}, {Morris}, {Mulone}, {Munoz},
  {Muraveva}, {Murphy}, {Musella}, {Noval}, {Ord{\'e}novic}, {Orr{\`u}},
  {Osinde}, {Pagani}, {Pagano}, {Palaversa}, {Palicio}, {Panahi}, {Pawlak},
  {Pe{\~n}alosa Esteller}, {Penttil{\"a}}, {Piersimoni}, {Pineau}, {Plachy},
  {Plum}, {Poggio}, {Poretti}, {Poujoulet}, {Pr{\v{s}}a}, {Pulone}, {Racero},
  {Ragaini}, {Rainer}, {Raiteri}, {Rambaux}, {Ramos}, {Ramos-Lerate}, {Re
  Fiorentin}, {Regibo}, {Reyl{\'e}}, {Ripepi}, {Riva}, {Rixon}, {Robichon},
  {Robin}, {Roelens}, {Rohrbasser}, {Romero-G{\'o}mez}, {Rowell}, {Royer},
  {Rybicki}, {Sadowski}, {Sagrist{\`a} Sell{\'e}s}, {Sahlmann}, {Salgado},
  {Salguero}, {Samaras}, {Sanchez Gimenez}, {Sanna}, {Santove{\~n}a},
  {Sarasso}, {Schultheis}, {Sciacca}, {Segol}, {Segovia}, {S{\'e}gransan},
  {Semeux}, {Shahaf}, {Siddiqui}, {Siebert}, {Siltala}, {Slezak}, {Smart},
  {Solano}, {Solitro}, {Souami}, {Souchay}, {Spagna}, {Spoto}, {Steele},
  {Steidelm{\"u}ller}, {Stephenson}, {S{\"u}veges}, {Szabados}, {Szegedi-Elek},
  {Taris}, {Tauran}, {Taylor}, {Teixeira}, {Thuillot}, {Tonello}, {Torra},
  {Torra}, {Turon}, {Unger}, {Vaillant}, {van Dillen}, {Vanel}, {Vecchiato},
  {Viala}, {Vicente}, {Voutsinas}, {Weiler}, {Wevers}, {Wyrzykowski}, {Yoldas},
  {Yvard}, {Zhao}, {Zorec}, {Zucker}, {Zurbach}, \& {Zwitter}}]{gaia2021}
---. 2021, \aap, 649, A1

\bibitem[{{Garro} {et~al.}(2021){Garro}, {Minniti}, {G{\'o}mez},
  {Alonso-Garc{\'\i}a}, {Palma}, {Smith}, \& {Ripepi}}]{Garro2021}
{Garro}, E.~R., {Minniti}, D., {G{\'o}mez}, M., {et~al.} 2021, \aap, 649, A86

\bibitem[{{Garro} {et~al.}(2022){Garro}, {Minniti}, {G{\'o}mez},
  {Alonso-Garc{\'\i}a}, {Ripepi}, {Fern{\'a}ndez-Trincado}, \& {Vivanco
  C{\'a}diz}}]{Garro2022}
---. 2022, \aap, 658, A120

\bibitem[{{Garro} {et~al.}(2020){Garro}, {Minniti}, {G{\'o}mez},
  {Alonso-Garc{\'\i}a}, {Barb{\'a}}, {Barbuy}, {Clari{\'a}}, {Chen{\'e}},
  {Dias}, {Hempel}, {Ivanov}, {Lucas}, {Majaess}, {Mauro}, {Moni Bidin},
  {Palma}, {Pullen}, {Saito}, {Smith}, {Surot}, {Ram{\'\i}rez Alegr{\'\i}a},
  {Rejkuba}, {Ripepi}, \& {Fern{\'a}ndez Trincado}}]{Garro2020}
---. 2020, \aap, 642, L19

\bibitem[{{Gatto} {et~al.}(2021){Gatto}, {Ripepi}, {Bellazzini}, {Tosi},
  {Tortora}, {Cignoni}, {Spavone}, {Dall'ora}, {Clementini}, {Cusano}, {Longo},
  {Musella}, {Marconi}, \& {Schipani}}]{Gatto2021}
{Gatto}, M., {Ripepi}, V., {Bellazzini}, M., {et~al.} 2021, Research Notes of
  the American Astronomical Society, 5, 159

\bibitem[{{Gran} {et~al.}(2022){Gran}, {Zoccali}, {Saviane}, {Valenti},
  {Rojas-Arriagada}, {Contreras Ramos}, {Hartke}, {Carballo-Bello},
  {Navarrete}, {Rejkuba}, \& {Olivares Carvajal}}]{Gran2022}
{Gran}, F., {Zoccali}, M., {Saviane}, I., {et~al.} 2022, \mnras, 509, 4962

\bibitem[{{Hao} {et~al.}(2022{\natexlab{a}}){Hao}, {Xu}, {Wu}, {Lin}, {Bian},
  {Li}, \& {Liu}}]{hao22c}
{Hao}, C.~J., {Xu}, Y., {Wu}, Z.~Y., {et~al.} 2022{\natexlab{a}}, \aap, 668,
  A13

\bibitem[{{Hao} {et~al.}(2022{\natexlab{b}}){Hao}, {Xu}, {Wu}, {Lin}, {Liu}, \&
  {Li}}]{hao22}
---. 2022{\natexlab{b}}, \aap, 660, A4

\bibitem[{{He} {et~al.}(2023){He}, {Liu}, {Luo}, {Wang}, \& {Jiang}}]{he23a}
{He}, Z., {Liu}, X., {Luo}, Y., {Wang}, K., \& {Jiang}, Q. 2023, \apjs, 264, 8

\bibitem[{{He} {et~al.}(2022{\natexlab{a}}){He}, {Wang}, {Luo}, {Li}, {Liu}, \&
  {Jiang}}]{he22b}
{He}, Z., {Wang}, K., {Luo}, Y., {et~al.} 2022{\natexlab{a}}, \apjs, 262, 7

\bibitem[{{He} {et~al.}(2022{\natexlab{b}}){He}, {Li}, {Zhong}, {Liu}, {Bai},
  {Qin}, {Jiang}, {Zhang}, \& {Chen}}]{he22a}
{He}, Z., {Li}, C., {Zhong}, J., {et~al.} 2022{\natexlab{b}}, \apjs, 260, 8

\bibitem[{{He} {et~al.}(2021{\natexlab{a}}){He}, {Xu}, {Hao}, {Wu}, \&
  {Li}}]{he21}
{He}, Z.-H., {Xu}, Y., {Hao}, C.-J., {Wu}, Z.-Y., \& {Li}, J.-J.
  2021{\natexlab{a}}, Research in Astronomy and Astrophysics, 21, 093

\bibitem[{{He} {et~al.}(2021{\natexlab{b}}){He}, {Xu}, \& {Hou}}]{he211}
{He}, Z.-H., {Xu}, Y., \& {Hou}, L.-G. 2021{\natexlab{b}}, Research in
  Astronomy and Astrophysics, 21, 009

\bibitem[{{Huchra} {et~al.}(1999){Huchra}, {Vogeley}, \& {Geller}}]{Huchra99}
{Huchra}, J.~P., {Vogeley}, M.~S., \& {Geller}, M.~J. 1999, \apjs, 121, 287

\bibitem[{{Hunt} \& {Reffert}(2021)}]{Hunt21}
{Hunt}, E.~L., \& {Reffert}, S. 2021, \aap, 646, A104

\bibitem[{{Hunt} \& {Reffert}(2023)}]{hunt23}
---. 2023, arXiv e-prints, arXiv:2303.13424

\bibitem[{{Jaehnig} {et~al.}(2021){Jaehnig}, {Bird}, \&
  {Holley-Bockelmann}}]{Jaehnig21}
{Jaehnig}, K., {Bird}, J., \& {Holley-Bockelmann}, K. 2021, \apj, 923, 129

\bibitem[{{Kharchenko} {et~al.}(2013){Kharchenko}, {Piskunov}, {Schilbach},
  {R{\"o}ser}, \& {Scholz}}]{Kharchenko13}
{Kharchenko}, N.~V., {Piskunov}, A.~E., {Schilbach}, E., {R{\"o}ser}, S., \&
  {Scholz}, R.~D. 2013, \aap, 558, A53

\bibitem[{{Kounkel} \& {Covey}(2019)}]{Kounkel19}
{Kounkel}, M., \& {Covey}, K. 2019, \aj, 158, 122

\bibitem[{{Kounkel} {et~al.}(2020){Kounkel}, {Covey}, \& {Stassun}}]{Kounkel20}
{Kounkel}, M., {Covey}, K., \& {Stassun}, K.~G. 2020, \aj, 160, 279

\bibitem[{{Kuhn} {et~al.}(2019){Kuhn}, {Hillenbrand}, {Sills}, {Feigelson}, \&
  {Getman}}]{Kuhn19}
{Kuhn}, M.~A., {Hillenbrand}, L.~A., {Sills}, A., {Feigelson}, E.~D., \&
  {Getman}, K.~V. 2019, \apj, 870, 32

\bibitem[{{Laevens} {et~al.}(2014){Laevens}, {Martin}, {Sesar}, {Bernard},
  {Rix}, {Slater}, {Bell}, {Ferguson}, {Schlafly}, {Burgett}, {Chambers},
  {Denneau}, {Draper}, {Kaiser}, {Kudritzki}, {Magnier}, {Metcalfe}, {Morgan},
  {Price}, {Sweeney}, {Tonry}, {Wainscoat}, \& {Waters}}]{Laevens2014}
{Laevens}, B. P.~M., {Martin}, N.~F., {Sesar}, B., {et~al.} 2014, \apjl, 786,
  L3

\bibitem[{{Laevens} {et~al.}(2015){Laevens}, {Martin}, {Bernard}, {Schlafly},
  {Sesar}, {Rix}, {Bell}, {Ferguson}, {Slater}, {Sweeney}, {Wyse}, {Huxor},
  {Burgett}, {Chambers}, {Draper}, {Hodapp}, {Kaiser}, {Magnier}, {Metcalfe},
  {Tonry}, {Wainscoat}, \& {Waters}}]{Laevens2015}
{Laevens}, B. P.~M., {Martin}, N.~F., {Bernard}, E.~J., {et~al.} 2015, \apj,
  813, 44

\bibitem[{{Li} \& {Mao}(2023)}]{li2023}
{Li}, Z., \& {Mao}, C. 2023, \apjs, 265, 3

\bibitem[{{Li} {et~al.}(2022){Li}, {Deng}, {Chi}, {Chen}, {Liu}, {Yan}, {Chen},
  {Guo}, \& {Xia}}]{li22}
{Li}, Z., {Deng}, Y., {Chi}, H., {et~al.} 2022, \apjs, 259, 19

\bibitem[{{Lindegren} {et~al.}(2021){Lindegren}, {Bastian}, {Biermann},
  {Bombrun}, {de Torres}, {Gerlach}, {Geyer}, {Hern{\'a}ndez}, {Hilger},
  {Hobbs}, {Klioner}, {Lammers}, {McMillan}, {Ramos-Lerate},
  {Steidelm{\"u}ller}, {Stephenson}, \& {van Leeuwen}}]{Lindegren21}
{Lindegren}, L., {Bastian}, U., {Biermann}, M., {et~al.} 2021, \aap, 649, A4

\bibitem[{{Liu} \& {Pang}(2019)}]{Liu19}
{Liu}, L., \& {Pang}, X. 2019, \apjs, 245, 32

\bibitem[{{Lu{\'\i}sa Buzzo} {et~al.}(2021){Lu{\'\i}sa Buzzo}, {Cortesi},
  {Forbes}, {Brodie}, {Couch}, {Barbosa}, {de Brito Silva}, {Coelho},
  {Chies-Santos}, {Escudero}, {Sesto}, {Men{\'e}ndez-Delmestre},
  {Gol{\c{c}}alves}, {Bom}, {Alvarez-Candal}, {Smith Castelli}, {Schoenell},
  {Kanaan}, {Ribeirto}, \& {Mendes de Oliveira}}]{Buzzo2021}
{Lu{\'\i}sa Buzzo}, M., {Cortesi}, A., {Forbes}, D.~A., {et~al.} 2021, arXiv
  e-prints, arXiv:2111.14993

\bibitem[{{Magrini} {et~al.}(2009){Magrini}, {Sestito}, {Randich}, \&
  {Galli}}]{Magrini09}
{Magrini}, L., {Sestito}, P., {Randich}, S., \& {Galli}, D. 2009, \aap, 494, 95

\bibitem[{{McConnachie}(2012)}]{McConnachie12}
{McConnachie}, A.~W. 2012, \aj, 144, 4

\bibitem[{{Minniti} {et~al.}(2021{\natexlab{a}}){Minniti},
  {Fern{\'a}ndez-Trincado}, {G{\'o}mez}, {Smith}, {Lucas}, \& {Contreras
  Ramos}}]{Minniti2021b}
{Minniti}, D., {Fern{\'a}ndez-Trincado}, J.~G., {G{\'o}mez}, M., {et~al.}
  2021{\natexlab{a}}, \aap, 650, L11

\bibitem[{{Minniti} {et~al.}(2011){Minniti}, {Hempel}, {Toledo}, {Ivanov},
  {Alonso-Garc{\'\i}a}, {Saito}, {Catelan}, {Geisler}, {Jord{\'a}n},
  {Borissova}, {Zoccali}, {Kurtev}, {Carraro}, {Barbuy}, {Clari{\'a}},
  {Rejkuba}, {Emerson}, \& {Moni Bidin}}]{Minniti2011}
{Minniti}, D., {Hempel}, M., {Toledo}, I., {et~al.} 2011, \aap, 527, A81

\bibitem[{{Minniti} {et~al.}(2017{\natexlab{a}}){Minniti}, {Palma},
  {D{\'e}k{\'a}ny}, {Hempel}, {Rejkuba}, {Pullen}, {Alonso-Garc{\'\i}a},
  {Barb{\'a}}, {Barbuy}, {Bica}, {Bonatto}, {Borissova}, {Catelan},
  {Carballo-Bello}, {Chene}, {Clari{\'a}}, {Cohen}, {Contreras Ramos}, {Dias},
  {Emerson}, {Froebrich}, {Buckner}, {Geisler}, {Gonzalez}, {Gran}, {Hajdu},
  {Irwin}, {Ivanov}, {Kurtev}, {Lucas}, {Majaess}, {Mauro}, {Moni-Bidin},
  {Navarrete}, {Ram{\'\i}rez Alegr{\'\i}a}, {Saito}, {Valenti}, \&
  {Zoccali}}]{Minniti2017A}
{Minniti}, D., {Palma}, T., {D{\'e}k{\'a}ny}, I., {et~al.} 2017{\natexlab{a}},
  \apjl, 838, L14

\bibitem[{{Minniti} {et~al.}(2017{\natexlab{b}}){Minniti}, {Geisler},
  {Alonso-Garc{\'\i}a}, {Palma}, {Beam{\'\i}n}, {Borissova}, {Catelan},
  {Clari{\'a}}, {Cohen}, {Contreras Ramos}, {Dias}, {Fern{\'a}ndez-Trincado},
  {G{\'o}mez}, {Hempel}, {Ivanov}, {Kurtev}, {Lucas}, {Moni-Bidin}, {Pullen},
  {Ram{\'\i}rez Alegr{\'\i}a}, {Saito}, \& {Valenti}}]{Minniti2017}
{Minniti}, D., {Geisler}, D., {Alonso-Garc{\'\i}a}, J., {et~al.}
  2017{\natexlab{b}}, \apjl, 849, L24

\bibitem[{{Minniti} {et~al.}(2021{\natexlab{b}}){Minniti}, {Ripepi},
  {Fern{\'a}ndez-Trincado}, {Alonso-Garc{\'\i}a}, {Smith}, {Lucas},
  {G{\'o}mez}, {Pullen}, {Garro}, {Vivanco C{\'a}diz}, {Hempel}, {Rejkuba},
  {Saito}, {Palma}, {Clari{\'a}}, {Gregg}, \& {Majaess}}]{Minniti2021}
{Minniti}, D., {Ripepi}, V., {Fern{\'a}ndez-Trincado}, J.~G., {et~al.}
  2021{\natexlab{b}}, \aap, 647, L4

\bibitem[{{Mu{\~n}oz} {et~al.}(2012){Mu{\~n}oz}, {Geha}, {C{\^o}t{\'e}},
  {Vargas}, {Santana}, {Stetson}, {Simon}, \& {Djorgovski}}]{Munoz2012}
{Mu{\~n}oz}, R.~R., {Geha}, M., {C{\^o}t{\'e}}, P., {et~al.} 2012, \apjl, 753,
  L15

\bibitem[{{Obasi} {et~al.}(2021){Obasi}, {G{\'o}mez}, {Minniti}, \&
  {Alonso-Garc{\'\i}a}}]{Obasi2021}
{Obasi}, C., {G{\'o}mez}, M., {Minniti}, D., \& {Alonso-Garc{\'\i}a}, J. 2021,
  \aap, 654, A39

\bibitem[{{Palma} {et~al.}(2019){Palma}, {Minniti}, {Alonso-Garc{\'\i}a},
  {Crestani}, {Netzel}, {Clari{\'a}}, {Saito}, {Dias},
  {Fern{\'a}ndez-Trincado}, {Kammers}, {Geisler}, {G{\'o}mez}, {Hempel}, \&
  {Pullen}}]{Palma2019}
{Palma}, T., {Minniti}, D., {Alonso-Garc{\'\i}a}, J., {et~al.} 2019, \mnras,
  487, 3140

\bibitem[{{Perren} {et~al.}(2022){Perren}, {Pera}, {Navone}, \&
  {V{\'a}zquez}}]{Perren22}
{Perren}, G.~I., {Pera}, M.~S., {Navone}, H.~D., \& {V{\'a}zquez}, R.~A. 2022,
  \aap, 663, A131

\bibitem[{{Perryman} {et~al.}(2001){Perryman}, {de Boer}, {Gilmore}, {H{\o}g},
  {Lattanzi}, {Lindegren}, {Luri}, {Mignard}, {Pace}, \& {de
  Zeeuw}}]{Perryman01}
{Perryman}, M.~A.~C., {de Boer}, K.~S., {Gilmore}, G., {et~al.} 2001, \aap,
  369, 339

\bibitem[{{Piatti} {et~al.}(2023){Piatti}, {Illesca}, {Massara}, {Chiarpotti},
  {Rold{\'a}n}, {Mor{\'o}n}, \& {Bazzoni}}]{Piatti23}
{Piatti}, A.~E., {Illesca}, D. M.~F., {Massara}, A.~A., {et~al.} 2023, \mnras,
  518, 6216

\bibitem[{{Qin} {et~al.}(2023){Qin}, {Zhong}, {Tang}, \& {Chen}}]{qin23}
{Qin}, S., {Zhong}, J., {Tang}, T., \& {Chen}, L. 2023, \apjs, 265, 12

\bibitem[{{Qin} {et~al.}(2021){Qin}, {Li}, {Chen}, \& {Zhong}}]{Qin20}
{Qin}, S.-M., {Li}, J., {Chen}, L., \& {Zhong}, J. 2021, Research in Astronomy
  and Astrophysics, 21, 045

\bibitem[{{Riello} {et~al.}(2021){Riello}, {De Angeli}, {Evans}, {Montegriffo},
  {Carrasco}, {Busso}, {Palaversa}, {Burgess}, {Diener}, {Davidson}, {Rowell},
  {Fabricius}, {Jordi}, {Bellazzini}, {Pancino}, {Harrison}, {Cacciari}, {van
  Leeuwen}, {Hambly}, {Hodgkin}, {Osborne}, {Altavilla}, {Barstow}, {Brown},
  {Castellani}, {Cowell}, {De Luise}, {Gilmore}, {Giuffrida}, {Hidalgo},
  {Holland}, {Marinoni}, {Pagani}, {Piersimoni}, {Pulone}, {Ragaini}, {Rainer},
  {Richards}, {Sanna}, {Walton}, {Weiler}, \& {Yoldas}}]{Riello21}
{Riello}, M., {De Angeli}, F., {Evans}, D.~W., {et~al.} 2021, \aap, 649, A3

\bibitem[{{Ryu} \& {Lee}(2018)}]{Ryu2018}
{Ryu}, J., \& {Lee}, M.~G. 2018, \apjl, 863, L38

\bibitem[{{Salaris} {et~al.}(2004){Salaris}, {Weiss}, \&
  {Percival}}]{Salaris04}
{Salaris}, M., {Weiss}, A., \& {Percival}, S.~M. 2004, \aap, 414, 163

\bibitem[{{Sim} {et~al.}(2019){Sim}, {Lee}, {Ann}, \& {Kim}}]{Sim19}
{Sim}, G., {Lee}, S.~H., {Ann}, H.~B., \& {Kim}, S. 2019, Journal of Korean
  Astronomical Society, 52, 145

\bibitem[{{Swift}(1888)}]{Swift1888}
{Swift}, L. 1888, Astronomische Nachrichten, 120, 33

\bibitem[{{Vasiliev} \& {Baumgardt}(2021)}]{Vasiliev21}
{Vasiliev}, E., \& {Baumgardt}, H. 2021, \mnras, 505, 5978

\bibitem[{{Villanova} {et~al.}(2019){Villanova}, {Monaco}, {Geisler},
  {O'Connell}, {Minniti}, {Assmann}, \& {Barb{\'a}}}]{Villanova2019}
{Villanova}, S., {Monaco}, L., {Geisler}, D., {et~al.} 2019, \apj, 882, 174

\bibitem[{{Willman} {et~al.}(2005){Willman}, {Blanton}, {West}, {Dalcanton},
  {Hogg}, {Schneider}, {Wherry}, {Yanny}, \& {Brinkmann}}]{Willman2005}
{Willman}, B., {Blanton}, M.~R., {West}, A.~A., {et~al.} 2005, \aj, 129, 2692

\end{thebibliography}

\end{CJK*}

\end{document}